\RequirePackage{lineno}
\documentclass[aps,prc,twocolumn,superscriptaddress,showpacs,floatfix]{revtex4} 
\usepackage{color}
\usepackage{subfigure}
\usepackage{amsmath,epsfig}
\usepackage{graphicx}
\usepackage{epstopdf}
\usepackage{dcolumn}
\usepackage{bm}
\usepackage{multirow}
\usepackage{hhline}
\usepackage{afterpage}
\usepackage{hyperref} 

\def	\Tkin 	{T_{\rm kin}}
\def	\sNN 	{\sqrt{s_{NN}}}

\begin{document}


\title{Bulk Properties of the Medium Produced in Relativistic Heavy-Ion Collisions from the Beam Energy Scan Program}
\affiliation{AGH University of Science and Technology, FPACS, Cracow 30-059, Poland}
\affiliation{Argonne National Laboratory, Argonne, Illinois 60439}
\affiliation{Brookhaven National Laboratory, Upton, New York 11973}
\affiliation{University of California, Berkeley, California 94720}
\affiliation{University of California, Davis, California 95616}
\affiliation{University of California, Los Angeles, California 90095}
\affiliation{Central China Normal University, Wuhan, Hubei 430079}
\affiliation{University of Illinois at Chicago, Chicago, Illinois 60607}
\affiliation{Creighton University, Omaha, Nebraska 68178}
\affiliation{Czech Technical University in Prague, FNSPE, Prague, 115 19, Czech Republic}
\affiliation{Nuclear Physics Institute AS CR, 250 68 Prague, Czech Republic}
\affiliation{Frankfurt Institute for Advanced Studies FIAS, Frankfurt 60438, Germany}
\affiliation{Institute of Physics, Bhubaneswar 751005, India}
\affiliation{Indiana University, Bloomington, Indiana 47408}
\affiliation{Alikhanov Institute for Theoretical and Experimental Physics, Moscow 117218, Russia}
\affiliation{University of Jammu, Jammu 180001, India}
\affiliation{Joint Institute for Nuclear Research, Dubna, 141 980, Russia}
\affiliation{Kent State University, Kent, Ohio 44242}
\affiliation{University of Kentucky, Lexington, Kentucky, 40506-0055}
\affiliation{Lamar University, Physics Department, Beaumont, Texas 77710}
\affiliation{Institute of Modern Physics, Chinese Academy of Sciences, Lanzhou, Gansu 730000}
\affiliation{Lawrence Berkeley National Laboratory, Berkeley, California 94720}
\affiliation{Lehigh University, Bethlehem, PA, 18015}
\affiliation{Max-Planck-Institut fur Physik, Munich 80805, Germany}
\affiliation{Michigan State University, East Lansing, Michigan 48824}
\affiliation{National Research Nuclear University MEPhI, Moscow 115409, Russia}
\affiliation{National Institute of Science Education and Research, Bhubaneswar 751005, India}
\affiliation{National Cheng Kung University, Tainan 70101 }
\affiliation{Ohio State University, Columbus, Ohio 43210}
\affiliation{Institute of Nuclear Physics PAN, Cracow 31-342, Poland}
\affiliation{Panjab University, Chandigarh 160014, India}
\affiliation{Pennsylvania State University, University Park, Pennsylvania 16802}
\affiliation{Institute of High Energy Physics, Protvino 142281, Russia}
\affiliation{Purdue University, West Lafayette, Indiana 47907}
\affiliation{Pusan National University, Pusan 46241, Korea}
\affiliation{Rice University, Houston, Texas 77251}
\affiliation{University of Science and Technology of China, Hefei, Anhui 230026}
\affiliation{Shandong University, Jinan, Shandong 250100}
\affiliation{Shanghai Institute of Applied Physics, Chinese Academy of Sciences, Shanghai 201800}
\affiliation{State University Of New York, Stony Brook, NY 11794}
\affiliation{Temple University, Philadelphia, Pennsylvania 19122}
\affiliation{Texas A\&M University, College Station, Texas 77843}
\affiliation{University of Texas, Austin, Texas 78712}
\affiliation{University of Houston, Houston, Texas 77204}
\affiliation{Tsinghua University, Beijing 100084}
\affiliation{University of Tsukuba, Tsukuba, Ibaraki, Japan,}
\affiliation{Southern Connecticut State University, New Haven, CT, 06515}
\affiliation{United States Naval Academy, Annapolis, Maryland, 21402}
\affiliation{Valparaiso University, Valparaiso, Indiana 46383}
\affiliation{Variable Energy Cyclotron Centre, Kolkata 700064, India}
\affiliation{Warsaw University of Technology, Warsaw 00-661, Poland}
\affiliation{Wayne State University, Detroit, Michigan 48201}
\affiliation{World Laboratory for Cosmology and Particle Physics (WLCAPP), Cairo 11571, Egypt}
\affiliation{Yale University, New Haven, Connecticut 06520}

\author{L.~Adamczyk}\affiliation{AGH University of Science and Technology, FPACS, Cracow 30-059, Poland}
\author{J.~K.~Adkins}\affiliation{University of Kentucky, Lexington, Kentucky, 40506-0055}
\author{G.~Agakishiev}\affiliation{Joint Institute for Nuclear Research, Dubna, 141 980, Russia}
\author{M.~M.~Aggarwal}\affiliation{Panjab University, Chandigarh 160014, India}
\author{Z.~Ahammed}\affiliation{Variable Energy Cyclotron Centre, Kolkata 700064, India}
\author{N.~N.~Ajitanand}\affiliation{State University Of New York, Stony Brook, NY 11794}
\author{I.~Alekseev}\affiliation{Alikhanov Institute for Theoretical and Experimental Physics, Moscow 117218, Russia}\affiliation{National Research Nuclear University MEPhI, Moscow 115409, Russia}
\author{D.~M.~Anderson}\affiliation{Texas A\&M University, College Station, Texas 77843}
\author{R.~Aoyama}\affiliation{University of Tsukuba, Tsukuba, Ibaraki, Japan,}
\author{A.~Aparin}\affiliation{Joint Institute for Nuclear Research, Dubna, 141 980, Russia}
% ...

\collaboration{STAR Collaboration}\noaffiliation


\begin{abstract}
We present measurements of bulk properties of the matter produced 
in Au+Au collisions at $\sqrt{s_{NN}}=$ 7.7, 11.5, 19.6, 27, and 39
GeV using identified hadrons ($\pi^\pm$, $K^\pm$, $p$ and $\bar{p}$) 
from the STAR experiment
in the Beam Energy Scan (BES) Program at the Relativistic Heavy Ion Collider (RHIC). 
Midrapidity ($|y|<$0.1) results for multiplicity densities $dN/dy$,
average transverse momenta $\langle p_T \rangle$ and particle ratios are
presented. The chemical and kinetic freeze-out dynamics at these energies are
discussed and presented as a function of collision
centrality and energy. 
These results constitute the systematic measurements of bulk
properties of matter formed in heavy-ion collisions over a broad range
of energy (or baryon chemical
potential) at RHIC. 
\end{abstract}

\maketitle

\section{Introduction}

\begin{figure}[htb]
\begin{center}
\includegraphics[width=8.cm]{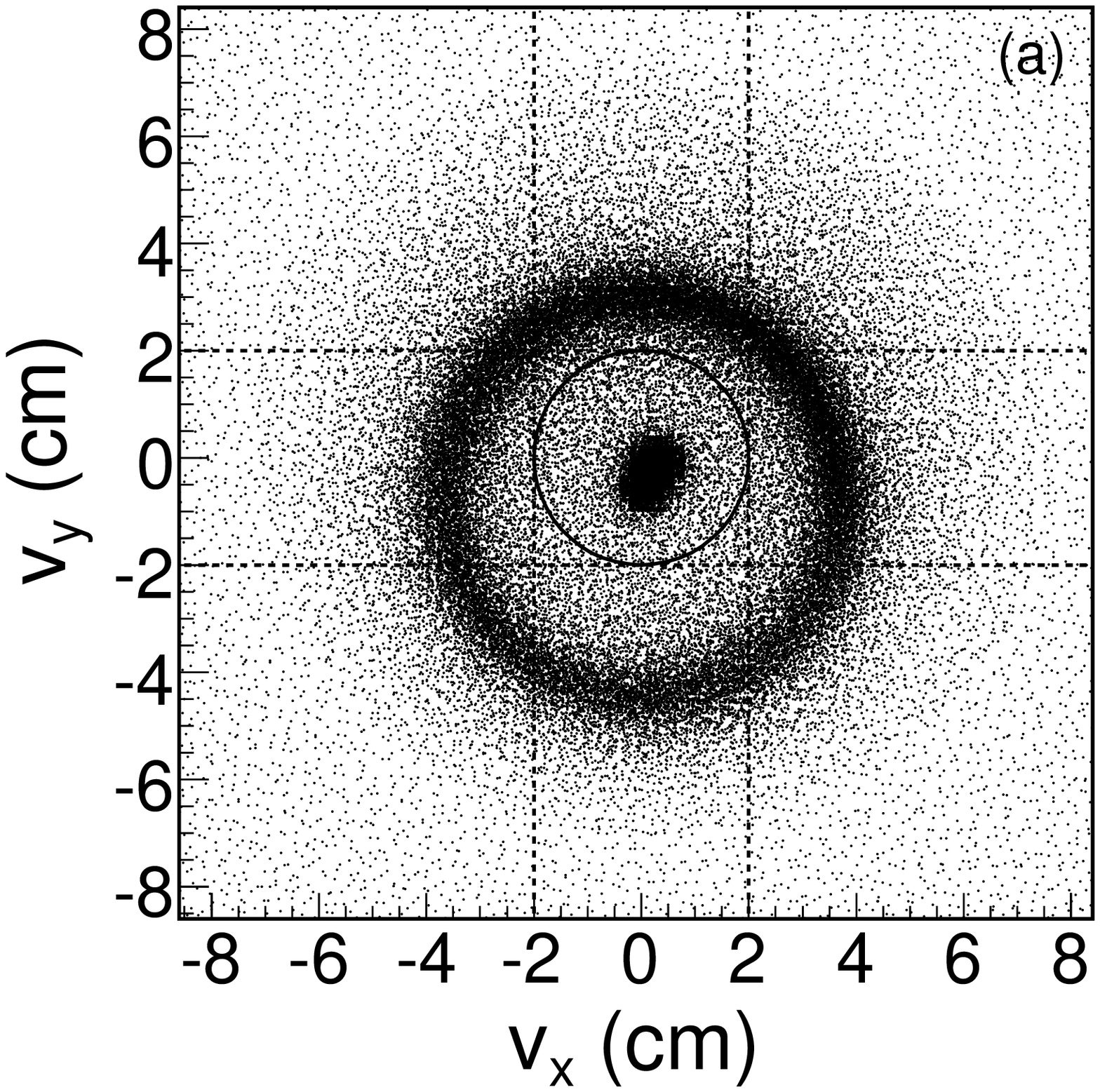}
\includegraphics[width=8.cm]{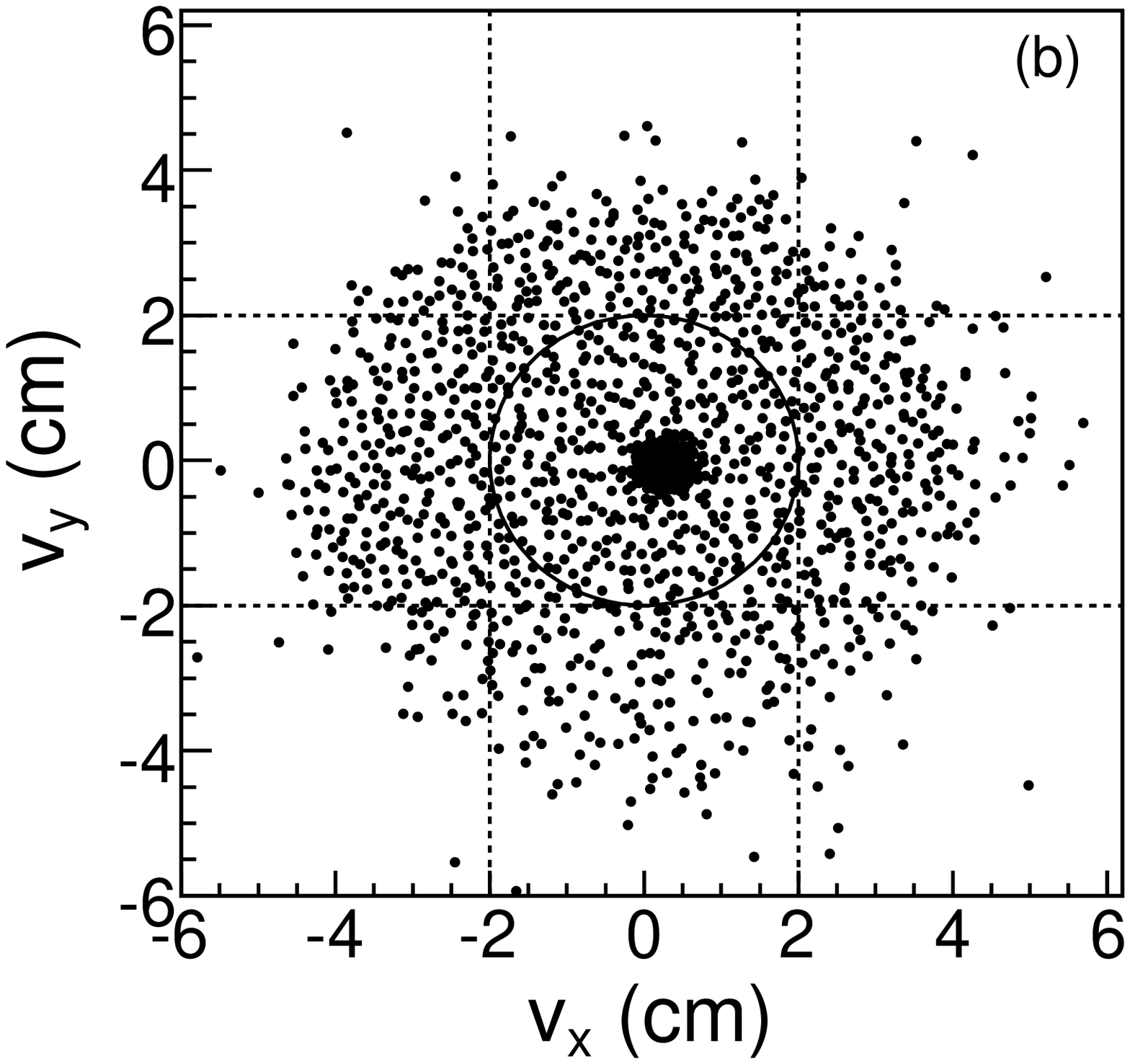}
\vspace{-0.5cm}
\caption{ 
The $x$ and $y$ positions of the reconstructed event vertices in Au+Au collisions 
at (a) $\sqrt{s_{NN}}$ = 7.7 GeV and (b)~$\sqrt{s_{NN}}$ = 39 GeV.
The events involving beam-pipe interactions are rejected by applying 
a cut of less than 2 cm on the transverse radial position of the event vertex. See text for more details.
}
\label{vrad_4}
\end{center}
\vspace{-0.5cm}
\end{figure}

Exploring the Quantum Chromodynamics (QCD) phase diagram is 
one important goal of high-energy heavy-ion collision 
experiments~\cite{Arsene:2004fa,Adcox:2004mh,Back:2004je,Adams:2005dq}.       
The QCD phase diagram is usually plotted as temperature ($T$) versus baryon chemical 
potential ($\mu_{\mathrm {B}}$). Assuming a thermalized system is created in heavy-ion 
collisions, both of these quantities can be varied by 
changing the collision
energy~\cite{Cleymans:1999st,Becattini:2005xt,Andronic:2005yp}. 
Theory suggests that the phase diagram includes
a possible transition from a high energy density and 
high temperature phase called Quark Gluon Plasma (QGP) phase, dominated by partonic degrees of freedom, 
to a phase where the relevant degrees of freedom are hadronic~\cite{Rajagopal:2000wf,Laermann:2003cv,Stephanov:2007fk}. 
Several observations at the top RHIC energy i.e. at $\sqrt{s_{NN}} =$ 200 GeV  
have been associated with the existence of a phase with partonic degrees of freedom 
in the early stages of heavy-ion 
collisions~\cite{Adams:2003im,Adams:2003kv,Arsene:2004fa,Adcox:2004mh,Back:2004je,Adams:2005dq,Abelev:2006jr,Abelev:2007ra,Abelev:2007rw,Abelev:2008ae}. 
Examples of such observations include the suppression of high transverse momentum 
($p_{T}$) hadron production in Au+Au collisions relative to scaled $p$+$p$ 
collisions~\cite{Adams:2003im,Adams:2003kv,Arsene:2004fa,Adcox:2004mh,Back:2004je,Adams:2005dq,Abelev:2006jr,Abelev:2007ra}, large elliptic flow ($v_{2}$) for hadrons 
with light, as well as heavier strange valence quarks, and differences between 
baryon and meson $v_{2}$ at intermediate $p_{T}$ in Au+Au collisions~\cite{Adams:2005zg}. 

Lattice QCD calculations indicate that 
a system produced at $\mu_{\mathrm {B}} =$ 0 MeV evolves through a rapid
crossover at the parton-hadron phase transition~\cite{Aoki:2006we,Cheng:2007jq}. Calculations 
from lattice QCD~\cite{Ejiri:2008xt} and from several QCD-based models~\cite{Asakawa:1989bq,Barducci:1989wi,Barducci:1989eu,Stephanov:2004wx} 
suggest that for a system created in collisions corresponding 
to larger values of $\mu_{\mathrm {B}}$,
the transition is first order. The point in the ($T$, $\mu_{\mathrm {B}}$)
plane where the first order phase transition ends, is the QCD critical point~\cite{Fodor:2004nz,Gavai:2008zr}. 

Searching for the critical point and phase boundary in the QCD
phase diagram is currently a focus of experimental and theoretical nuclear
physics research.
To this end, RHIC has undertaken the first phase of
the BES Program~\cite{Abelev:2009bw,Mohanty:2009vb,Aggarwal:2010cw,Kumar:2012fb,Kumar:2013cqa}.
The idea is to vary
the collision energy, thereby scanning
the phase diagram from the top
RHIC energy (lower $\mu_B$) to the lowest possible energy (higher
$\mu_B$), to look for the signatures of the QCD phase boundary
and the QCD critical point. 
To look for the phase boundary, we study the established
signatures of the QGP formation at 200 GeV as a function of beam energy. Turn-off of
these signatures at a particular energy would suggest that a partonic
medium is no longer formed at that energy. 
Near the critical point, there would be enhanced
fluctuations 
in multiplicity distributions of conserved quantities (net-charge,
net-baryon number, and net-strangeness)~\cite{Stephanov:2008qz,Aggarwal:2010wy,Karsch:2010ck,Gupta:2011wh}. 
These observables would suggest the existence of a critical point if they were to show
large fluctuations or divergence from a baseline in a limited
collision energy region. 

However, before looking for these signatures, it is important to know
the ($T,\mu_B$) region of the phase diagram we can access. 
The spectra of produced particles and their yield ratios allow us to
infer the $T$ and $\mu_B$ values at freeze-out.  
In
addition, bulk properties such as $dN/dy$, $\langle p_T \rangle$,
particle ratios, and freeze-out properties may provide
insight into the particle production mechanisms at these energies. The systematic
study of these bulk properties may reveal the evolution and change in
behavior of the system formed in heavy-ion collisions as a function of collision energy.

\begin{figure}[htb]
\begin{center}
\includegraphics[width=8cm]{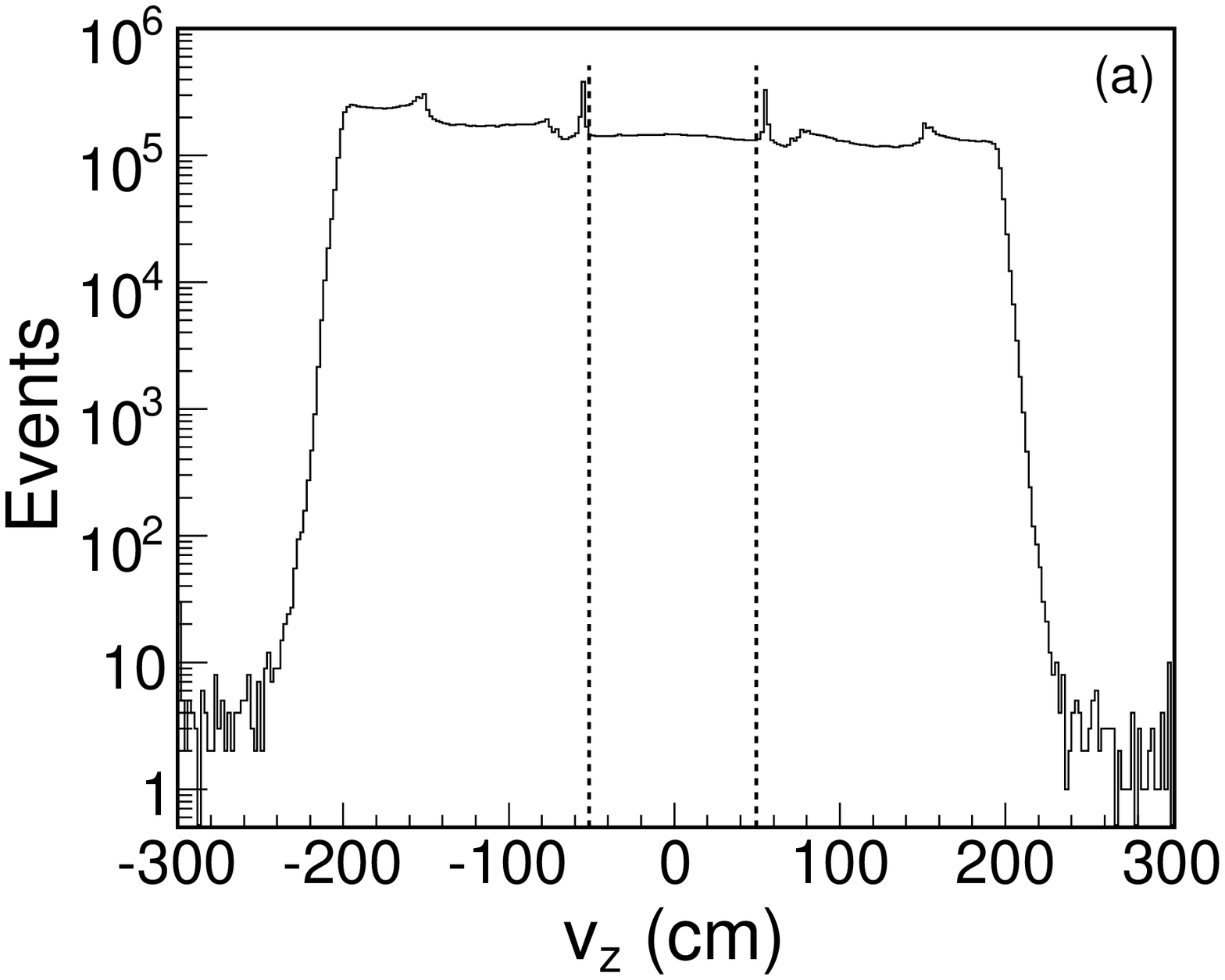}
\includegraphics[width=8cm]{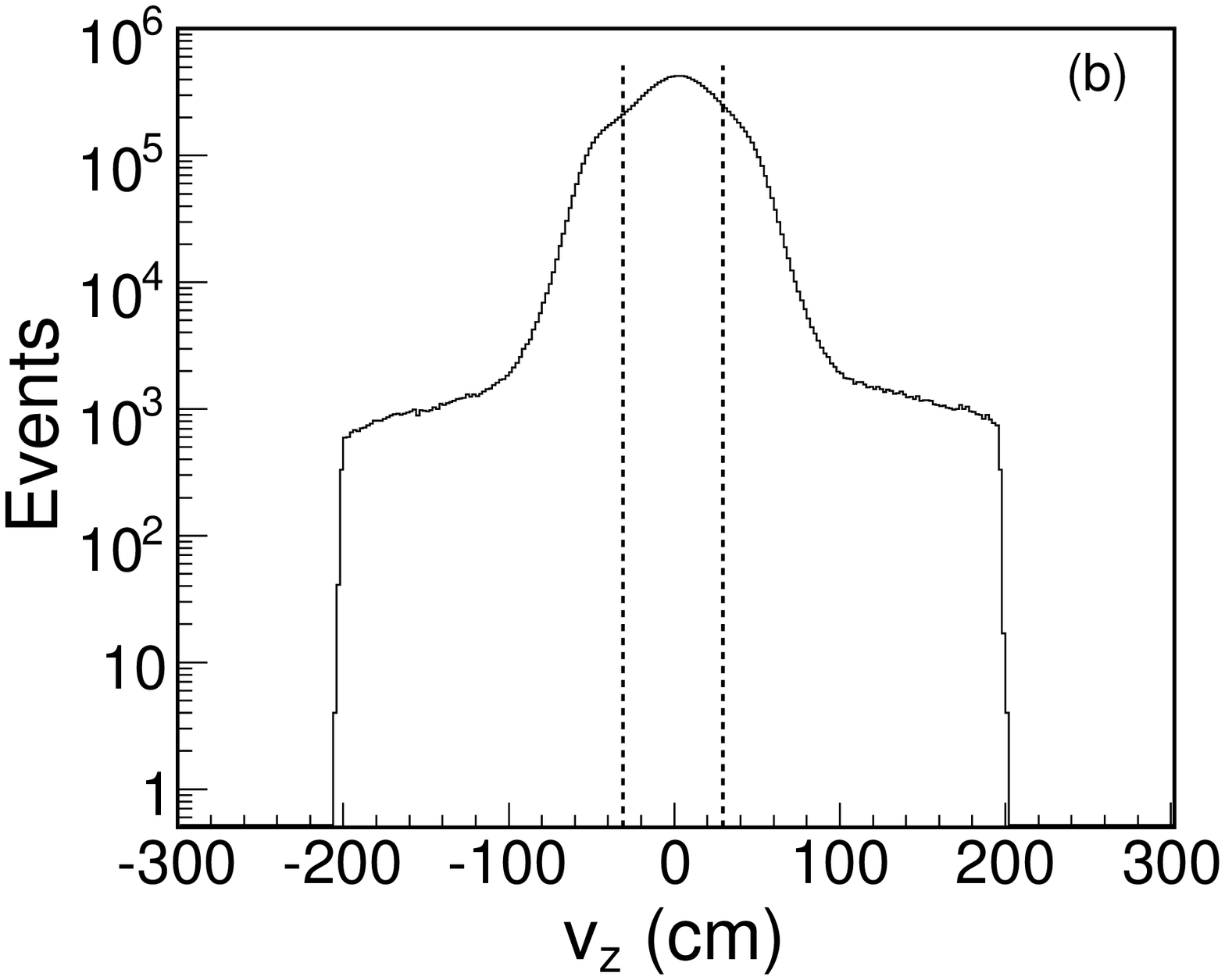}
\vspace{-0.5cm}
\caption{ 
Distributions of
the $z$-position of the reconstructed primary vertex ($V_{z}$) in Au+Au collisions 
at (a) $\sqrt{s_{NN}}$ = 7.7 and (b) $\sqrt{s_{NN}}$ = 39 GeV. 
}
\label{vz}
\end{center}
\vspace{-0.5cm}
\end{figure}

\section{Experiment and Data Analysis}
\subsection{STAR experiment}
The results presented here are based on data taken 
with the STAR experiment~\cite{Ackermann:2002ad} 
in Au+Au collisions at $\sqrt{s_{NN}}$ = 7.7, 11.5, 19.6, 27 and 39
GeV.  The 7.7, 11.5 and 39 GeV data were collected in the year 2010,
while the 19.6 and 27 GeV data were collected in the year 2011.
These data sets were taken with a minimum bias trigger, which 
was defined using a coincidence of hits in the
zero degree calorimeters (ZDCs)~\cite{Adler:2000bd}, vertex position detectors (VPDs)~\cite{Llope:2003ti}, and/or beam-beam
counters (BBCs)~\cite{Bieser:2002ah,2008AIPC..980..390W}. 

The main detectors used to obtain
the results on $p_{\mathrm {T}}$ spectra, yields and particle ratios
for charged hadrons are the Time Projection Chamber 
(TPC)~\cite{Anderson:2003ur} and Time-Of-Flight detectors (TOF)~\cite{Llope:2005yw}.
The TPC is the primary tracking device at STAR. 
It is 4.2 m long  and 4 m in diameter. 
It covers about $\pm1$ units of pseudorapidity ($\eta$) and the full azimuthal angle. 
The sensitive volume of the TPC contains P10 gas (10\% methane, 
90\% argon) regulated at 2 mbar  above atmospheric pressure. 
The TPC resides in a nearly constant magnetic field of 0.5 Tesla
oriented in the longitudinal ($z$) direction.
The TPC data is used to determine particle trajectories, thereby their momenta, and particle types through ionization energy 
loss ($dE/dx$). The TOF is based on Multi-gap Resistive Plate Chamber (MRPC)
technology  and is used to identify particles at relatively high momenta. 
The details of the design and other characteristics of the STAR detectors 
can be found in Ref.~\cite{Ackermann:2002ad}.

\subsection{Event Selection}
The primary vertex for each event is determined by finding the
most probable point of common origin of the tracks measured by the TPC. 
Figure~\ref{vrad_4} shows, as examples, the transverse $x,y$ positions of the primary vertices in 7.7 and 39 GeV Au+Au collisions.
In order to reject background events which involve interactions with
the beam pipe of radius 3.95 cm, 
the event vertex radius
(defined as $\sqrt{V_{x}^{2} + V_{y}^{2}}$ where 
$V_{x}$ and $V_{y}$ are the vertex positions along the $x$ and $y$ directions) 
is required to be  within 2 cm of the center of STAR (see Fig.~\ref{vrad_4}). The
ring in Fig.~\ref{vrad_4} (a) corresponds to
collisions between the beam nuclei and the beam pipe.
This type of background is more significant in low energy data.  

The distributions of the 
primary vertex position along the longitudinal (beam) direction ($V_{z}$) are
shown in Fig.~\ref{vz} for 7.7 and 39 GeV. The lower energy vertex distribution is flat near zero while that at 39
GeV is peaked.
The wide $z$-vertex distribution
at lower energies is due to the fact that the beams are more difficult to focus at
lower energies. 
The $V_z$ distributions for other BES energies are also flattened
relative to higher energies. 
Only those events which have a $V_{z}$ within 50 cm of the nominal collision 
point (center of the detector) are selected for the 7.7 GeV analysis,
while for the other data sets, events within 30 cm were selected
for the analysis. 
\begin{table}
\caption{Total number of events analyzed for various energies obtained
  after all the event selection cuts are applied.
\label{table_events}}
\begin{center}
\begin{tabular}{c|c}
\hline
$\sqrt{s_{NN}}$ (GeV) & No. of events (million) \\ 
\hline
7.7   &    4 \\ 
11.5 &    8 \\ 
19.6 &    17.3 \\
27    &    33 \\
39    &    111\\
\hline
\end{tabular}
\end{center}
\end{table}
These values are chosen in order to achieve uniform detector performance 
and sufficient statistical significance of the measured
observables. 
Table~\ref{table_events} shows the total number of events that are used
for the analysis at each energy after the above-mentioned event selection cuts.

\subsection{Centrality Selection}
Centralities in Au+Au collisions at $\sqrt{s_{NN}}$ = 7.7--39 GeV are defined 
by using the number of primary charged-particle tracks reconstructed in the TPC 
over the full azimuth and pseudorapidity $|\eta| < 0.5$. 
This is generally called the ``reference multiplicity'' in STAR. 
For each energy, a correction is applied to the standard 
definition by: removing bad 
runs, applying acceptance and efficiency corrections to reference multiplicity for 
different $z$-vertex positions, and performing corrections
for trigger inefficiencies (only important for low reference multiplicity events)
for different $z$-vertices. 

The centrality classes are obtained as fractions of the reference
multiplicity distribution. The events are divided into
following centrality classes 0--5\%, 5--10\%, 10--20\%, 20--30\%, 30--40\%, 40--50\%,
50--60\%, 60--70\%, and 70--80\%.
The mean values of  the number of participating nucleons $\langle N_{\rm
  part} \rangle$ 
corresponding to these centrality classes are evaluated  using a Glauber model and are given in
Table~\ref{table_nparts} for various energies. More details on centrality 
and $\langle N_{\rm part} \rangle$ values estimations can be found in Refs.~\cite{Abelev:2008ab,Abelev:2009bw}.
\begin{table}
\caption{\label{table_nparts}
The average number of participating nucleons ($\langle N_{\rm{part}} \rangle$) 
for various collision centralities in Au+Au collisions at
$\sqrt{s_{NN}}$ = 7.7--39 GeV. The numbers in parentheses represent
the uncertainties.}
\begin{center}
\begin{tabular}{l|l|l|l|l|l}
\hline
 & \multicolumn{5}{l} {~~~~~~~~~~~~~~~~~~~~~~~~~~~~$\langle N_{\rm {part}} \rangle$} \\
\cline{2-6} \\  [-0.3cm]
\% cross &~~7.7 & ~11.5 & ~19.6 &~~27   & ~~39\\ 
 section  &~ GeV& ~GeV & ~GeV  &~GeV & ~GeV\\ [0.15cm]
\hline
0-5      &  337 (2)	& 338  (2)     &  338 (2)     &  343 (2)  &   342 (2)	    \\ 
5-10    &  290 (6)	& 291 (6)	    &  289 (6)      &  299 (6)  &    294 (6)     \\ 
10-20  &  226 (8)	& 226 (8)	    &  225 (9)      &  234 (9)   &  230 (9)	   \\ 
20-30  &  160 (10)   & 160 (9)   &  158 (10)    &  166 (11) &  162 (10)	  \\
30-40  &  110 (11)   & 110 (10)   &  108 (11)    &  114 (11) &  111 (11)	\\
40-50  &  72 (10)	& 73 (10)	    &  71 (10)      &  75 (10)   &  74 (10)	   \\
50-60  &    45 (9)	& 45 (9)	    &  44 (9)        &  47 (9)     &  46 (9)	   \\
60-70  &    26 (7)	& 26 (7)	    &  26 (7)        &  27 (8)      & 26 (7)	   \\
70-80  &    14 (4)	& 14 (6)	    &  14 (5)        &  14 (6)      & 14 (5)   \\ 
\hline 
\end{tabular}
\end{center}
\end{table}

\begin{table}
\begin{center}
\caption{ Track selection criteria at all energies.
\label{table_tcuts}}
\vspace{.2cm}
\begin{center}
\begin{tabular}{c|c|c|c|c}
\hline
$|y|$       & DCA       & No. of &  No. of fit points & No. of \\
       &        & fit points &  $\overline{\rm{No.~of~possible~hits}}$ & $dE/dx$ points\\
\hline
$<$ 0.1  &  $\le$ 3 cm  & $\ge$ 25      & $\ge$ 0.52           &  $\ge$ 15  \\
\hline
\end{tabular}
\end{center}
\end{center}
\end{table}

\begin{figure}[htb]
\begin{center}
\includegraphics[width=9.5cm]{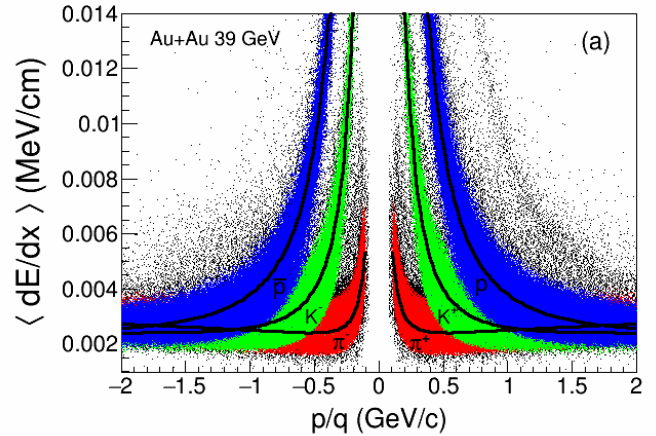}
\includegraphics[width=9.5cm]{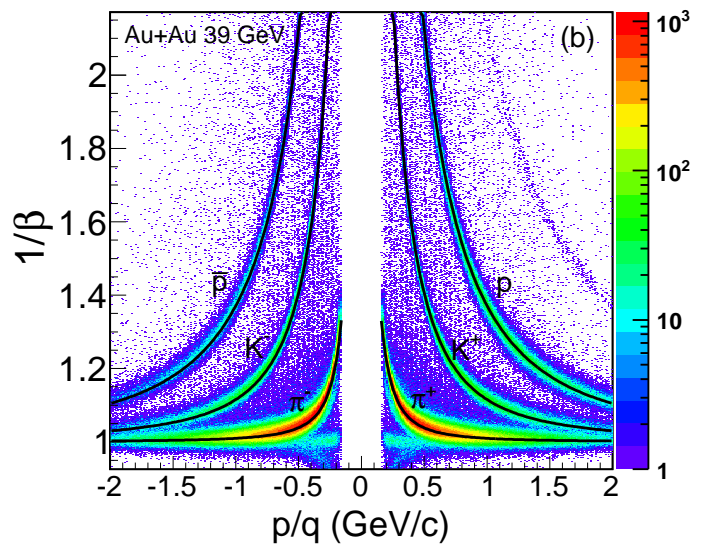}
\vspace{-0.5cm}
\caption{(Color online) (a)
The $\langle dE/dx \rangle$ of charged tracks at midrapidity ($|y|<0.1$) plotted as function of
rigidity $(p/q)$ in Au+Au collisions at $\sqrt{s_{NN}} =$ 39 GeV. 
The various bands correspond to different particles such as $\pi^\pm$,
$K^\pm$, $p$ and $\bar{p}$.
The curves represent the
Bichsel~\cite{Bichsel:2006cs} expectation values of the corresponding particles. 
(b) 1/$\beta$ from TOF vs. rigidity at same energy. The
curves, from low to up, show the expected mean values of pions, kaons,
and (anti-) protons, respectively.
}
\label{dedx}
\end{center}
\end{figure}
\begin{figure*}[htb]
\begin{center}
\includegraphics[width=15.cm]{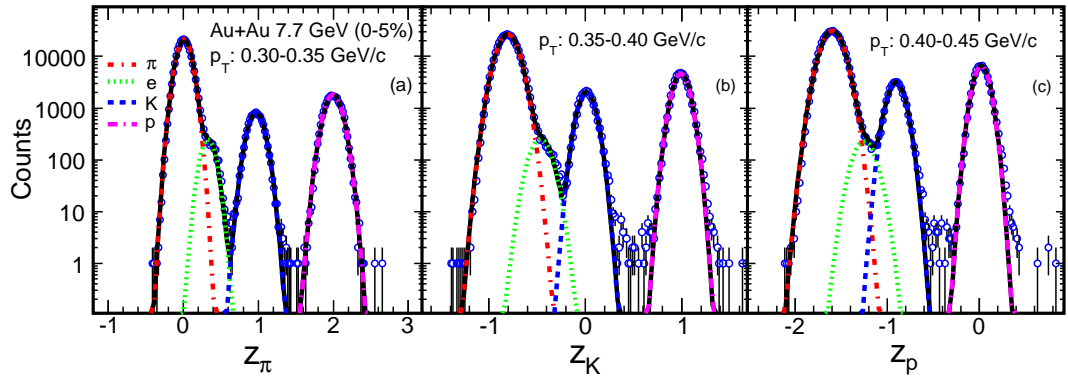}
\vspace{-0.8cm}
\caption{ (Color online) 
The $z_{\pi}$, $z_{K}$, and $z_{p}$ distributions for positively
charged hadrons ($\pi$, $K$, and $p$) at midrapidity ($|y|<0.1$) in the TPC for various $p_T$ ranges in Au+Au collisions at $\sqrt{s_{NN}}$ = 7.7 GeV. The 
curves are Gaussian fits representing contributions from pions
(dash-dotted, red), electrons (dotted, green), kaons (dashed, blue),
and protons (long dash-dotted, magenta). 
Uncertainties are statistical only.
}
\label{nsigma}
\end{center}
\end{figure*}
\begin{figure*}[htb]
\begin{center}
\includegraphics[width=15.cm]{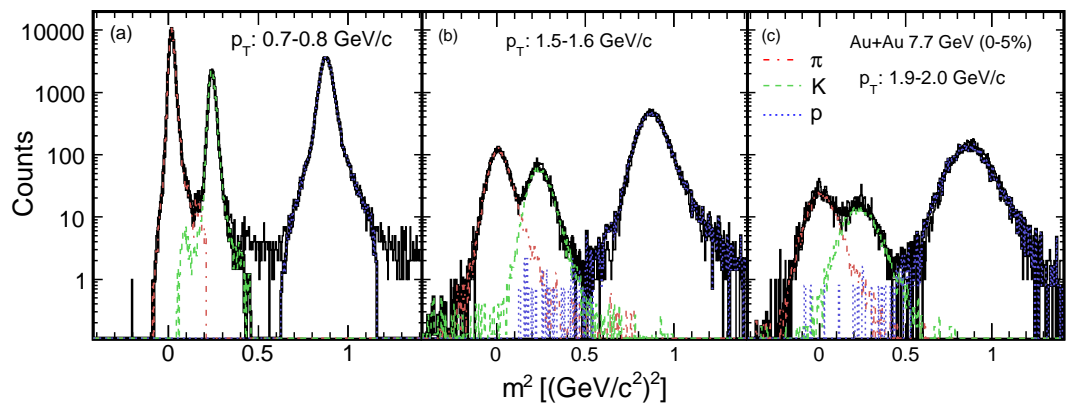}
\vspace{-0.8cm}
\caption{(Color online)
The $m^{2}$ distributions for positively charged hadrons used to extract
raw yields for pions, kaons, and protons in $|y|<$0.1 for Au+Au collisions at 7.7 GeV at 
three different $p_T$ ranges. The curves are predicted $m^2$ fits
representing contributions from pions (dash-dotted, red), 
kaons (dashed, green), and protons (dotted, blue).
}
\label{fig_m2}
\end{center}
\end{figure*}

\subsection{Track Selection}
Track selection criteria for  all analyses are presented in Table~\ref{table_tcuts}.
In order to suppress admixture of tracks from secondary vertices, a
requirement of less than 3 cm is placed on
the distance of closest approach (DCA) between each track and the
event vertex. 
Tracks must have at least 25 points  used in track fitting out of
the maximum of 45 hits possible in the TPC. 
To prevent multiple counting of split tracks, at least 52\% of the
total possible fit points are required.  
This is a standard cut in STAR analysis, but does not impose further
cut beyond the stricter cut of 25 points implemented for track fitting used here.
A condition
is placed on the number of $dE/dx$ points used to derive $dE/dx$
values. The results presented here are within rapidity
$|y|<$0.1
and have the same track cuts for all energies.

\begin{figure}
\begin{center}
\includegraphics[width=8cm]{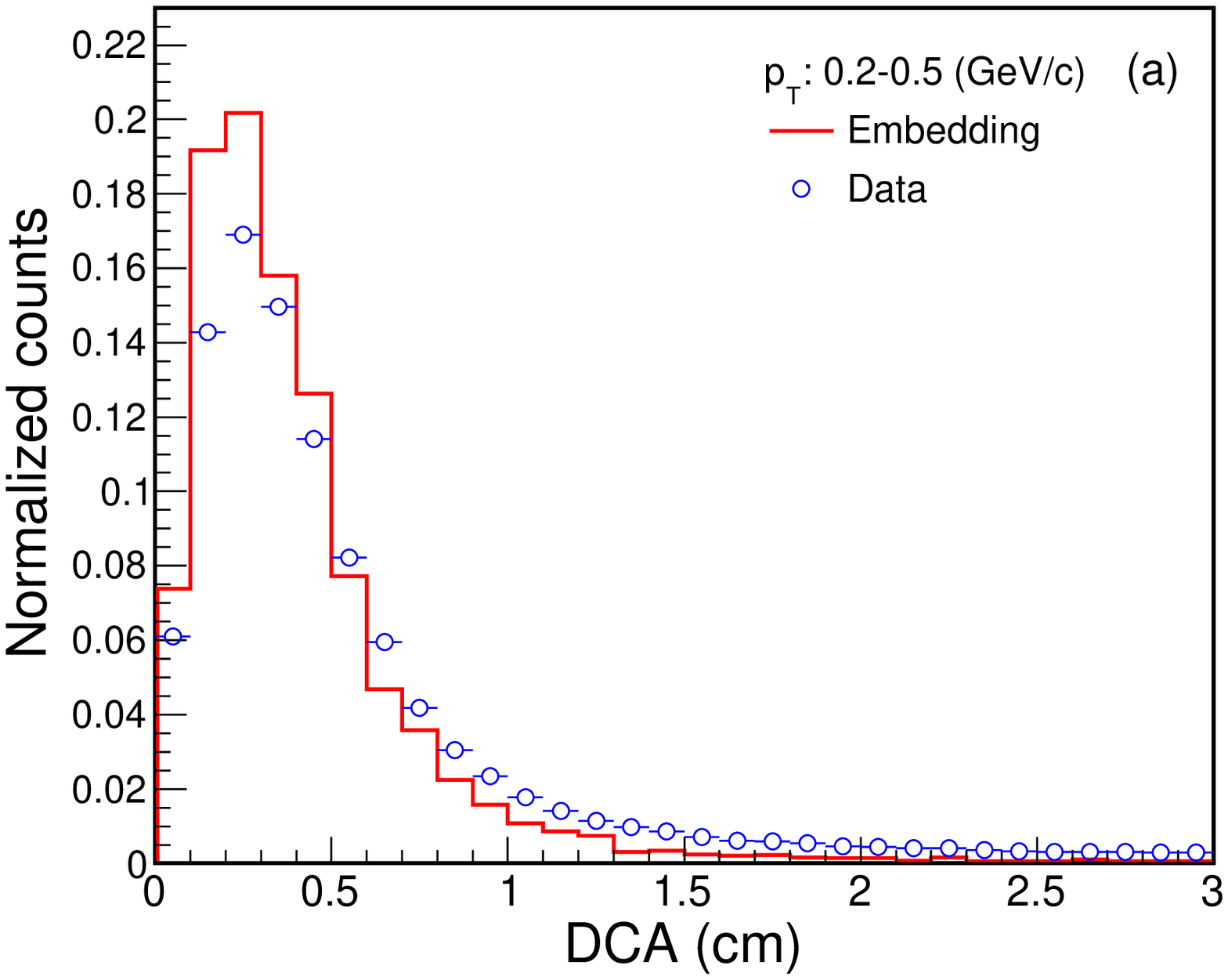}
\includegraphics[width=8cm]{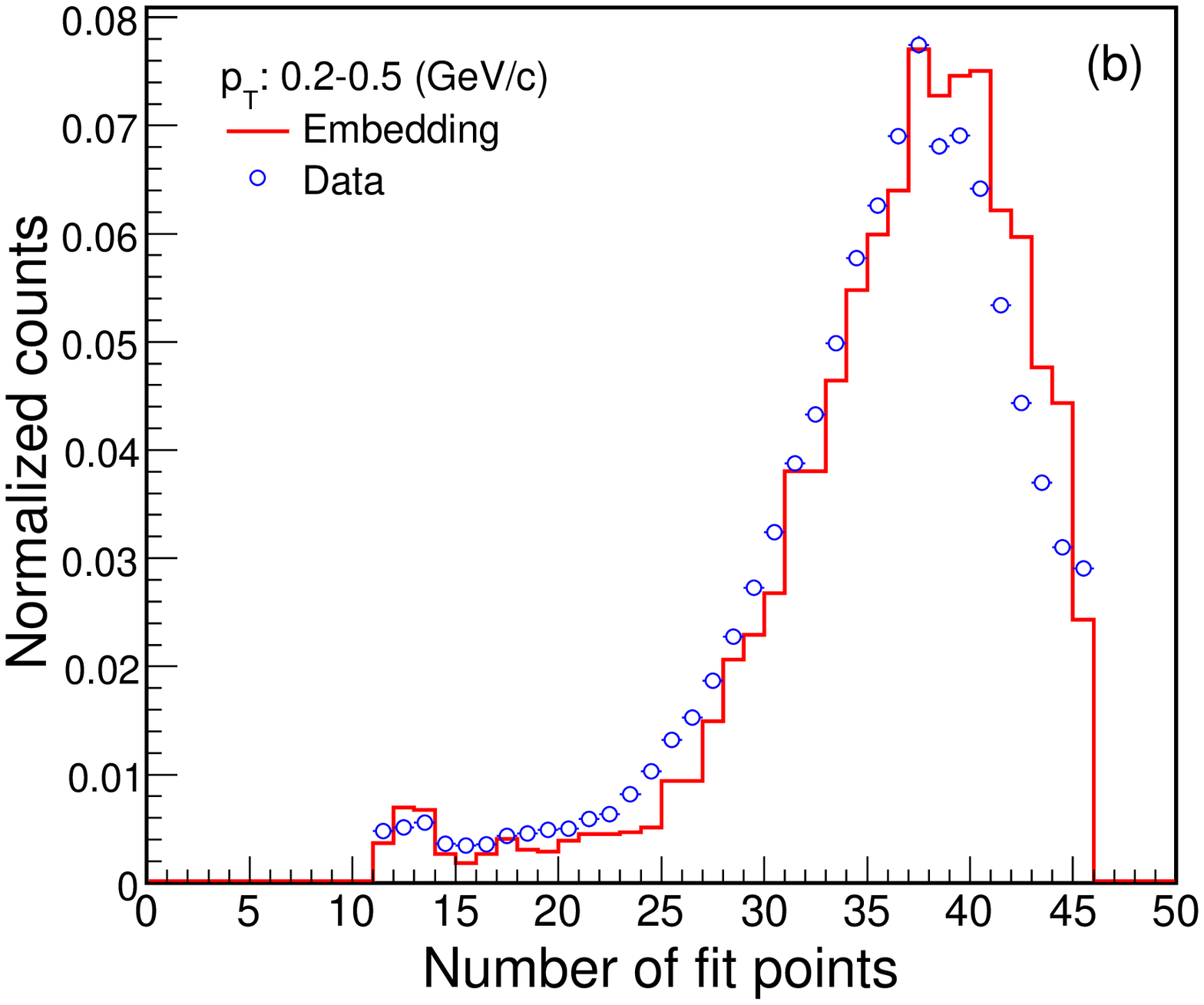}
\vspace{-0.5cm}
\caption{ (Color online)
(a) Distribution of distance of closest approach of pion
tracks to the primary vertex. 
The embedded tracks are compared to the ones in 
real data at 0.2 $< p_{T} < 0.5$ GeV/$c$ at midrapidity ($|y|<$0.1) in Au+Au collisions
at $\sqrt{s_{NN}}$ = 7.7 GeV. 
(b) Comparison between the distributions 
of number of fit points for pions from embedding and from real data for  
0.2 $< p_{T} < 0.5$ GeV/$c$ at midrapidity in Au+Au collisions
at $\sqrt{s_{NN}}$ = 7.7 GeV. 
}
\label{dca}
\end{center}
\end{figure}

\begin{figure*}
\begin{center}
\includegraphics[width=15cm]{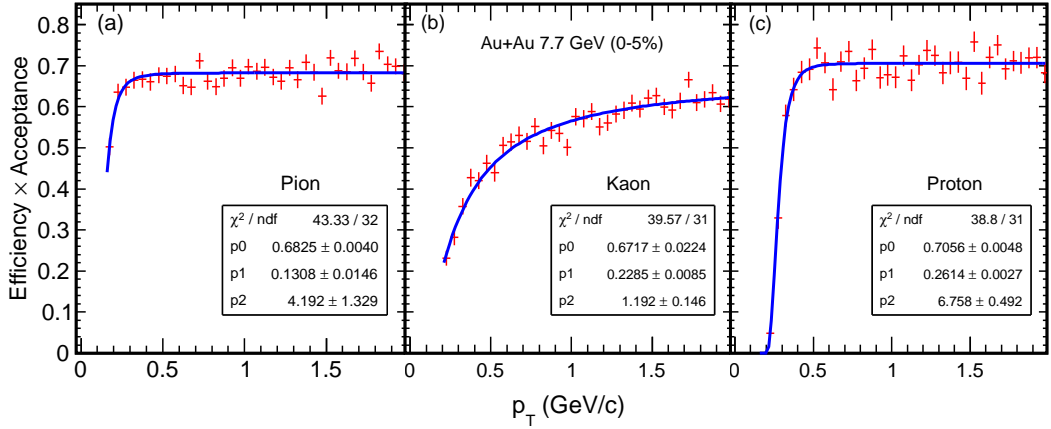}
\vspace{-0.8cm}
\caption{ (Color online)
Efficiency $\times$ acceptance obtained from Monte Carlo embedding
for reconstructed (a) pions, (b) kaons, and (c) protons in
the TPC as a function of $p_{T}$ at midrapidity ($|y|<$0.1) for 
0--5\% Au+Au collisions at $\sqrt{s_{NN}}$ = 7.7 GeV. The curves
represent the functional fit of the form $p_0\exp[-(p_1/p_T)^{p_2}]$.
}
\label{eff}
\end{center}
\end{figure*}
\begin{figure*}
\begin{center}
\includegraphics[width=15cm]{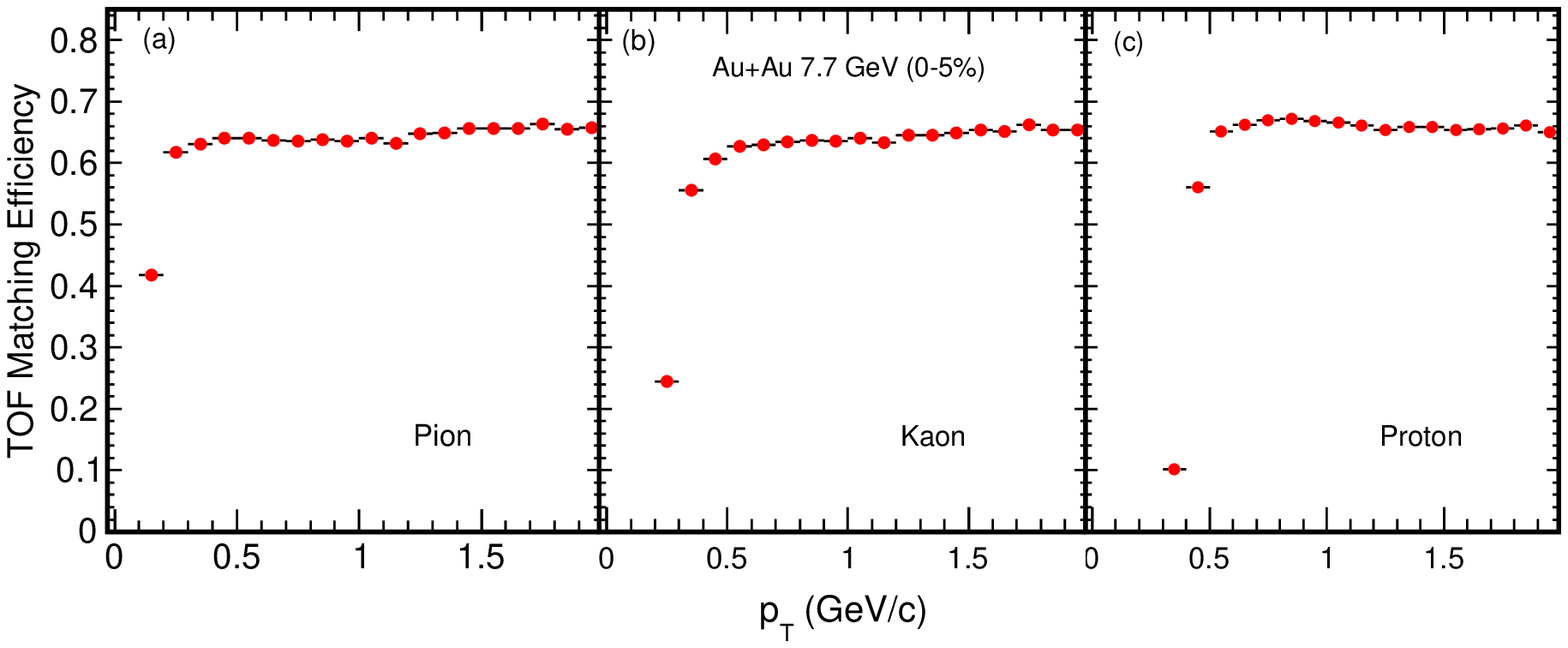}
\vspace{-0.8cm}
\caption{ (Color online)
TOF matching efficiency for (a) pions, (b) kaons, and
(c) protons as a function of $p_{T}$ at midrapidity ($|y|<$0.1)  for 0--5\% Au+Au
collisions at $\sqrt{s_{NN}}$ = 7.7 GeV. 
}
\label{fig_tofmatch}
\end{center}
\end{figure*}

\subsection{Particle Identification}

Particle identification is accomplished in the TPC by measuring the $dE/dx$. 
Figure~\ref{dedx} (a) shows the average $dE/dx$ of 
measured charged particles
plotted
as a function of ``rigidity'' (i.e. momentum/charge) of 
the particles. The curves represent the
Bichsel~\cite{Bichsel:2006cs} expectation values. 
 It can be seen that the TPC can identify pions ($\pi^\pm$),
kaons ($K^\pm$), and protons ($p$) and anti-protons ($\bar{p}$) at low
momentum as illustrated by the color bands. We note that the color bands are only used for illustration here.
The quantitative technique to extract particle yields is discussed in detail later.

For higher momentum, we use time-of-flight information to identify
particles. The TOF particle identification for this analysis is used above $p_T =$ 0.4 GeV/$c$. 
Figure~\ref{dedx} (b) shows
the inverse of particle velocity in unit of the speed of light
$1/\beta$, as a function of rigidity.
The expectation values for pions, kaons, and protons
are shown as the curves. 
As seen in the figure, there is a band representing
$1/\beta<$ 1 or $\beta>$1 at low momentum. This non-physical band 
is the result of a charged hadron and a photon converted electron hitting
in the same TOF cluster.  
The conversion may happen in the TPC Outer Field Cage or TOF tray box. 
Due to high occupancy, these TOF hits are accidentally matched to
hadron tracks in the TPC, resulting in the wrong time of flight. 
They have a negligible effect on charged hadron yields.

The $\langle dE/dx \rangle$ distribution for a fixed particle type
is not Gaussian~\cite{AguilarBenitez:1991yy}. 
It has been shown that a better Gaussian variable, for a given
 particle type, is the $z$-variable~\cite{AguilarBenitez:1991yy}, 
defined as
\begin{equation}
\label{eqnsigma}
z_{X}= \ln \left( \frac{\langle dE/dx \rangle}{\langle dE/dx \rangle_{X}^{B}} \right)
\end{equation}
where $X$
is the particle type ($e^{\pm},\pi^{\pm},K^{\pm}$, $p$, or $\bar{p}$)
and $\langle dE/dx \rangle_{X}^{B}$ is the corresponding 
Bichsel function
~\cite{Bichsel:2006cs}. 
The most probable value of $z_{X}$ for the particle $X$ is 0. 

\begin{figure*}
\begin{center}
\includegraphics[width=15cm]{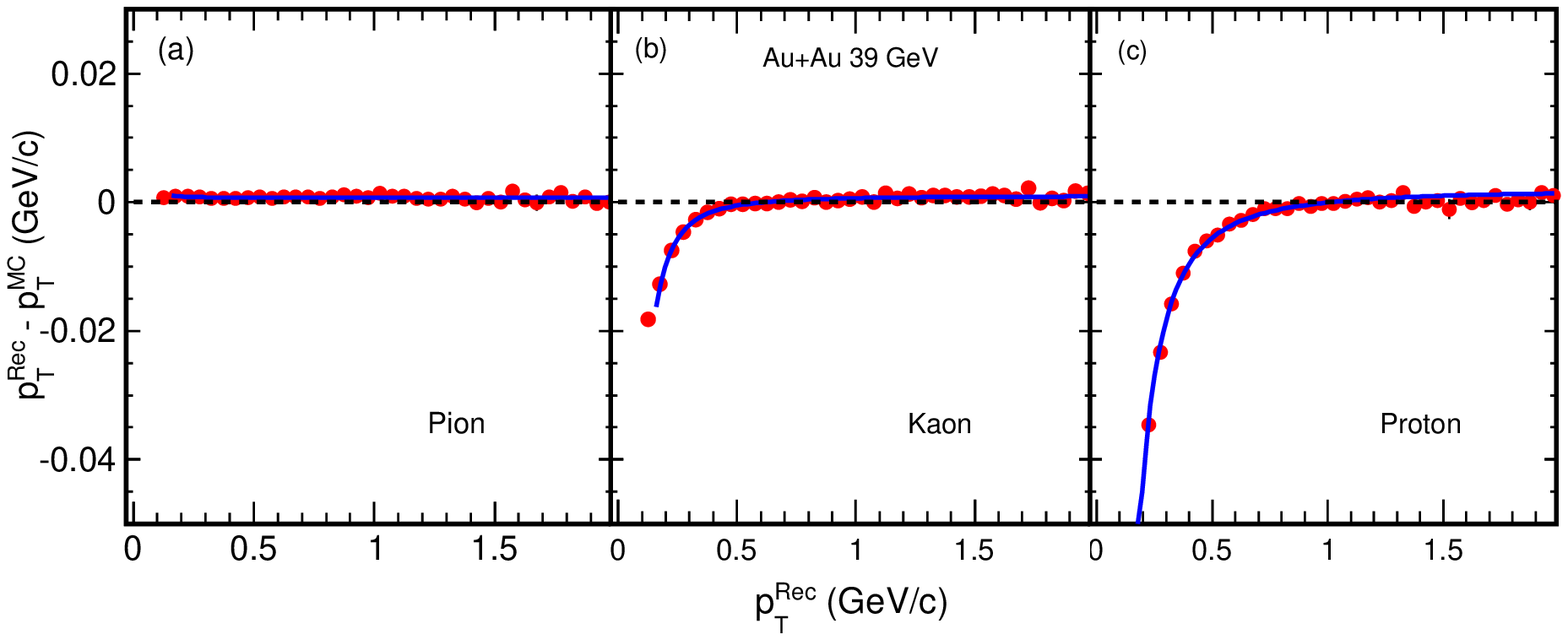}
\vspace{-0.8cm}
\caption{(Color online) The $p_{T}$ difference between reconstructed and embedded tracks
  plotted as a function of the $p_{T}$ of the reconstructed track for
  (a) pions, (b) kaons,
and (c) protons at midrapidity ($|y|<0.1$)  in Au+Au collisions at
$\sqrt{s_{NN}}=$ 39 GeV. This difference is due to particle energy
loss in the detector material, which is already corrected in the tracking algorithm for pions, but only partially for kaons and protons.
}
\label{eloss}
\end{center}
\vspace{-0.5cm}
\end{figure*}

The $z_{X}$ distribution is constructed for a given particle type in a given $p_T$ bin within $|y|<$ 0.1.
Figure~\ref{nsigma} shows the $z_{\pi}$, $z_{K}$, and $z_{p}$
distributions for positively charged particles at different $p_T$ bins
in central Au+Au collisions at $\sqrt{s_{NN}}=$ 7.7 GeV. 
To extract the raw yields in a given $p_{T}$ bin, a multi-Gaussian fit
is applied to the $z_{X}$ distributions as shown in Fig.~\ref{nsigma}.
The Gaussian area corresponding to the particle of interest
(i.e. the Gaussian with centroid at zero)
gives the yield
of that particle in the given $p_{T}$ bin. 
At low $p_T$, the peaks of pion, kaon, and proton distributions are
well separated. However, at higher $p_T$ these distributions start to
overlap. In the overlap $p_T$ region, the sigma of the Gaussian fits 
are constrained by the values from the lower $p_T$ bins. 
Further details on extraction of raw yields for identified hadrons from $z$
distributions can be found in Ref.~\cite{Abelev:2008ab}.

The raw yields from the TOF are obtained using the variable mass-square
($m^{2}$), given by
\begin{equation}
\label{msquare}
m^{2} = p^{2} \left( \frac {c^{2} T^{2}}{L^{2}} -1 \right ),
\end{equation}
 where, $p$, $T$, $L$, and $c$ are the momentum, time-of-travel by the
 particle, path length, and speed of light, respectively. 
 The $m^{2}$ distributions are obtained for rapidity $|y|<$ 0.1 for
 all particles in different $p_T$ ranges as shown by the black histograms in Fig.~\ref{fig_m2}. Since the $m^{2}$ distributions
 are not exactly Gaussian, 
we use the predicted $m^2$ distributions to fit these
distributions to extract the raw yields.
The predicted $m^2$ distributions can be obtained using 
\begin{equation}
\label{eqnmasssq}
m^{2}_{\rm{predicted}} = p^{2} \left( \frac {c^2T^{2}_{\rm{predicted}}}{L^{2}} -1 \right ).
\end{equation} 
Here $T_{\rm{predicted}}$ is the predicted time-of-flight based on the
random shift to the expected time-of-flight distributions for a given
particle, i.e. $T_{\rm{predicted}} =  T_{\rm{expected}} + t_{\rm{random}}$,
where $t_{\rm{random}}$ represents the Gaussian random time shift based on the
$\Delta T (=T_{\rm{measured}}-T_{\rm{expected}})$ distribution 
for a given $dE/dx$ identified hadron. 
Here, $T_{\rm{measured}}$ represents the experimentally
measured time-of-flight 
and $T_{\rm{expected}}$ is the expected time-of-flight for a given
hadron obtained using its known mass in  Eq.~(\ref{msquare}).
The $m^2_{\rm{predicted}}$ distributions are fitted to measured $m^2$ distributions, simultaneously for
 pions (dash-dotted, red), 
kaons (dashed, green), and protons (dotted, blue) as
shown in Fig.~\ref{fig_m2}. Using $\chi^2$-minimization, 
the raw yield for a given hadron in a given $p_T$ range is
obtained. 
\begin{figure}
\begin{center}
\includegraphics[width=9cm]{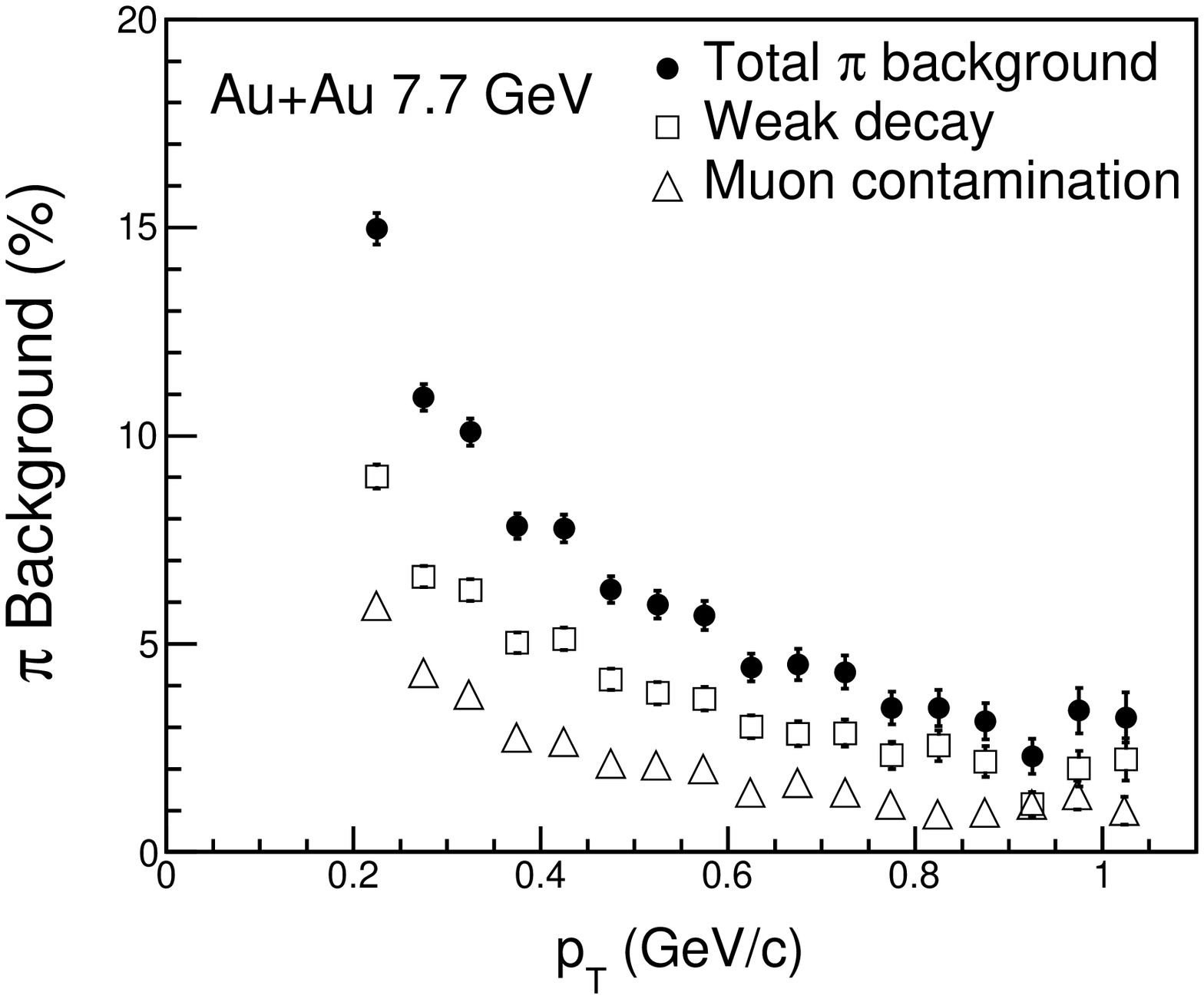}
\caption{Percentage of pion background contribution estimated from
  HIJING+GEANT simulation as a function of $p_{T}$ at midrapidity ($|y|<0.1$) in 0--5\% Au+Au collisions at $\sqrt{s_{NN}}$ = 7.7 GeV.
The contributions from different sources are shown separately,  as
well as the total background.
}
\label{fig_pibkg}
\end{center}
\vspace{-0.5cm}
\end{figure}
\begin{figure}
\begin{center}
\includegraphics[width=9.cm]{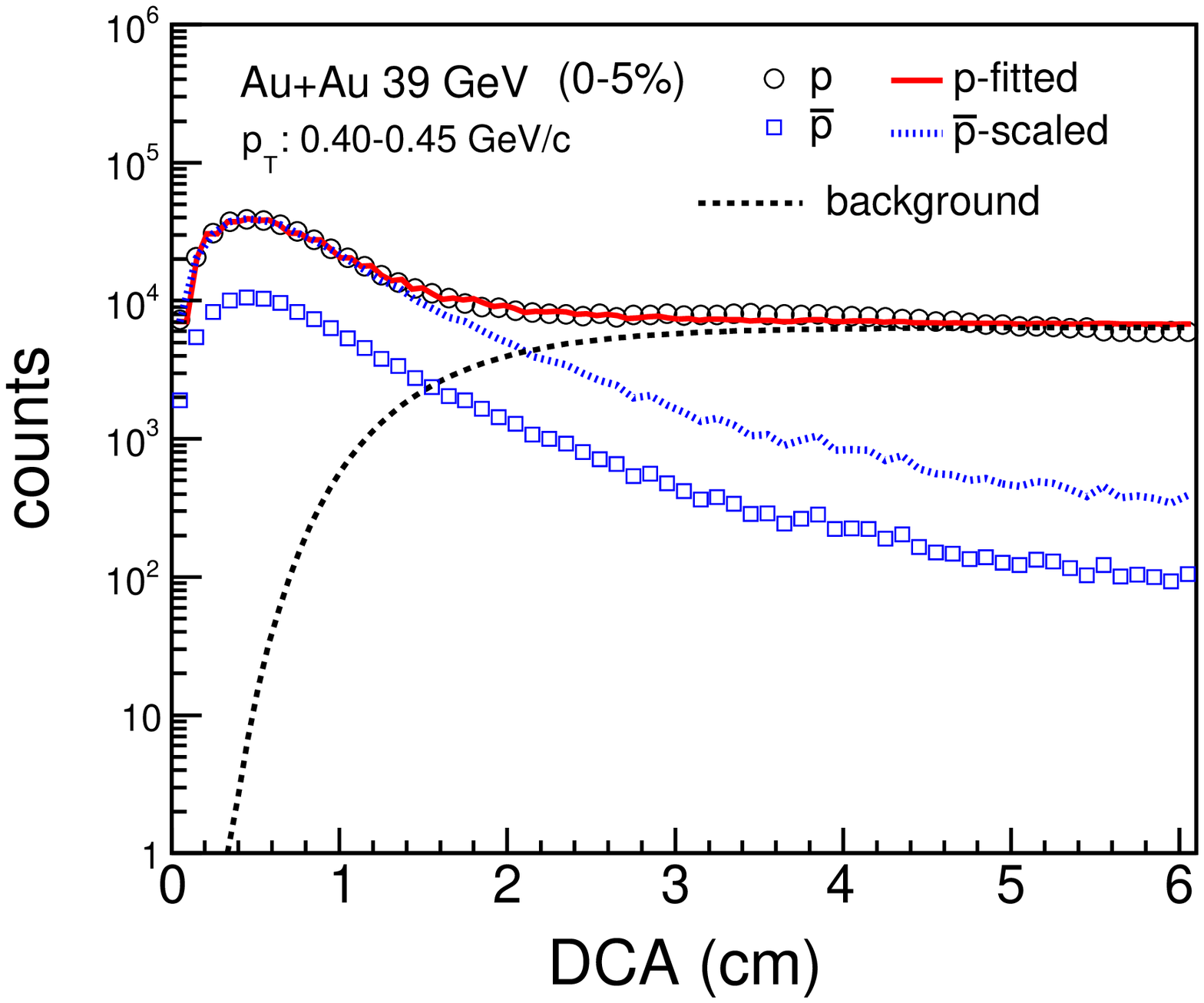}
\caption{ (Color online) 
The DCA distributions of protons and antiprotons for 0.40
$<p_T<$ 0.45 GeV/$c$ at midrapidity ($|y|<0.1$) in 0--5\% Au+Au collisions at $\sqrt{s_{NN}}=$ 39 GeV. The dashed curve is the fitted proton background; the dotted
histogram is the $\bar{p}$ distribution scaled by the $p/\bar{p}$
ratio; and the solid histogram is the fit given by
Eq.~(\ref{eq_bkgd}). 
}
\label{fig_peloss}
\end{center}
\vspace{-0.5cm}
\end{figure}

\section{Correction Factors} 
\subsection{TPC Tracking Efficiency and Acceptance}
The principal correction to the raw spectra 
accounts for the detector acceptance and for the efficiency 
of reconstructing particle tracks. 
These effects are determined together by embedding Monte Carlo
tracks simulated using the GEANT~\cite{Fine:2000qx} model of the STAR detector
into 
real events at the raw data level. One important requirement is to have a
match in the distributions of reconstructed embedded tracks and real data tracks 
for quantities reflecting track quality and used for track selection.
Figure~\ref{dca} shows the comparisons of DCA 
and 
number of fit points (for embedded pions) distributions, respectively, 
in the low $p_T$ range $0.2 <  p_{T} < 0.5$ GeV/$c$.
Similar  agreement as in Fig.~\ref{dca} is observed between embedded tracks and real data 
in other measured $p_{T}$ ranges and beam energies for all the
identified hadrons presented here. The ratio  
of the distribution of reconstructed and original Monte Carlo tracks
as a function of $p_{T}$ gives the efficiency $\times$ acceptance correction 
factor for the rapidity interval studied. The typical efficiency
$\times$ acceptance factors for pions, kaons and
protons at midrapidity ($|y| < 0.1$) in 0--5\% 
Au+Au collisions at $\sqrt{s_{NN}}=$ 7.7 GeV are shown in Fig.~\ref{eff}.
The raw yields are scaled by the inverse of the
efficiency$\times$acceptance factors to obtain the corrected yields.

\subsection{TOF Matching Efficiency}
The TPC and the TOF are separate detectors. While the TPC identifies
low-$p_T$ ($<$ 1 GeV/$c$) particles well, the TOF gives better particle identification
than the TPC at higher momenta. 
However, not all TPC tracks give a hit in the TOF, so there is an extra
correction called the TOF matching efficiency correction needed for the spectra
obtained using the TOF detector. 
This is done with a data driven technique. The TOF matching efficiency
for a given particle species is defined as the ratio of the number of
tracks detected in the TOF to the number of the total tracks in the TPC
within the same
acceptance. Figure~\ref{fig_tofmatch} represents the typical TOF matching
efficiencies for pions, kaons, and protons
for 0--5\% Au+Au collisions at $\sqrt{s_{NN}}=$ 7.7 GeV. The
raw yields obtained from the TOF are scaled by the inverse of the TOF matching
efficiency to obtain the corrected yields.

\subsection{Energy Loss Correction}
Low momentum particles lose significant energy while traversing the detector material.
The track reconstruction algorithm takes into account the
Coulomb scattering and energy loss assuming the
pion mass for each particle. Therefore, a correction for the
energy loss by heavier particles is needed. 
This correction is obtained from embedding Monte Carlo simulations,
in which the $p_{T}$ difference between reconstructed and embedded tracks is
plotted as a function of $p_{T}$ of the reconstructed track. 

Figure~\ref{eloss} shows the energy loss
as a function of $p_{T}$ for  pions, kaons, and protons. 
The curves represent the function fitted to the data points~\cite{Abelev:2008ab}
\begin{equation}
\label{eqneloss}
f(p_{T}) = A_{e} + B_{e}\left(1 + \frac{C_{e}}{p_{T}^{2}}\right)^{D_{e}},
\end{equation}
where $A_{e}$, $B_{e}$, $C_{e}$ and $D_{e}$ are the fit parameters.
Table~\ref{table_eloss} shows the values of these parameters
obtained for kaons and protons.
The errors on some fit parameters are large but they do not affect the
correction factors as only the mean values of parameters are used to
estimate the $p_T$ dependence of the energy loss effect.
The energy loss for a given particle is
independent of beam energy and collision centrality.
For the results presented here, the track $p_{T}$ is corrected for this 
energy loss effect. 

\begin{table}
\caption{The values of energy loss parameters for kaons and protons
 obtained using Eq.~\ref{eqneloss}.
\label{table_eloss}}
\begin{center}
\begin{tabular}{c|c|c}
\hline
Values & Kaons & Protons \\ 
\hline
$A_e$  &  $(9.7 \pm 1.0) \times 10^{-4}$ & $(1.2 \pm 0.1) \times 10^{-3}$  \\ [0.02cm]
$B_e$  &   $(-2.8 \pm 8.3) \times 10^{-6}$  & $(-7.2 \pm 2.3) \times10^{-6}$\\[0.02cm] 
$C_e$  & $90 \pm 70$ &   $ 98 \pm 88$\\[0.02cm]
$D_e$  & $1.07 \pm 0.04$  &  $1.13 \pm 0.02$\\[0.02cm]
$\chi^2/NDF$  &   43/36  &  43/34   \\
\hline
\end{tabular}
\end{center}
\end{table}

\subsection{Pion Background Subtraction}
The charged pion spectra are corrected for feed-down contribution from
weak decays, muon contamination, and background pions
produced in the detector materials. These corrections are
obtained from Monte Carlo simulations of HIJING events at $\sqrt{s_{NN}}$ = 7.7--39 GeV, 
with the STAR geometry for these data and a realistic description of the detector
response implemented in GEANT. The simulated events are reconstructed
in the same way as the real data. The weak-decay daughter
pions are mainly from $K^{0}_{S}$ and $\Lambda$, and are identified by the parent particle 
information accessible from the simulation. 
The muons from pion decays can 
be misidentified as primordial pions 
due to their similar masses.
This contamination is obtained from Monte Carlo simulations by
identifying the decay, which is accessible in the
simulation. 
The weak-decay pion background
and muon contamination obtained from the simulation are shown in Fig.~\ref{fig_pibkg}, 
as a function of simulated pion $p_{T}$ for 0--5\% central Au+Au collisions at 
$\sqrt{s_{NN}}$ = 7.7 GeV.  The total pion background contribution
from weak-decays decreases with increasing $p_T$. This
contribution has been estimated for beam energies $\sqrt{s_{NN}} =$ 7.7--39 GeV.  
The background percentage for different energies and centralities
is of similar order.
The final pion spectra at different energies are corrected for this background effect. 

\subsection{Proton Background Subtraction}
The STAR experiment has previously observed that proton yields have significant 
contamination from secondary protons, due to interactions of
energetic particles produced in collisions with detector materials~\cite{Adams:2003ve,Abelev:2008ab}. 
As these secondary, so-called knock-out protons are produced away from the primary interaction point, they 
appear as a long tail in the DCA distribution of protons. 

To estimate this proton background, a comparison between the shapes of DCA distributions of
protons and anti-protons 
is done~\cite{Adams:2003ve,Abelev:2008ab}. 
Figure~\ref{fig_peloss} shows the DCA distributions of protons and antiprotons for 0.40
$<p_T<$ 0.45 (GeV/$c$) in 0--5\% Au+Au collisions at $\sqrt{s_{NN}}=$ 39 GeV. The protons and antiprotons are selected using a $dE/dx$ 
cut of $|n_{{\sigma}_p}| < 2$, where
$n_{{\sigma}_p}=z_p/\sigma_p$, $\sigma_p$ being the relative $dE/dx$ resolution of
the TPC which is track length dependent.
The long and flat DCA tail in the proton distribution comes
mainly from knock-out background protons. 
Antiprotons do not have this background and hence no flat tail
in their DCA distributions. 
To correct for the knock-out background protons, DCA dependence at DCA
$<$ 3 cm is needed for knock-out protons.  
It is
obtained from MC simulation~\cite{Adams:2003ve,Abelev:2008ab} and 
is given by
\begin{equation}
\label{eq_bkgd1}
N_p^{\rm{bkgd}} ({\rm{DCA}}) \propto [1-\exp(-{\rm{DCA}}/{\rm{DCA}_{0})}]^{\alpha},
\end{equation}
where ${\rm{DCA}_{0}}$ and $\alpha$ are fit parameters. It is assumed that the
shape of the background-subtracted proton DCA distribution is
identical to that of the anti-proton. This distribution can be fit by
\begin{equation}
\label{eq_bkgd}
N_p ({\rm{DCA}}) = N_{\bar{p}}({\rm{DCA}})/r_{\bar{p}/p} + F N_p^{\rm{bkgd}} ({\rm{DCA}})
\end{equation}
Here, $r_{\bar{p}/p}$ and $F$ are the fit parameters. We used this
functional form to fit the proton DCA distributions for every
$p_T$ bin in each centrality at each energy to obtain the fraction
of proton background.

The proton background fraction decreases with
increasing $p_T$.
The fraction of proton
background increases from central to peripheral collisions.
In Au+Au collisions at $\sqrt{s_{NN}}=$ 39 GeV, the background fraction at $p_T$=0.40--0.45 GeV/$c$ is about 15\% 
for 0--5\% centrality and 30\% for 70--80\%, while at the lowest energy ($\sqrt{s_{NN}}=$ 7.7
GeV), it is 2\% and 10\% for 0--5\% and 70--80\% centralities,
respectively. 
The reason for variation of proton background fraction with centrality
may be that the ratio of proton multiplicity to total particle multiplicity
shows centrality dependence. 
The proton background fraction as a function of $p_T$ is subtracted from the proton raw
yields for each centrality and collision energy studied.

It may be noted that the results presented here for BES energies
correspond to inclusive protons and anti-protons similar to those at
higher RHIC energies~\cite{Abelev:2008ab} as the feed-down correction
has large uncertainty and is very model dependent. 
The analysis cut (DCA $<$ 3 cm) used for the identified
particle studies rejects only a negligible fraction of
daughter protons from the hyperon decays~\cite{Adams:2003xp,Aggarwal:2010pj}. Therefore, the (anti-) protons
yields presented here are truly inclusive. 

\begin{figure*}
\begin{center}
\includegraphics[width=15cm]{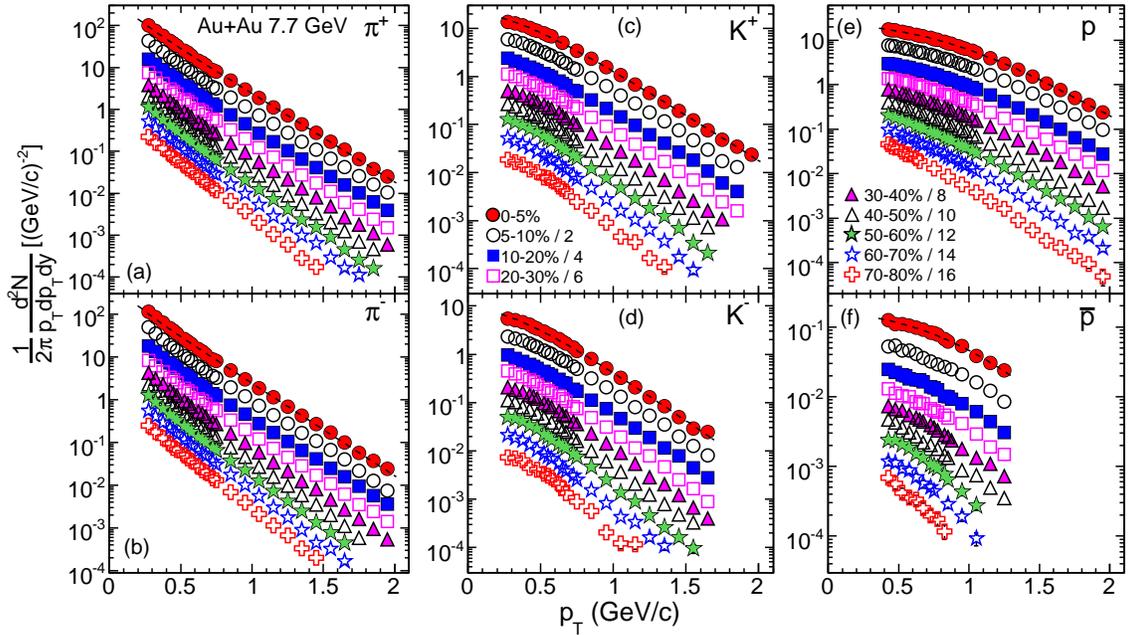}
\vspace{-0.5cm}
\caption{ (Color online)
Mid-rapidity ($|y|<0.1$) transverse momentum spectra for 
(a) $\pi^{+}$, (b) $\pi^{-}$, (c) $K^{+}$, (d) $K^{-}$, (e) $p$, and (f) $\bar{p}$ 
in Au+Au collisions at $\sqrt{s_{NN}}$
= 7.7 GeV for different centralities. 
The spectra for centralities other than 0--5\% are
scaled for clarity as shown in figure. The curves represent the Bose-Einstein,  
$m_T$-exponential, and double-exponential function fits to 0--5\%
central data
for pions, kaons, and (anti-) protons, respectively. 
The uncertainties are statistical and systematic added in quadrature.}
\label{ptspectra_7}
\end{center}
\end{figure*}
\begin{figure*}
\begin{center}
\includegraphics[width=15cm]{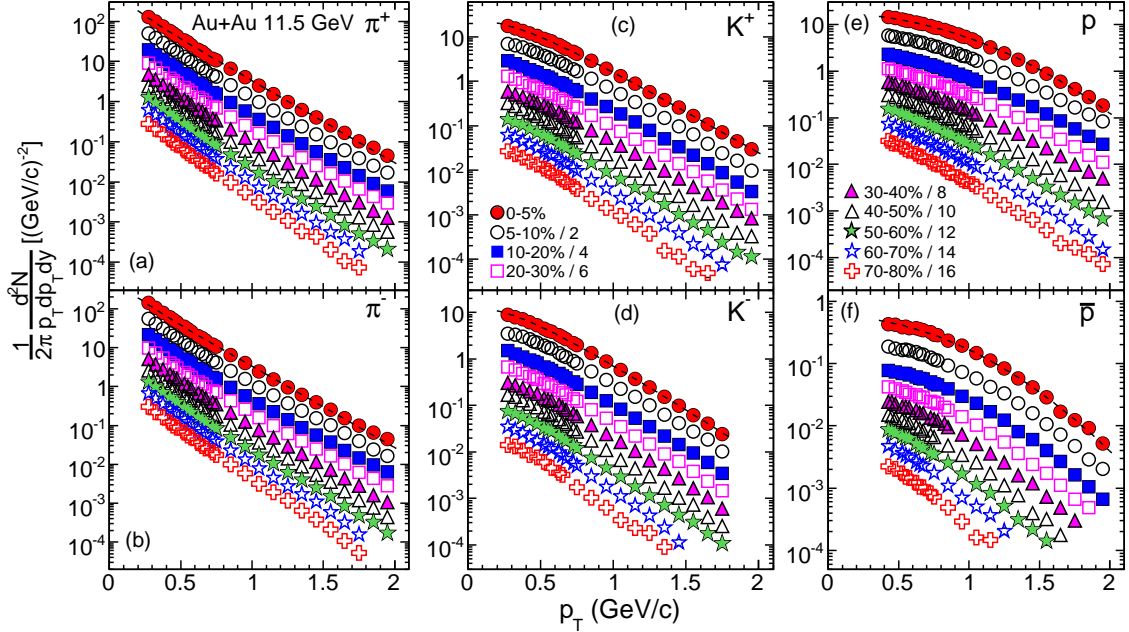}
\vspace{-0.5cm}
\caption{
Same as Fig.~\ref{ptspectra_7} but for Au+Au collisions at $\sqrt{s_{NN}} =$ 11.5 GeV.
}
\label{ptspectra_11}
\end{center}
\end{figure*}
\section{Systematic Uncertainties}
Systematic uncertainties on the spectra are estimated by varying 
cuts, and by assessing the purity of identified hadron samples
from $dE/dx$ measurements. 
Details of various sources
of systematic uncertainties on the pion, kaon and proton/anti-proton  yields 
in Au+Au collisions at $\sqrt{s_{NN}}$ = 7.7 GeV are given
below. The systematic uncertainties for other energies are estimated in a similar manner and are of a similar order.

The systematic uncertainties are estimated, by varying the $V_{z}$ range (from $|V_z|<50$ cm to
$|V_z|<30$ cm). The track cuts are also varied such as the DCA (from 3 cm to 2 cm),
number of fit points (from 25 to 20), number of $dE/dx$ points (from 15 to 10), 
PID cut i.e. $|n_{\sigma}|$, for the purity of a
hadron used to obtain predicted $m^2$ distributions (from
$|n_{\sigma}|<2$ to $|n_{\sigma}|<1$), and range of
Gaussian fits to normalized $dE/dx$ distributions.
Combined systematic uncertainties due to
all these analysis cut variations are of the order of 4\%, 3\%, and 6\% for pions,
kaons, and protons, respectively. 
The systematic uncertainty due to 
track reconstruction efficiency and acceptance estimates is of the
order of 5\% which is obtained by varying parameters in the MC simulation.

The $p_T$-integrated particle yield $dN/dy$ and 
$\langle p_{T} \rangle$ 
are obtained from the $p_T$ spectra using data in the measured $p_{T}$ ranges and extrapolations assuming
certain functional forms for the unmeasured $p_{T}$
ranges~\cite{Abelev:2008ab}. 
The percentage 
contribution to the yields from extrapolation are about 20-30\%.
The extrapolation of yields to the unmeasured regions in $p_{T}$ is an
additional source of 
systematic error. This is estimated by comparing 
the extrapolations using different fit functions to the $p_{T}$ spectra. 
For pions, the default function used to obtain $dN/dy$ is the
Bose-Einstein function and the systematic error is obtained by
changing the functional form to the $p_T$-exponential function.
For kaons, the $m_T$-exponential function is used for
$dN/dy$ and the Boltzmann function for the systematic error.
Here $m_T$ = $\sqrt{m^2+p_T^2}$ represents the particle transverse mass. 
For protons, the double-exponential function is used for $dN/dy$ and the
$m_T$-exponential function is used to obtain the systematic error. The functional
form of these functions are 
\begin{itemize}
\item
Bose-Einstein: $\propto 1/(e^{m_T/T} - 1)$, 
\item
$p_T$-exponential: $\propto e^{-p_T/T}$, 
\item 
$m_T$-exponential: $\propto e^{-(m_T-m)/T}$, 
\item 
Boltzmann: $\propto m_T e^{-m_T/T}$, 
\item Double-exponential:
$A_1 e^{-p_T^2/T_1^2}+A_2 e^{-p_T^2/T_2^2}$. 
\end{itemize}
Systematic uncertainties due to extrapolation to unmeasured $p_T$ region is
estimated to be of the order of 6-9\% for pions, 4-8\% for kaons, and
10-12\% for protons and anti-protons. 

The systematic uncertainties arising due to the pion and proton background 
are also studied.  The systematic uncertainty due to pion background is found to be
negligible. However, the uncertainty due to
the proton background is about 6-7\% (39--19.6 GeV) and 2-4\% (7.7--11.5 GeV).
In addition, the systematic uncertainties due to
energy loss estimation (discussed previously) for kaons and protons
are found to be of the order of 3\% and 2\%, respectively. 

\begin{figure*}
\begin{center}
\includegraphics[width=15cm]{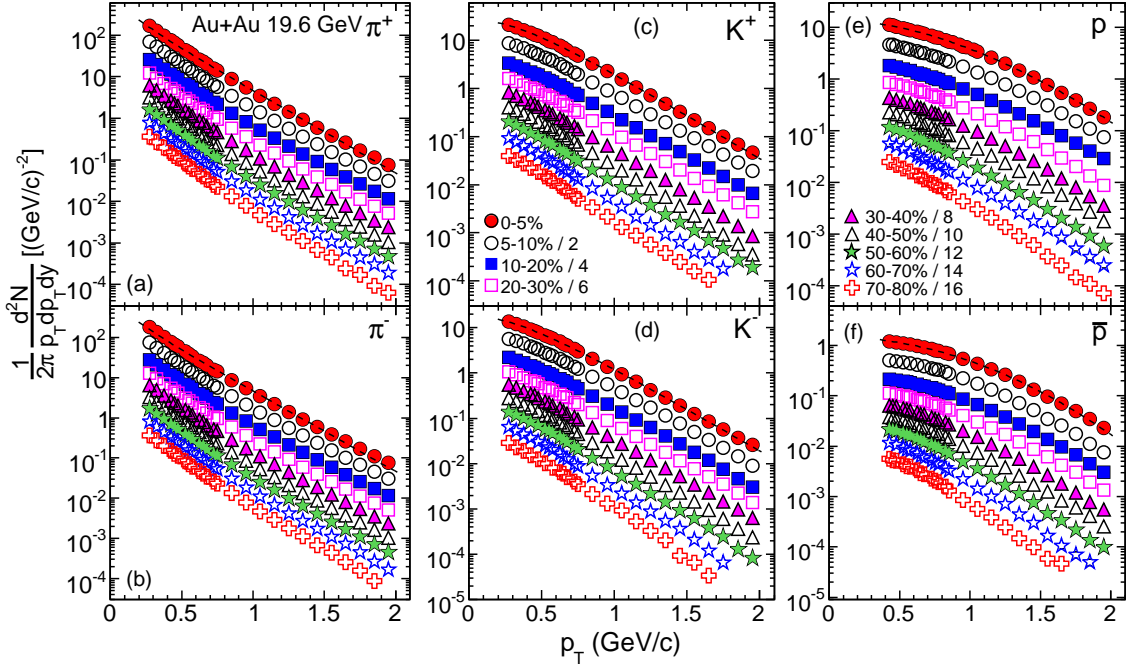}
\vspace{-0.5cm}
\caption{
Same as Fig.~\ref{ptspectra_7} but for Au+Au collisions at $\sqrt{s_{NN}} =$ 19.6 GeV.
}
\label{ptspectra_19}
\end{center}
\end{figure*}
\begin{figure*}
\begin{center}
\includegraphics[width=15cm]{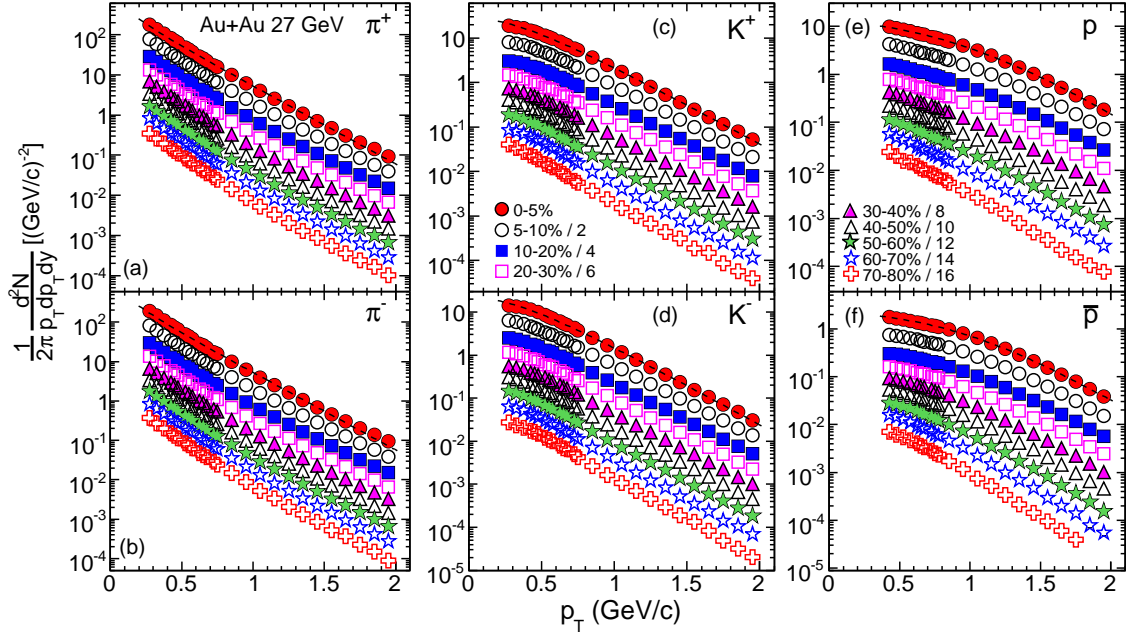}
\vspace{-0.5cm}
\caption{
Same as Fig.~\ref{ptspectra_7} but for Au+Au collisions at $\sqrt{s_{NN}} =$ 27 GeV.
}
\label{ptspectra_27}
\end{center}
\end{figure*}

The total systematic uncertainties are obtained by adding the contribution from
different sources in quadrature and are found to be of the order of
9-11\% for pions, 8-10\% for kaons, 
11-13\% for protons and 12-13\% for anti-protons for all energies. 
The results presented here are quadrature sums of the
systematic uncertainties and statistical uncertainties, the latter being negligible. Table~\ref{table_syserr} gives a summary of various
sources of systematic uncertainties for all energies. 
The systematic uncertainties
on particle ratios are obtained using the uncertainties on particle yields,
but excluding correlated uncertainties i.e. from efficiency. In
addition, the 
extrapolation and energy loss uncertainties are canceled to a large extent in the
antiparticle to particle ratios.
 The systematic uncertainties for $\langle p_T \rangle$ come mainly from the extrapolations as
discussed above.  The $\langle p_T \rangle$ also depends
on the range used
for fitting to the $p_T$ spectra. The variations
in the  $\langle p_T \rangle$ values due to different fitting
ranges are included in the systematic uncertainties. The total systematic uncertainties on $\langle
p_T \rangle$  for pions, kaons, and protons-antiprotons are 5-6\%, 4-6\%, and
6-11\%, respectively, across all beam energies.  

Chemical freeze-out parameters (chemical freeze-out temperature
$T_{\rm{ch}}$, $\mu_B$, $\mu_S$, $\gamma_S$, and radius $R$) are
extracted from the measured particle yields or ratios fitted in the
THERMUS model~\cite{Wheaton:2004qb}. The systematic
uncertainties on the yields are treated as independent, and are
propagated to the systematic uncertainties on chemical freeze-out
parameters. 
We have also estimated the effect of
correlated uncertainties in particle ratios used to extract the chemical freeze-out
parameters. The effect arises because the pion yield is used for
constructing many particle ratios. The effect of this on freeze-out
parameters is estimated by varying
the uncertainties on pion yields and extracting the freeze-out parameters
for a large sample of pion yields.  We have found that the effect is
within 3\% for the extracted  freeze-out parameters.

The kinetic freeze-out parameters are extracted from the simultaneous
fitting of $\pi^{\pm}$, $K^{\pm}$, and protons and antiprotons spectra
with the blast-wave model~\cite{Schnedermann:1993ws}. The extracted fit
parameters are kinetic freeze-out temperature $T_{\rm{kin}}$, average
radial flow velocity $\langle \beta \rangle$, and the flow velocity
profile exponent $n$. The point-to-point systematic uncertainties on the
spectra are included in the blast-wave fits. The measured pions
contain large contributions from resonance decays which vary as a
function of $p_T$. Since the default blast-wave model does not include
resonance decays, in order to reduce the systematic error due to resonance
decays, the low $p_T$ part ($<$ 0.5 GeV/$c$) of the pion spectra are
excluded from the blast-wave fit. The results from the blast-wave fits
are  sensitive to the range of $p_T$ used for fitting the
spectra. The effect on the extracted kinetic freeze-out
parameters due to different $p_T$ ranges used for fitting is 
estimated. These variations are included in the systematic uncertainties for
kinetic freeze-out parameters.  

The total systematic uncertainties reported in figures are
highly correlated among different
centralities. 
The results on $\pi^{\pm}$, $K^{\pm}$, protons and antiprotons 
particle spectra and yields have a correlated
uncertainty of 5\% from efficiency corrections.
This systematic uncertainty is canceled in particle ratios. 
The uncertainties from extrapolations to unmeasured $p_T$ regions 
are correlated between particle species  (see
Table~\ref{table_syserr}),
and are canceled in antiparticle to particle ratios.
Since the uncertainties of particle yields
and ratios are propagated in the extracted chemical and kinetic
freeze-out parameters, the freeze-out parameters also include the
corresponding correlated uncertainties.  
\begin{table}
\begin{center}
\caption{Sources of percentage systematic uncertainties for pions, kaons, and
  (anti-) protons  yields at all energies.
\label{table_syserr}}
\vspace{.2cm}
\begin{center}
\begin{tabular}{l|l|l|l}
\hline
Sources       & $\pi$      & $K$ &  $p$ ($\bar{p})$\\
\hline
Cuts            &   4\%  &  3\%      &   6\%  \\
Tracking eff. &   5\%  &  5\%      &   5\%  \\
Energy Loss &   --  &  3\%      &   2\%  \\
Extrapolation &   6--9\% &  4--8\%    &   10--12\% \\
Total  &   9--11\% &  8--10\%    &   11--13\%  \\
\hline
\end{tabular}
\end{center}
\end{center}
\end{table}

\section{Results and Discussions}

\begin{figure*}
\begin{center}
\includegraphics[width=15cm]{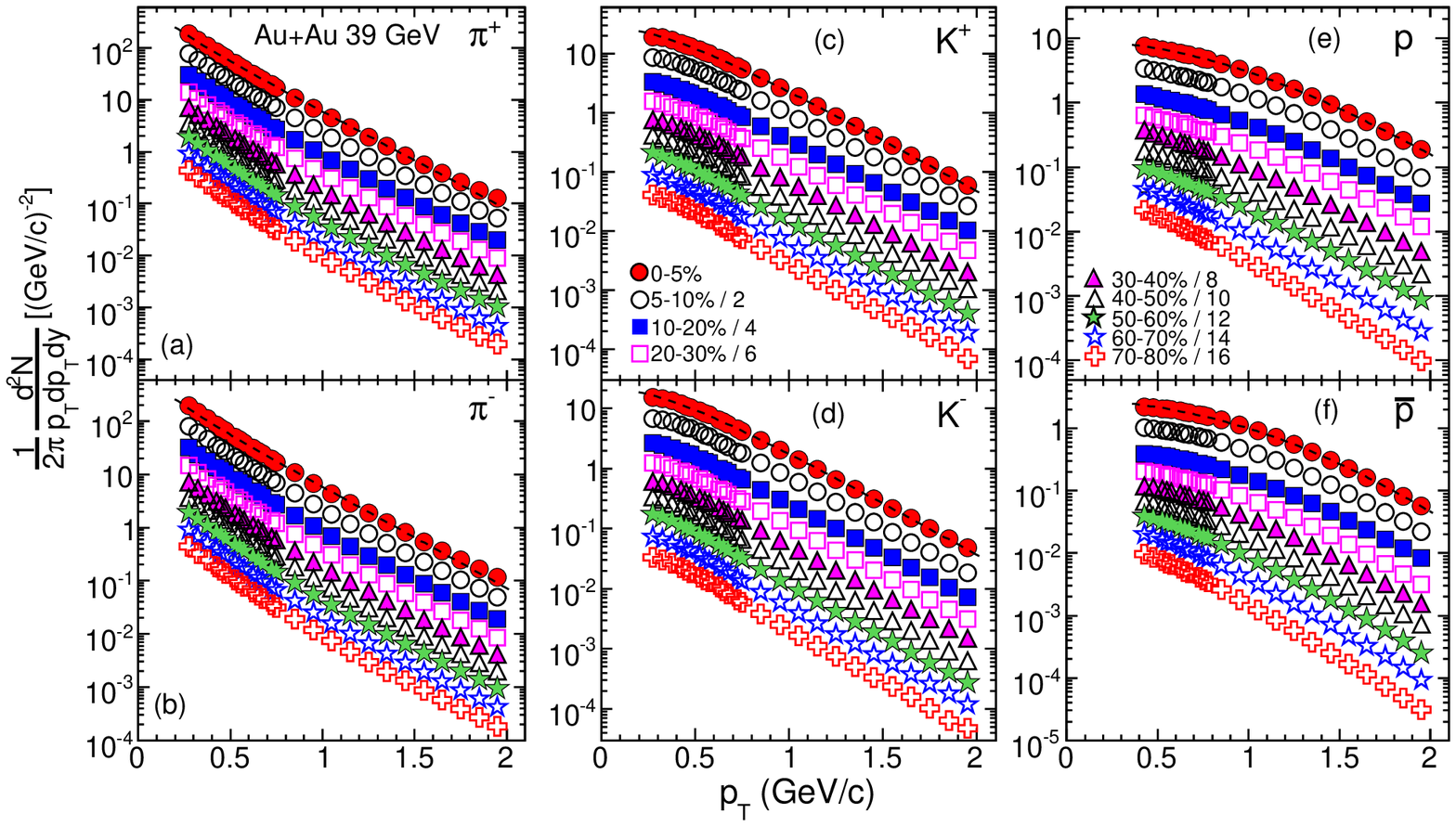}
\vspace{-0.5cm}
\caption{
Same as Fig.~\ref{ptspectra_7} but for Au+Au collisions at $\sqrt{s_{NN}} =$ 39 GeV.
}
\label{ptspectra_39}
\end{center}
\end{figure*}

\subsection{Transverse Momentum Spectra}
 Figure~\ref{ptspectra_7} shows the transverse momentum spectra 
for $\pi^{\pm}$, $K^{\pm}$, and $p$ ($\bar{p}$), 
in Au+Au collisions at $\sqrt{s_{NN}}$ = 7.7 GeV. The results 
are shown for 
the collision centrality classes of 0--5\%, 5--10\%, 10--20\%,
20-30\%, 30--40\%, 40--50\%, 50--60\%, 60--70\%, and 70--80\%. 
The $p_T$ spectra for 11.5, 19.6, 27, and 39 GeV are shown in 
Figs.~\ref{ptspectra_11}, ~\ref{ptspectra_19},
~\ref{ptspectra_27}, and ~\ref{ptspectra_39}, respectively.
The inverse slopes of
the identified hadron spectra follow the order $\pi$ $<$ $K$ $<$ $p$. The spectra can be 
further characterized by the $dN/dy$ and $\langle p_{T} \rangle$ or 
$\langle m_{T} \rangle - m$ for the produced hadrons, where $m$ is the mass of the hadron and 
$m_{T}$ is its transverse mass. 

\begin{figure*}
\begin{center}
\includegraphics[width=15cm]{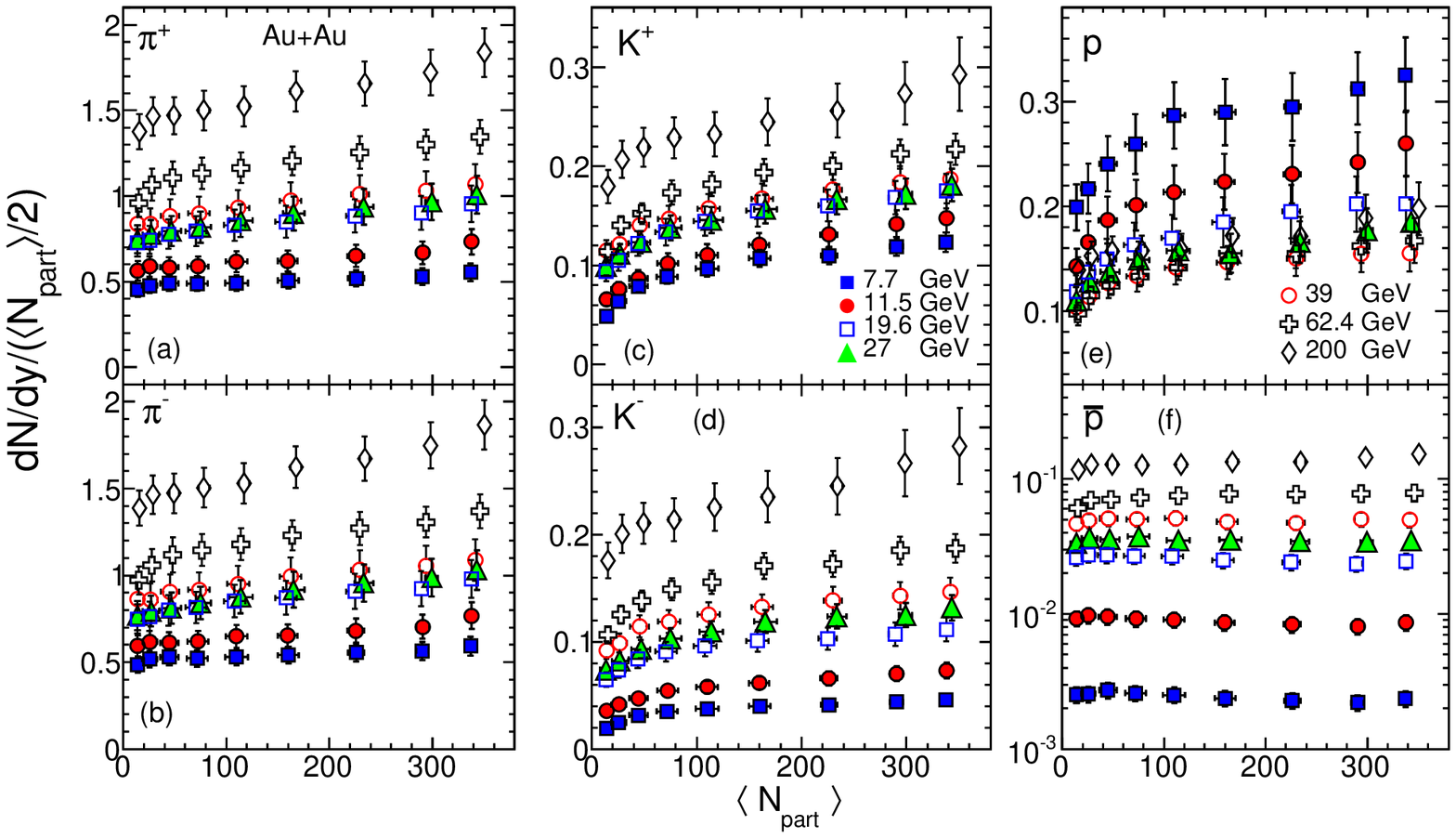}
\vspace{-0.4cm}
\caption{ (Color online) 
Centrality dependence of $dN/dy$ normalized by 
$\langle N_{\mathrm {part}} \rangle/2$ for  
(a) $\pi^{+}$, (b) $\pi^{-}$, (c) $K^{+}$, (d) $K^{-}$, (e) $p$, and (f) $\bar{p}$ 
at midrapidity ($|y|<0.1$) in Au+Au collisions at $\sqrt{s_{NN}}=$ 7.7, 11.5,
19.6, 27, and 39 GeV. Results are compared with published
results in Au+Au collisions at $\sqrt{s_{NN}} =$ 62.4 and 200
GeV~\cite{Adams:2003xp,Abelev:2008ab}. 
Errors shown are the quadrature sum of 
statistical and systematic uncertainties. For clarity, $\langle N_{\rm{part}}
\rangle$ uncertainties are not added in quadrature.
}
\label{dndycent}
\end{center}
\end{figure*}
\begin{figure*}
\begin{center}
\includegraphics[width=15cm]{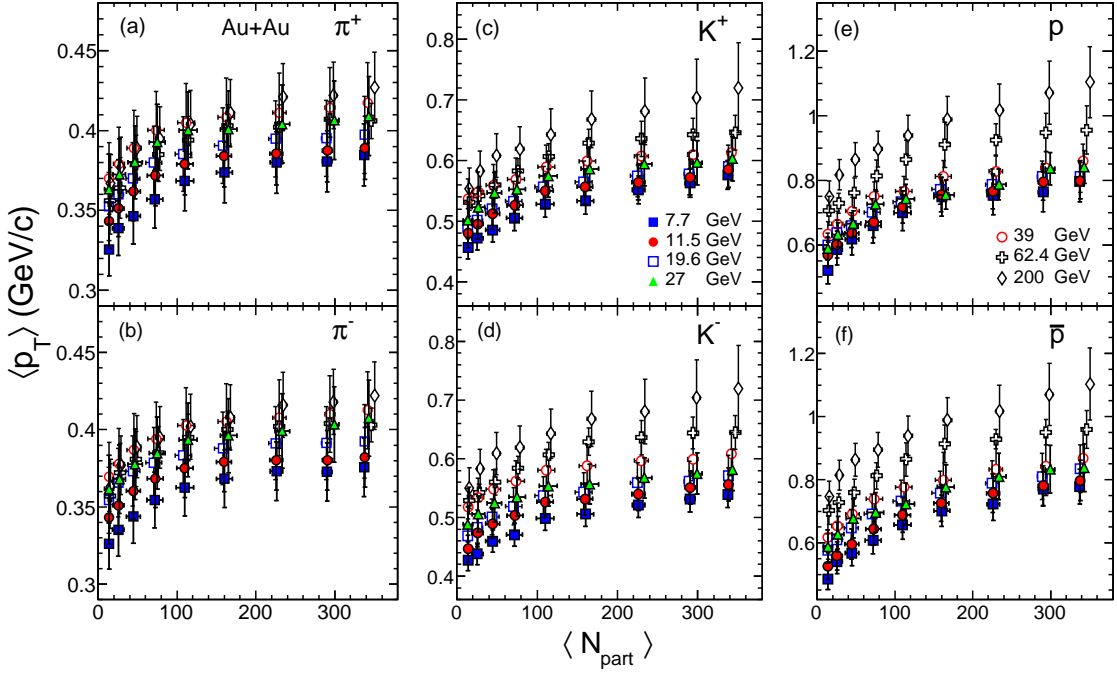}
\caption{ (Color online)
Centrality dependences of $\langle p_T \rangle$ for
(a) $\pi^{+}$, (b) $\pi^{-}$, (c) $K^{+}$, (d) $K^{-}$, (e) $p$, and (f) $\bar{p}$ 
at midrapidity ($|y|<0.1$) in Au+Au collisions at $\sqrt{s_{NN}}=$ 7.7, 11.5,
19.6, 27, and 39 GeV. Results are compared with  published
results in Au+Au collisions at $\sqrt{s_{NN}} =$ 62.4 and 
200 GeV~\cite{Adams:2003xp,Abelev:2008ab}. Errors
shown are quadrature sum of 
statistical and systematic uncertainties
where the latter dominates.
}
\label{meanpt62}
\end{center}
\end{figure*}
\begin{figure*}[htb]
\begin{center}
  \includegraphics[width=16cm]{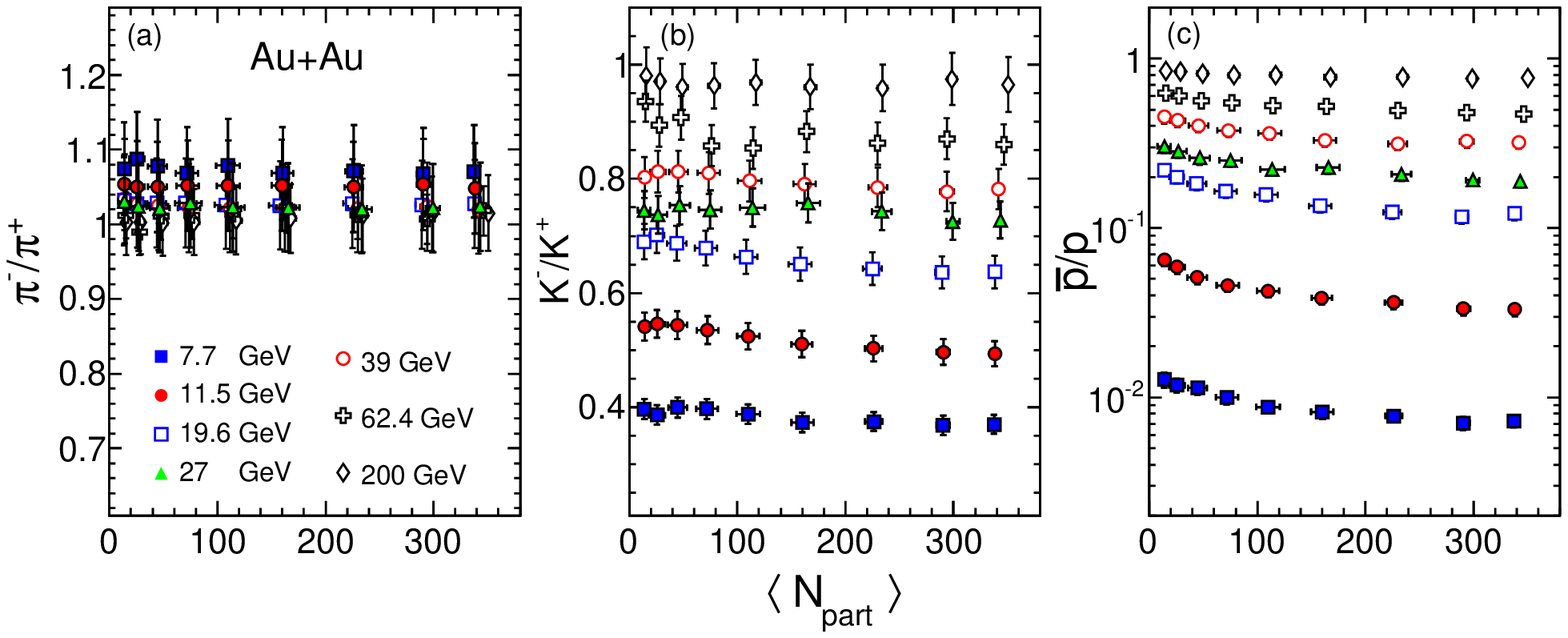}
\caption{ (Color online)
Variation of (a) $\pi^{-}$/$\pi^{+}$, (b) $K^{-}$/$K^{+}$, and
(c) $\bar{p}$/$p$
ratios as a function of $\langle N_{\mathrm {part}} \rangle$ at midrapidity ($|y|<0.1$)
in Au+Au collisions at all BES energies. Also shown for comparison 
are the corresponding results in Au+Au collisions at $\sqrt{s_{NN}}$ = 62.4 and 
200 GeV~\cite{Adams:2003im,Adams:2003kv,Abelev:2006jr,Abelev:2007ra,Abelev:2008ab}.
Errors shown are the quadrature sum of 
statistical and systematic uncertainties
where the latter dominates.
}
\label{centratio}
\end{center}
\end{figure*}
\begin{figure*}
\begin{center}
  \includegraphics[width=13cm]{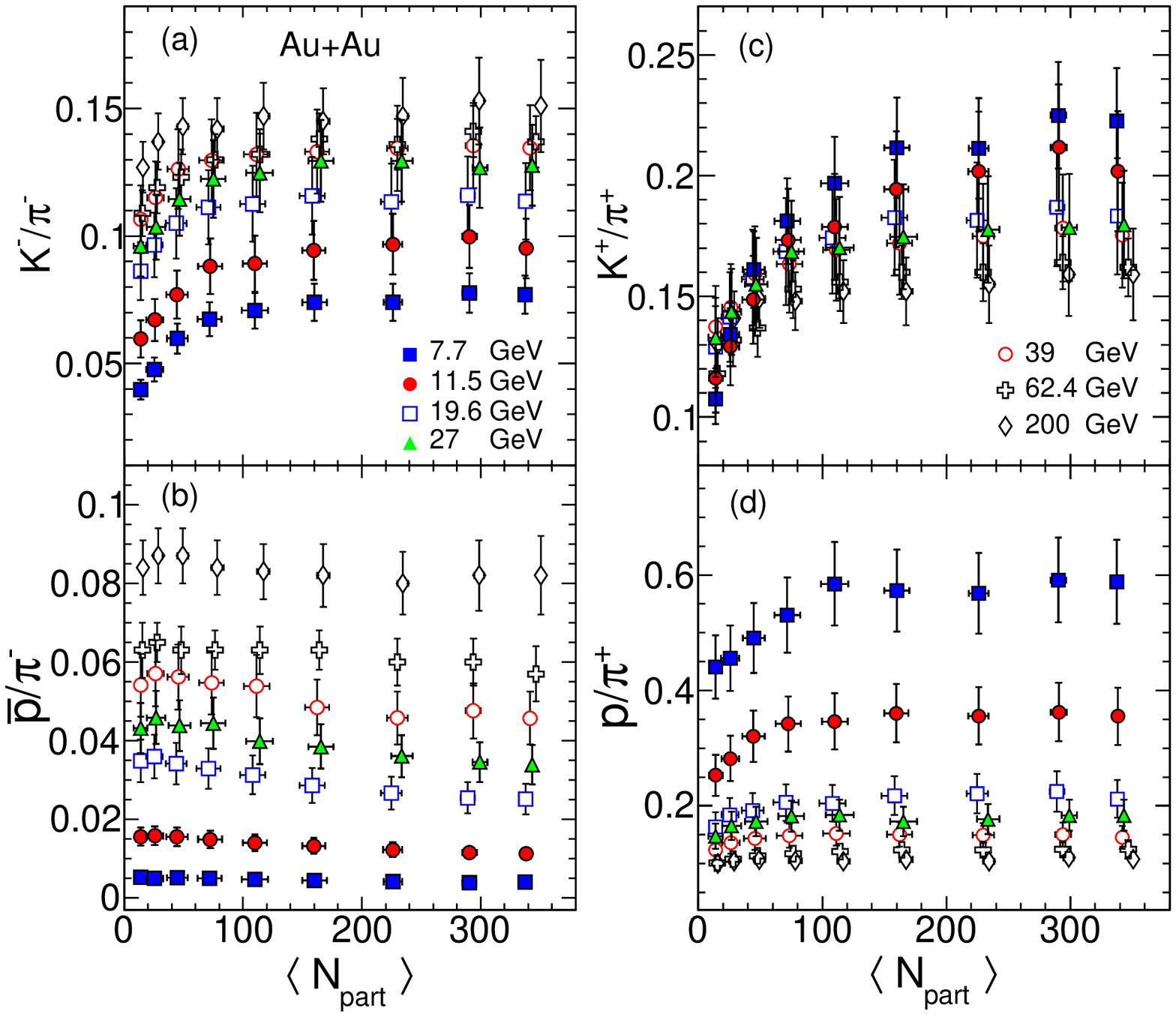}
\vspace{-0.4cm}
\caption{ (Color online) 
Variation of (a) $K^{-}$/$\pi^{-}$, (b) $\bar{p}$/$\pi^{-}$, 
(c) $K^{+}$/$\pi^{+}$, and (d) $p/\pi^{+}$ 
ratios as a function of $\langle N_{\mathrm {part}} \rangle$
at midrapidity ($|y|<0.1$) in Au+Au collisions at all BES energies. Also shown for comparison 
are the corresponding results in Au+Au collisions at $\sqrt{s_{NN}}$ = 62.4 and 
200 GeV~\cite{Adams:2003im,Adams:2003kv,Abelev:2006jr,Abelev:2007ra,Abelev:2008ab}.
Errors shown are the quadrature sum of statistical and systematic
uncertainties
where the latter dominates.}
\label{centratio_mix}
\end{center}
\end{figure*}

\subsection{Centrality Dependence of Particle Production}
\subsubsection{Particle yields ($dN/dy$)}

Figure~\ref{dndycent} shows the comparison of collision centrality 
dependence of  $dN/dy$ of $\pi^{\pm}$, $K^{\pm}$, $p$ and $\bar{p}$, normalized by $\langle N_{\mathrm {part}} \rangle$/2,
among the results at $\sqrt{s_{NN}}$ = 7.7, 11.5, 19.6, 27, and 39 GeV, and previously published results 
at $\sqrt{s_{NN}}$ = 62.4 and 200 GeV from the STAR experiment~\cite{Adams:2003im,Adams:2003kv,Adams:2003xp,Abelev:2006jr,Abelev:2007ra,Abelev:2008ab}. 
The yields of charged pions, kaons, and anti-protons decrease with
decreasing collision energy. However, the yield of protons is the highest for
the lowest energy of 7.7 GeV, which indicates the highest baryon density at
mid-rapidity at this energy. 
Proton yield decreases from 7.7 GeV through 11.5, 19.5, 27, and 39
GeV, lowest being at 39 GeV. Then it again increases at 62.4 GeV up to
200 GeV. 
The proton yields come from two mechanisms: 
pair production and baryon transport~\cite{Cleymans:2005zx}. The energy dependence trend
observed here
for the proton yield is due to interplay of these two mechanisms.
The collision centrality
dependence for the BES results is similar to that at higher beam
energies. The normalized yields decrease from central to peripheral collisions
for $\pi^{\pm}$, $K^{\pm}$, and $p$. However, the centrality dependence
of normalized yields for $\bar{p}$ is weak. The
$dN/dy$ values for $\pi^{\pm}$, $K^{\pm}$, $p$ and $\bar{p}$ in
different centralities at
various BES energies are listed in Table~\ref{tab:dndy}.

\subsubsection{Average Transverse Momentum $p_T$ ($\langle p_T \rangle$)}

Figure~\ref{meanpt62} shows the comparison of $\langle p_{T} \rangle$
as a function of $\langle N_{\mathrm {part}} \rangle$ for $\pi^{\pm}$,
$K^{\pm}$, $p$ and $\bar{p}$, in 
Au+Au collisions  at $\sqrt{s_{NN}}$ = 7.7, 11.5, 19.6, 
27, and 39 GeV. Results are compared with the published results in
Au+Au collisions at $\sqrt{s_{NN}}=$ 62.4 and 200 GeV ~\cite{Adams:2003im,Adams:2003kv,Adams:2003xp,Abelev:2006jr,Abelev:2007ra,Abelev:2008ab}.
The dependences of $\langle p_{T} \rangle$
on $\langle N_{\mathrm {part}} \rangle$ at BES energies
are similar to those at $\sqrt{s_{NN}} = $ 62.4 and 200 GeV. An increase in 
$\langle p_{T} \rangle$ with increasing hadron mass is observed 
at all BES energies.
A similar dependence is also observed for $\sqrt{s_{NN}} =$ 
62.4 and 200 GeV.
The mass dependence of $\langle p_{T} \rangle$
reflects collective expansion in the radial direction, although it
also includes the temperature component. 
The differences in central values of $\langle p_{T} \rangle$ between protons and pions/kaons are
smaller at lower energies compared to those at higher beam
energies. 
This suggests that the average collective velocity in the radial 
direction is smaller at lower energies. The
 $\langle p_{T} \rangle$ values for $\pi^{\pm}$, $K^{\pm}$, $p$ and $\bar{p}$ in
different centralities at
various BES energies are listed in Table~\ref{tab:meanpt}.

\subsubsection{Particle Ratios}
Figure~\ref{centratio} shows the various anti-particle to particle 
ratios ($\pi^{-}$/$\pi^{+}$, $K^{-}$/$K^{+}$, $\bar{p}/p$)
as a function of collision centrality expressed as 
$\langle N_{\mathrm {part}} \rangle$
in Au+Au collisions at all BES energies.
Corresponding results from Au+Au collisions at $\sqrt{s_{NN}} = $ 62.4 and 
200 GeV~\cite{,Adams:2003xp,Adams:2003im,Adams:2003kv,Abelev:2006jr,Abelev:2007ra,Abelev:2008ab} are also shown. 
The $\pi^{-}$/$\pi^{+}$ ratio is close to unity for most of the
energies. However, a slight energy dependence is observed for lower
energies. The lowest energy of 7.7 GeV has a larger $\pi^{-}/\pi^{+}$ ratio
than those at the other energies due to isospin and significant contributions
from resonance decays (such as $\Delta$ baryons).
The $K^{-}$/$K^{+}$ ratio increases with increasing
energy, and shows very little centrality dependence. The increase in
$K^{-}$/$K^{+}$ ratio with energy shows the increasing contribution to
kaon production due to pair production. However, at lower energies,
associated production dominates. Associated production refers to reactions 
such as $NN \rightarrow KYN$ and $\pi N \rightarrow KY$, where $N$ is
a nucleon and $Y$ a hyperon.
The $\bar{p}$/$p$ ratio increases with increasing energy. The ratio increases from central to
peripheral collisions. This increase in $\bar{p}$/$p$ ratio from
central to peripheral collisions reflects a higher baryon density
(baryon stopping) at mid-rapidity in central collisions compared to peripheral collisions.

Figure~\ref{centratio_mix} shows the centrality dependence of mixed
ratios ($K^{-}$/$\pi^{-}$, $K^{+}$/$\pi^{+}$, 
$\bar{p}$/$\pi^{-}$, and $p$/$\pi^{+}$). 
These results are also compared with 
corresponding results 
at $\sqrt{s_{NN}}$ = 62.4 and 200 GeV. 
The $K^{-}$/$\pi^{-}$ ratio increases with increasing energy, and also
increases from
peripheral to central collisions. However, the $K^{+}$/$\pi^{+}$ ratio
is  maximal at 7.7 GeV and then decreases with increasing energy. This
is due to the associated production dominance at lower energies as the
baryon stopping is large. 
The centrality dependence of $K^{+}$/$\pi^{+}$ is observed at all energies, i.e., the ratio increases from
peripheral to central collisions. This increase from peripheral to
central collisions is much greater at 7.7 GeV than at the higher BES energies. 
This may be due to large baryon stopping at midrapidity at the lower energy of 7.7 GeV. 
This baryon stopping is centrality dependent, i.e. higher in
more central collisions as also reflected by the $\bar{p}/{p}$ ratio.
The $\bar{p}/\pi^{-}$ ratio increases with increasing beam energy and
shows little centrality dependence.
The $p$/$\pi^{+}$ ratio decreases with increasing energy.
As discussed above, this 
is a consequence of the higher baryon stopping 
at lower energies. The ratio increases from peripheral to central
collisions and becomes almost constant after $\langle N_{\rm{part}}
\rangle >$ 100.

\subsection{Energy Dependence of Particle Production}
\subsubsection{Particle yields ($dN/dy$)}
\begin{figure*}[htb]
\begin{center}
\includegraphics[width=15cm]{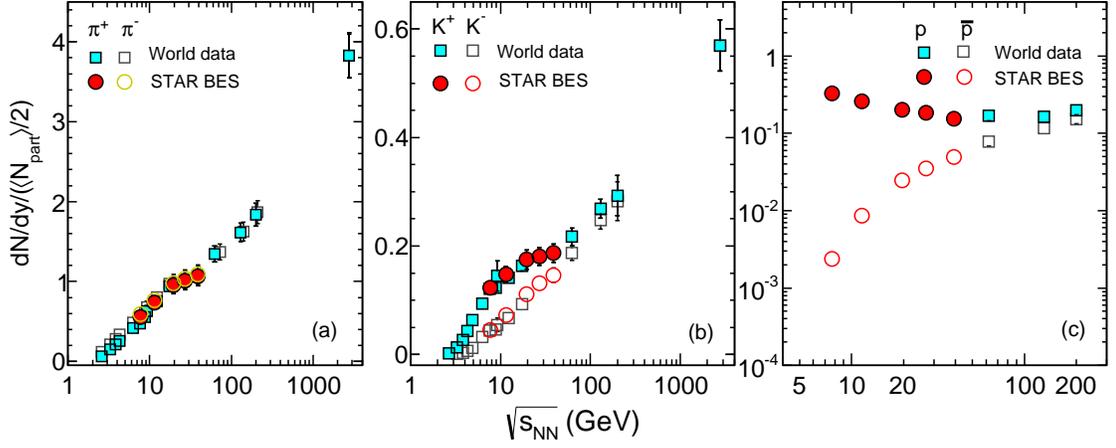}
\vspace{-0.4cm}
\caption{ (Color online)
The midrapidity ($|y|<0.1$) $dN/dy$ normalized by $\langle N_{\mathrm {part}}
\rangle$/2 as a function of $\sqrt{s_{NN}}$ for (a) $\pi^{\pm}$, (b)
$K^{\pm}$, and (c) $p$ and $\bar{p}$.
Results in 0--5\% Au+Au collisions at BES energies are compared to previous 
results from
AGS~\cite{Akiba:1996xf,Ahle:1998jc,Ahle:1999uy,Barrette:1999ry,Ahle:1999in,Ahle:1999va,Ahle:2000wq,Klay:2001tf},
SPS~\cite{Afanasiev:2002mx,Anticic:2004yj,Alt:2006dk,Alt:2007aa}, 
RHIC~\cite{Adler:2004zn,Abelev:2008ab,Abelev:2009bw}, and LHC~\cite{Abelev:2013vea}.
AGS results
correspond to 0--5\%, SPS to 0--7\%, top RHIC to 0--5\% (62.4 and 200
GeV) and 0-6\% (130 GeV), and LHC to 0--5\%
central collisions.
Errors shown are 
the quadrature sum of statistical and systematic uncertainties
where the latter dominates.
}
\label{dndetaen}
\end{center}
\end{figure*}
\begin{figure*}[htb]
\begin{center}
\includegraphics[width=15cm]{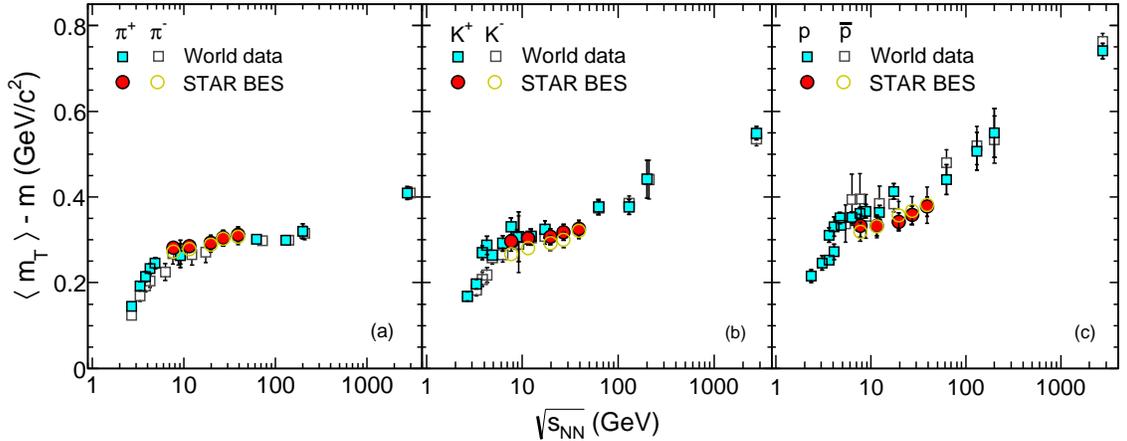}
\vspace{-0.5cm}
\caption{ (Color online)
$\langle m_{T} \rangle - m$ of (a) $\pi^{\pm}$, (b) $K^{\pm}$,
and (c) $p$ and $\bar{p}$ as a function
of $\sqrt{s_{NN}}$.
Midrapidity ($|y|<0.1$) results are shown for 0--5\% central 
Au+Au collisions at BES energies, and are compared to previous 
results from AGS~\cite{Akiba:1996xf,Ahle:1998jc,Ahle:1999uy,Barrette:1999ry,Ahle:1999in,Ahle:1999va,Ahle:2000wq,Klay:2001tf}, SPS~\cite{Afanasiev:2002mx,Anticic:2004yj,Alt:2006dk,Alt:2007aa}, RHIC~\cite{Abelev:2008ab}, and LHC~\cite{Abelev:2013vea}. AGS results
correspond to 0--5\%, SPS to 0--7\%, top RHIC to 0--5\% (62.4 and 200
GeV) and 0-6\% (130 GeV), and LHC to 0--5\%
central collisions.
The errors shown are the quadrature sum of statistical and systematic
uncertainties
where the latter dominates.
}
\label{energydndymt}
\end{center}
\end{figure*}
\begin{figure*}
\begin{center}
\includegraphics[width=15cm]{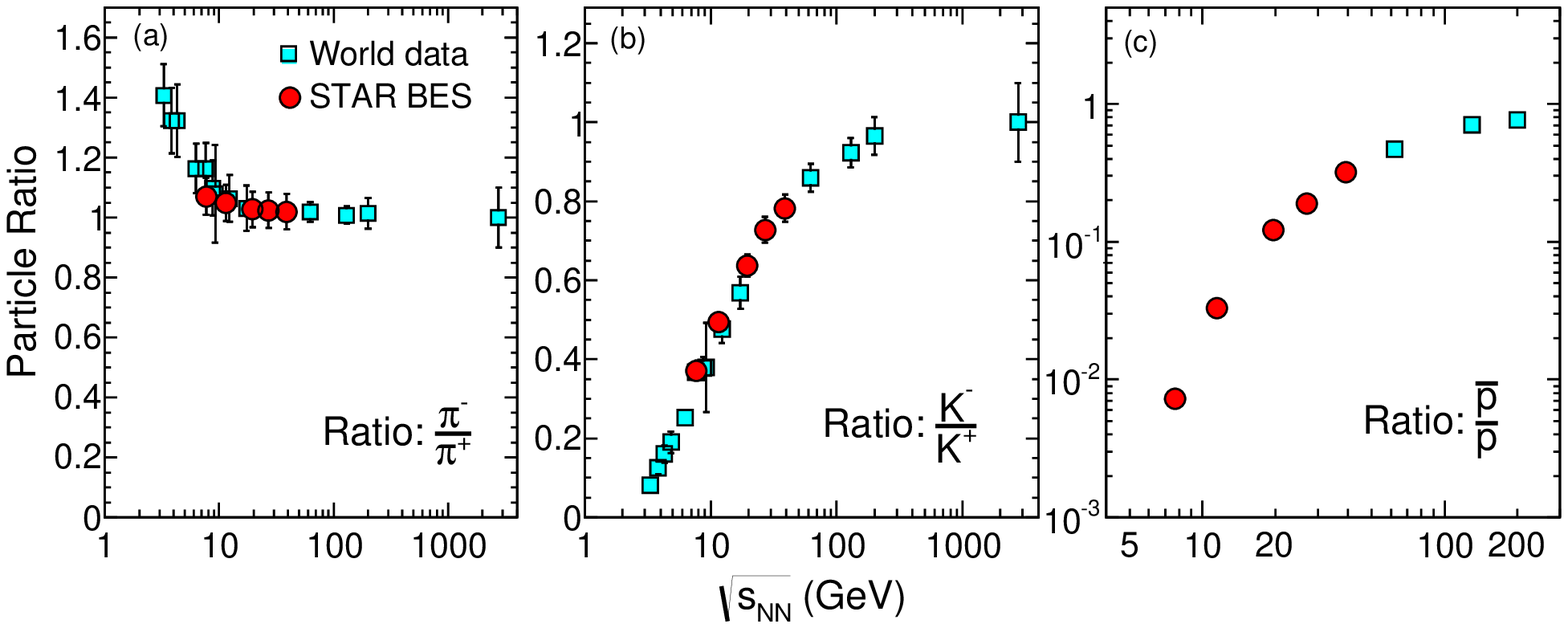}
\vspace{-0.5cm}
\caption{ (Color online) (a) 
$\pi^{-}/\pi^{+}$, (b) $K^{-}/K^{+}$, and (c) $\bar{p}/p$
ratios at midrapidity ($|y| < 0.1$) in central 0--5\% Au+Au collisions at 
$\sqrt{s_{NN}}$ = 7.7, 11.5, 19.6, 27, and 39 GeV, compared to
previous results from
AGS~\cite{Akiba:1996xf,Ahle:1998jc,Ahle:1999uy,Barrette:1999ry,Ahle:1999in,Ahle:1999va,Ahle:2000wq,Klay:2001tf},
SPS~\cite{Afanasiev:2002mx,Anticic:2004yj,Alt:2006dk,Alt:2007aa},
RHIC~\cite{Abelev:2008ab,Abelev:2009bw}, and LHC~\cite{Abelev:2013vea}. AGS results
correspond to 0--5\%, SPS to 0--7\%,  top RHIC to 0--5\% (62.4 and 200
GeV) and 0-6\% (130 GeV), and LHC to 0--5\%
central collisions.
Errors shown are the quadrature sum of statistical and systematic
uncertainties
where the latter dominates.
}
\label{ratioen}
\end{center}
\end{figure*}
\begin{figure}
\begin{center}
\includegraphics[width=9cm]{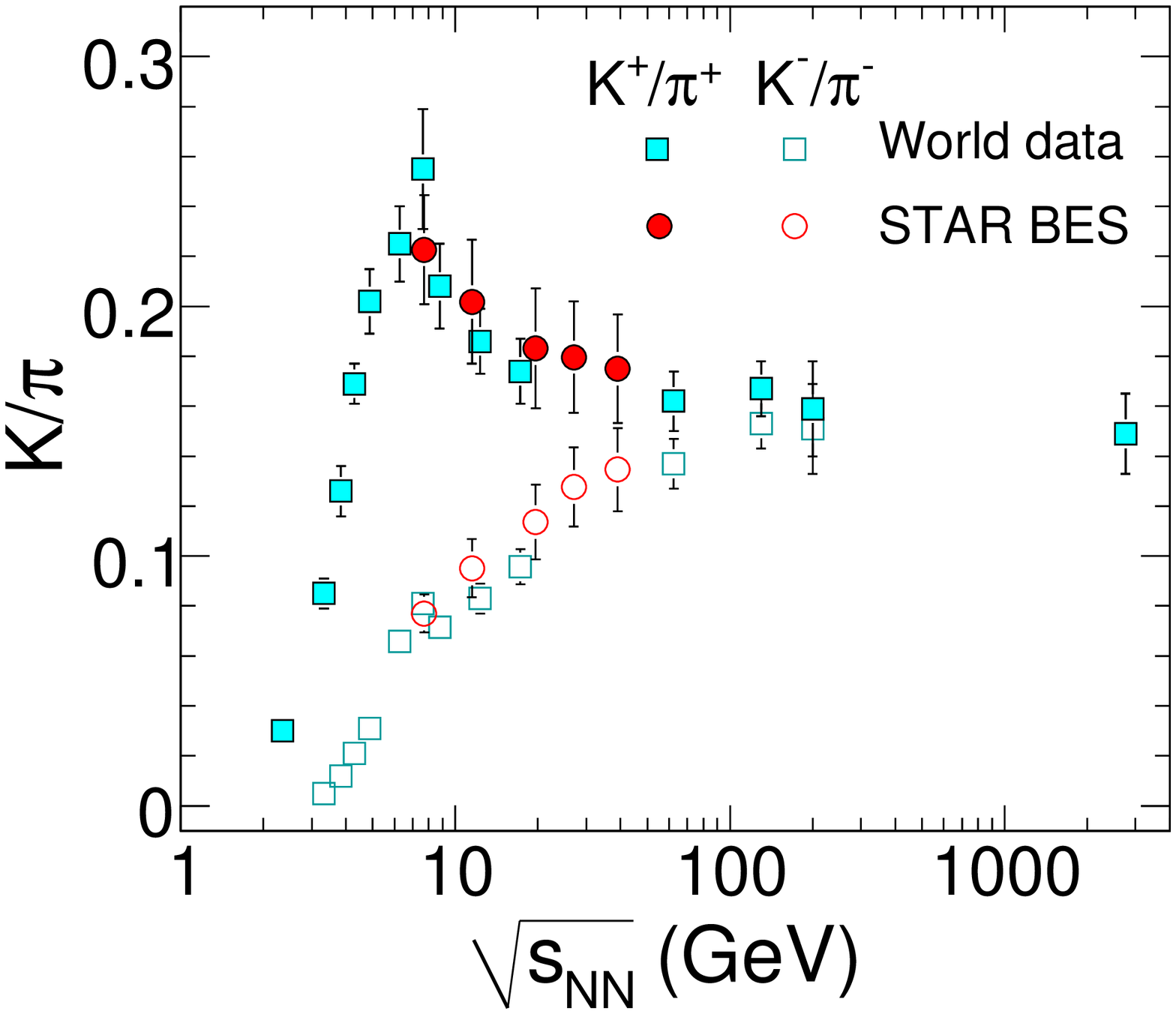}
\vspace{-0.7cm}
\caption{ (Color online)
$K/\pi$ ratio at midrapidity ($\mid y \mid < 0.1$) for central 0--5\% Au+Au collisions at 
$\sqrt{s_{NN}}$ = 7.7, 11.5, 19.6, 27, and 39 GeV, compared to
previous results from
AGS~\cite{Akiba:1996xf,Ahle:1998jc,Ahle:1999uy,Barrette:1999ry,Ahle:1999in,Ahle:1999va,Ahle:2000wq,Klay:2001tf},
SPS~\cite{Afanasiev:2002mx,Anticic:2004yj,Alt:2006dk,Alt:2007aa},
RHIC~\cite{Abelev:2008ab,Abelev:2009bw}, and LHC~\cite{Abelev:2013vea}. AGS results
correspond to 0--5\%, SPS to 0--7\%, top RHIC to 0--5\% (62.4 and 200
GeV) and 0-6\% (130 GeV), and LHC to 0--5\%
central collisions.
Errors shown are the quadrature sum of statistical and systematic
uncertainties
where the latter dominates.
}
\label{ratiok2pi}
\end{center}
\end{figure}
Figure~\ref{dndetaen} shows the $dN/dy$ of $\pi^{\pm}$, $K^{\pm}$, and
$p$/$\bar{p}$, at midrapidity normalized by
$\langle N_{\mathrm {part}} \rangle$/2 as a function of $\sqrt{s_{NN}}$.
The results from 0-5\% central Au+Au collisions at the BES are in agreement with the general energy dependence
trend observed at the
AGS~\cite{Akiba:1996xf,Ahle:1998jc,Ahle:1999uy,Barrette:1999ry,Ahle:1999in,Ahle:1999va,Ahle:2000wq,Klay:2001tf},
SPS~\cite{Afanasiev:2002mx,Anticic:2004yj,Alt:2006dk,Alt:2007aa},
RHIC~\cite{Adler:2004zn,Abelev:2008ab,Abelev:2009bw}, and LHC~\cite{Abelev:2013vea}. 
It may be noted that the energy dependence of pion yields show a linear
increase as a function of collision energy, but exhibit a kink
structure around 19.6 GeV. 
This may suggest a change in the particle production mechanism
around $\sqrt{s_{NN}} =$ 19.6 GeV. 

The energy dependence of kaon yields shows an interesting trend. There
is a significant difference between 
$K^{+}$ and $K^{-}$ production at beam energies 
from AGS to BES.
At these energies, $K^+$ production is a result of an interplay between
associated production and pair production,
while $K^-$ production is dominated by pair production.
The associated production
dominates at the low end of this range, while pair production becomes
more important at the upper end. 

The energy dependence of proton yields
reflects the increase in baryon density due to baryon stopping at
lower energies. At top RHIC energies, the proton and anti-proton
yields are of similar order, which is expected from pair production
mechanism. 
At lower energies, protons have a contribution due to baryon stopping
also, leading to higher yields at 7.7 GeV compared to 200 GeV.
The anti-proton
yields show an increase with increasing energy. 

\subsubsection{Mean Transverse Mass ($\langle m_T \rangle$)}
Figure~\ref{energydndymt} shows the energy dependence of $\langle
m_{T}\rangle - m$ for $\pi^{\pm}$, $K^{\pm}$, $p$, and $\bar{p}$. Results are shown for 0--5\% central 
Au+Au collisions at BES energies, and are compared to previous 
results from
AGS~\cite{Akiba:1996xf,Ahle:1998jc,Barrette:1999ry,Ahle:1999in,Ahle:1999va,Ahle:1999uy,Ahle:2000wq,Klay:2001tf},
SPS~\cite{Afanasiev:2002mx,Anticic:2004yj,Alt:2006dk,Alt:2007aa},
RHIC~\cite{Abelev:2008ab}, and LHC~\cite{Abelev:2013vea}. 
The $\langle m_{T}\rangle - m$
values increase with $\sqrt{s_{NN}}$ at lower AGS energies, stay
independent of $\sqrt{s_{NN}}$ at the higher SPS and BES energies,
then tend to rise further with 
increasing $\sqrt{s_{NN}}$ at the higher beam energies at RHIC and LHC. For a thermodynamic 
system, $\langle m_{T}\rangle - m$ can be an approximate representation of
the temperature of the system, and  $dN/dy$ $\propto$ $\ln(\sqrt{s_{NN}})$ 
may represent its 
entropy~\cite{Landau:1953gr,Hama:1983cv,VanHove:1982vk}. In such a
scenario, the energy dependence of $\langle m_{T}\rangle - m$ could reflect the characteristic 
signature of a first order 
phase transition, as proposed by Van Hove~\cite{VanHove:1982vk}. Then the constant value 
of $\langle m_{T}\rangle - m$ vs. $\sqrt{s_{NN}}$ around BES energies could be
interpreted as reflecting 
the formation of a mixed phase of a QGP and hadrons during the evolution of the heavy-ion
system. However, there could be several other effects to which $\langle m_{T}\rangle - m$ is 
sensitive, which also need to be understood for proper interpretation of the data~\cite{Mohanty:2003fy}. 

\subsubsection{Particle Ratios}
Figure~\ref{ratioen} shows the collision energy dependence of the 
particle ratios 
$\pi^{-}/\pi^{+}$, $K^{-}/K^{+}$, and $\bar{p}/p$, in central heavy-ion collisions. 
The new results from Au+Au collisions at BES energies
follow the $\sqrt{s_{NN}}$ trend established by previous measurements
from AGS~\cite{Akiba:1996xf,Ahle:1998jc,Barrette:1999ry,Ahle:1999in,Ahle:1999va,Ahle:1999uy,Ahle:2000wq,Klay:2001tf},
SPS~\cite{Afanasiev:2002mx,Anticic:2004yj,Alt:2006dk,Alt:2007aa},
RHIC~\cite{Abelev:2008ab}, and LHC~\cite{Abelev:2013vea}. 
The $p_{T}$-integrated $\pi^{-}/\pi^{+}$ ratios
at very low beam energies have values 
larger than unity, which is likely due to significant contributions from resonance 
decays (such as from $\Delta$ baryons). 
The $K^{-}/K^{+}$ ratios at BES energies 
are much less than unity, indicating a significant contribution to $K^+$ production from associated production at lower collision energies. With increasing $\sqrt{s_{NN}}$, the $K^{-}/K^{+}$ ratio
approaches unity, indicating dominance of kaon pair production.
The lower values of the $\bar{p}/p$ ratios at BES energies indicates large values of net-protons ($p - \bar{p}$) and  large baryon stopping  in these collisions. The $\bar{p}/p$ ratio increases with
increasing collision energy and approaches unity for top RHIC energies. This indicates that
at higher beam energies the collisions have a larger degree of transparency, and the $p$ ($\bar{p}$)
production at midrapidity is dominated by pair production.  

Figure~\ref{ratiok2pi} shows the energy dependence of $K/\pi$ particle
ratio. BES results are compared with those from AGS~\cite{Akiba:1996xf,Ahle:1998jc,Barrette:1999ry,Ahle:1999in,Ahle:1999va,Ahle:1999uy,Ahle:2000wq,Klay:2001tf},
SPS~\cite{Afanasiev:2002mx,Anticic:2004yj,Alt:2006dk,Alt:2007aa},
RHIC~\cite{Abelev:2008ab}, and LHC~\cite{Abelev:2013vea}.  
The $K/\pi$ ratio is of interest, as it reflects the strangeness
content relative to 
entropy in heavy-ion collisions. An enhancement 
in $K/\pi$ ratio in heavy-ion collisions 
 compared to $p+p$ collisions
has been taken previously as an indication of QGP formation~\cite{Alt:2007aa}.
The increase in $K^{+}/\pi^{+}$ ratio with beam energies up to 
$\sqrt{s_{NN}} =$ 7.7 GeV at SPS and the subsequent decrease and possible saturation 
with increasing beam energies has been a subject of intense theoretical debate~\cite{Afanasiev:2002mx,Anticic:2004yj,Nayak:2005tz,Cleymans:2005zx,Alt:2006dk,Tomasik:2006qs,Alt:2007aa,Andronic:2008gu,Rafelski:2008av,Chatterjee:2009km,Tawfik:2017ehz}. 
The discussions mainly focus on the question of the relevant degrees of freedom that are necessary 
to explain the energy dependence of the $K/\pi$ ratio. Our new results
from BES Au+Au collisions are found to be consistent with the
previously observed energy dependence. The peak position (usually 
called the ``horn'') in energy dependence of
$K^{+}/\pi^{+}$ has been suggested as the
signature of a phase transition from hadron gas to a QGP while going from lower to higher energies. However, various
models that do not include such a phase
transition could also explain this type of energy dependence of the
$K^{+}/\pi^{+}$ ratio. It may be noted that the peak position around
7.7 GeV corresponds to an energy where the maximum baryon density
is predicted to be achieved in heavy-ion collisions~\cite{Busza:1983rj,Cleymans:2005zx}.

\section{Freeze-out Parameters}
\begin{figure}
\begin{center}
\includegraphics*[width=9.4cm]{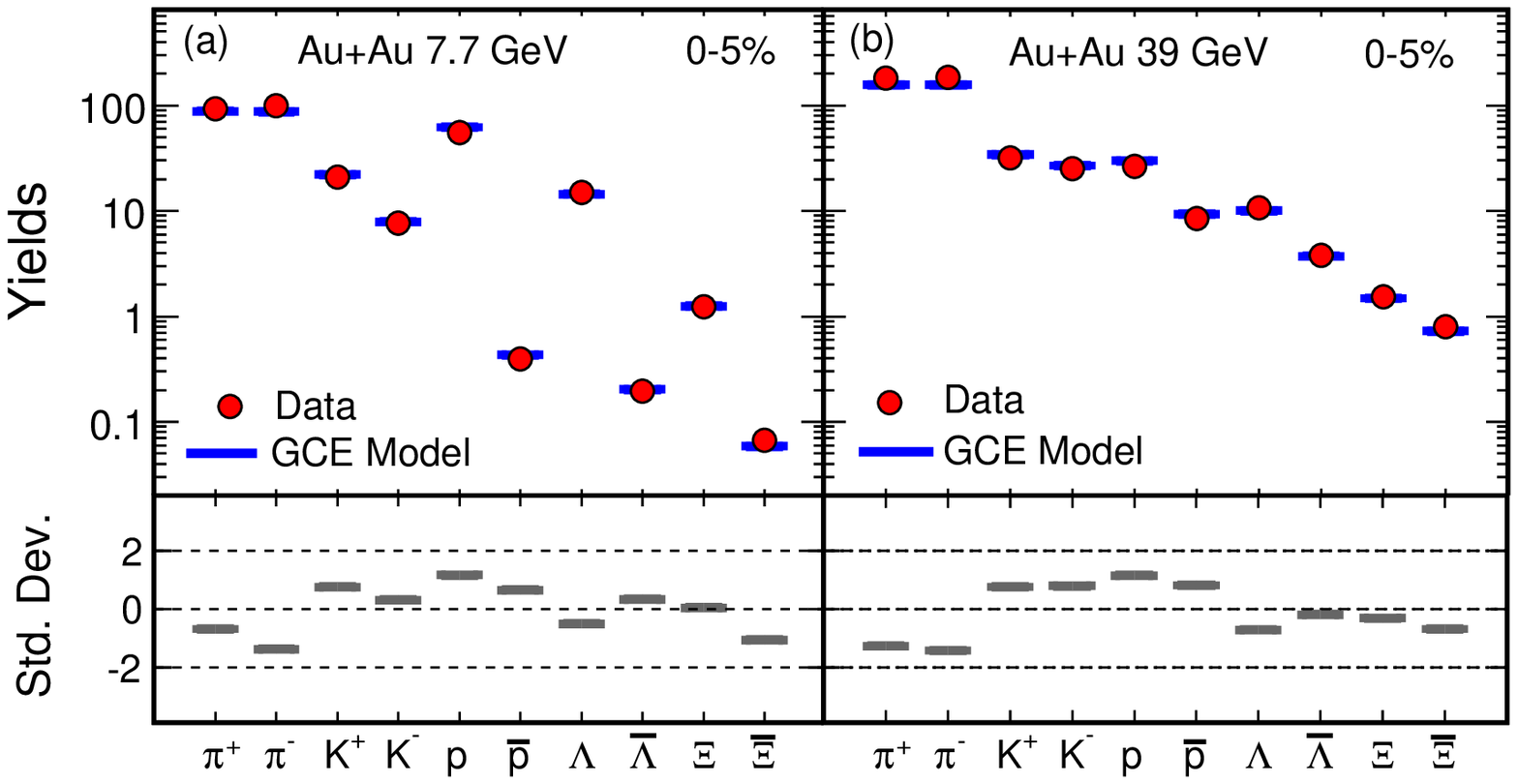}
\includegraphics*[width=9.4cm]{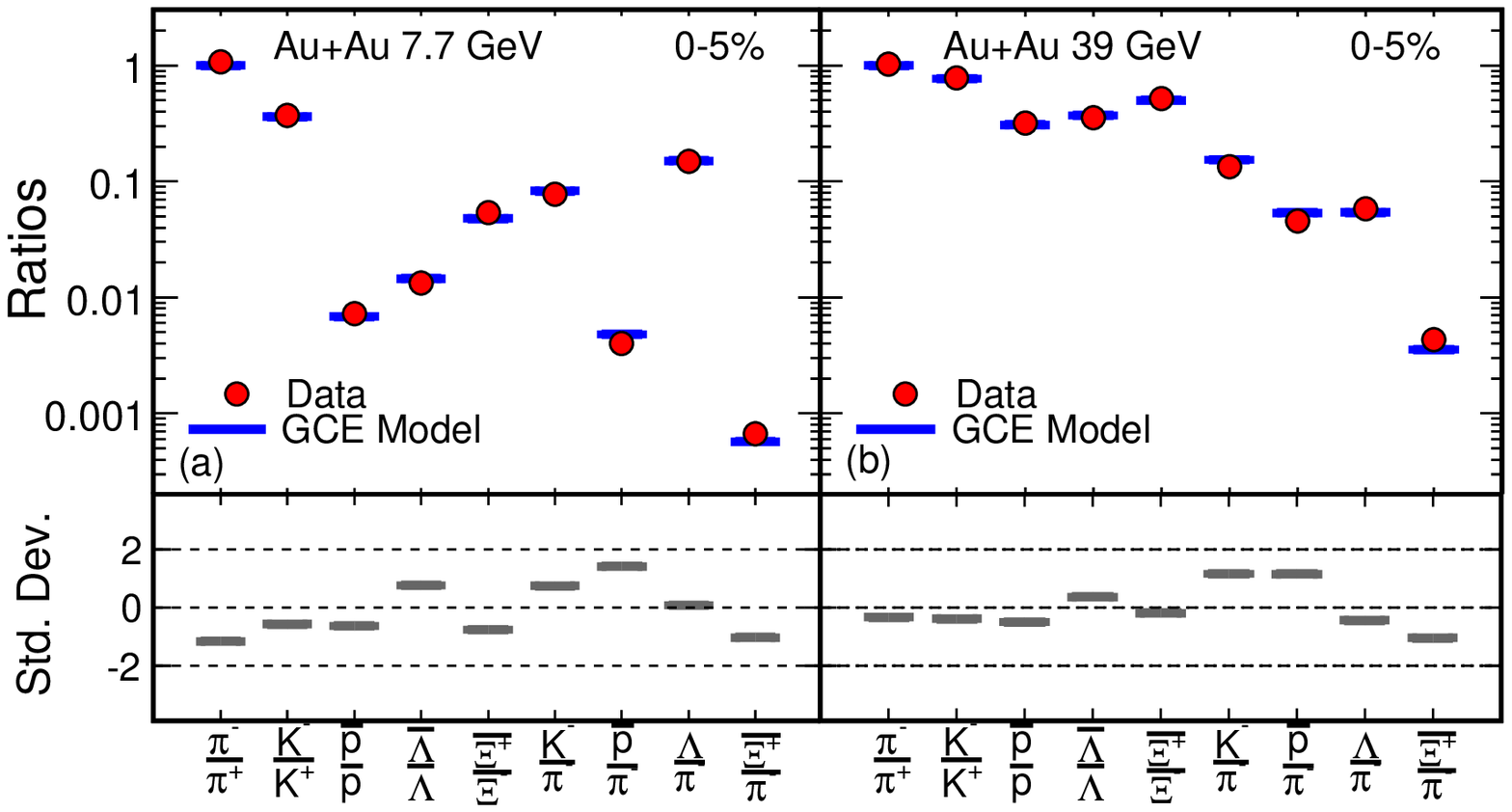}
\vspace{-0.6cm}
\caption{(Color online) The GCE model fits shown along with standard
  deviations for (a) Au+Au 7.7
  and (b) Au+Au 39 GeV in 0--5\% central collisions. Top panels are
  for the particle yields fit and lower panels are for the particle
  ratios fit. Uncertainties on experimental data represent statistical and
  systematic uncertainties added in quadrature. Here, the uncertainties
  are smaller than the symbol size.
}
\vspace{-0.5cm}
\label{fig:chem_fits}
\end{center}
\end{figure}
\begin{figure}
\begin{center}
\includegraphics*[width=9.4cm]{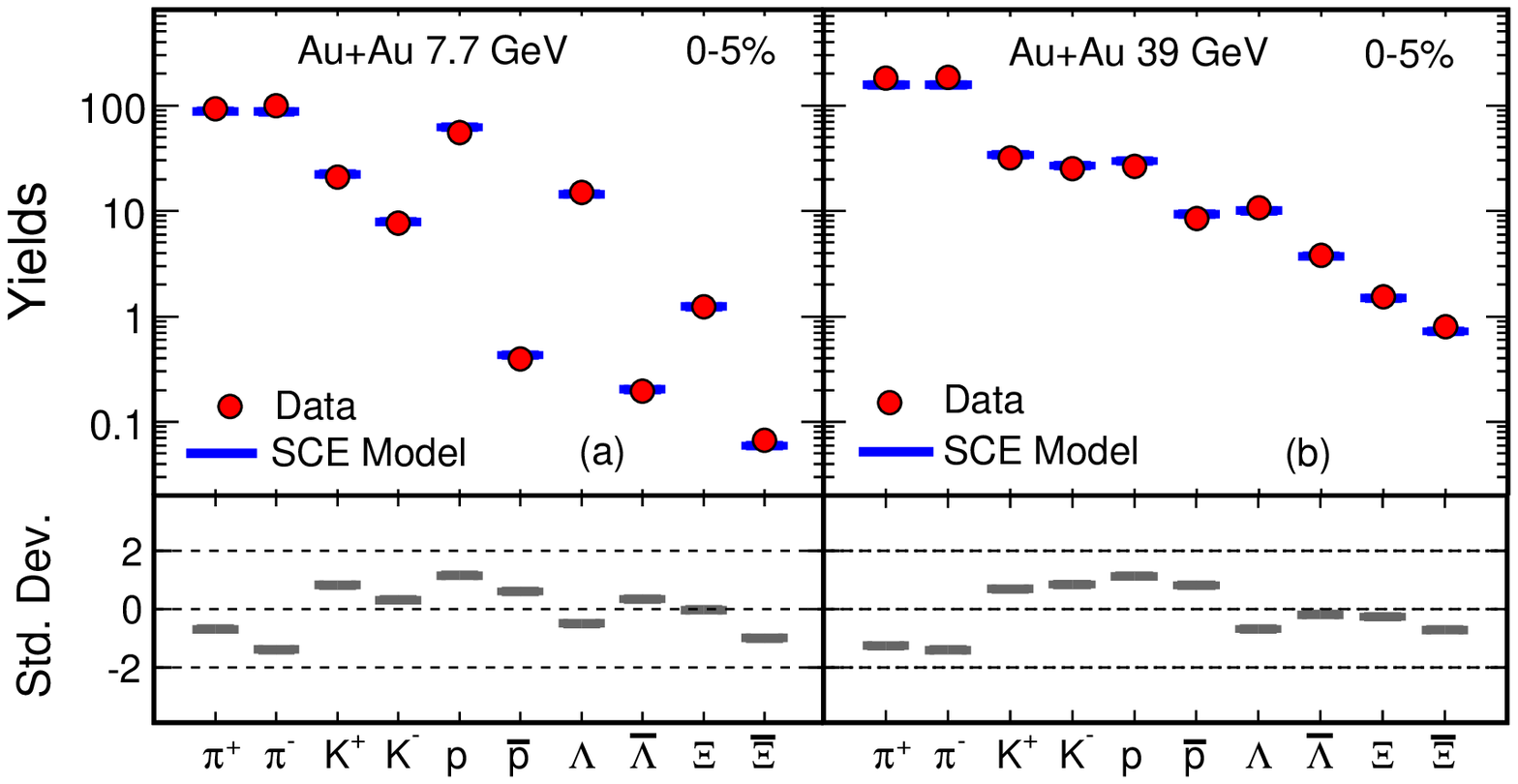}
\includegraphics*[width=9.4cm]{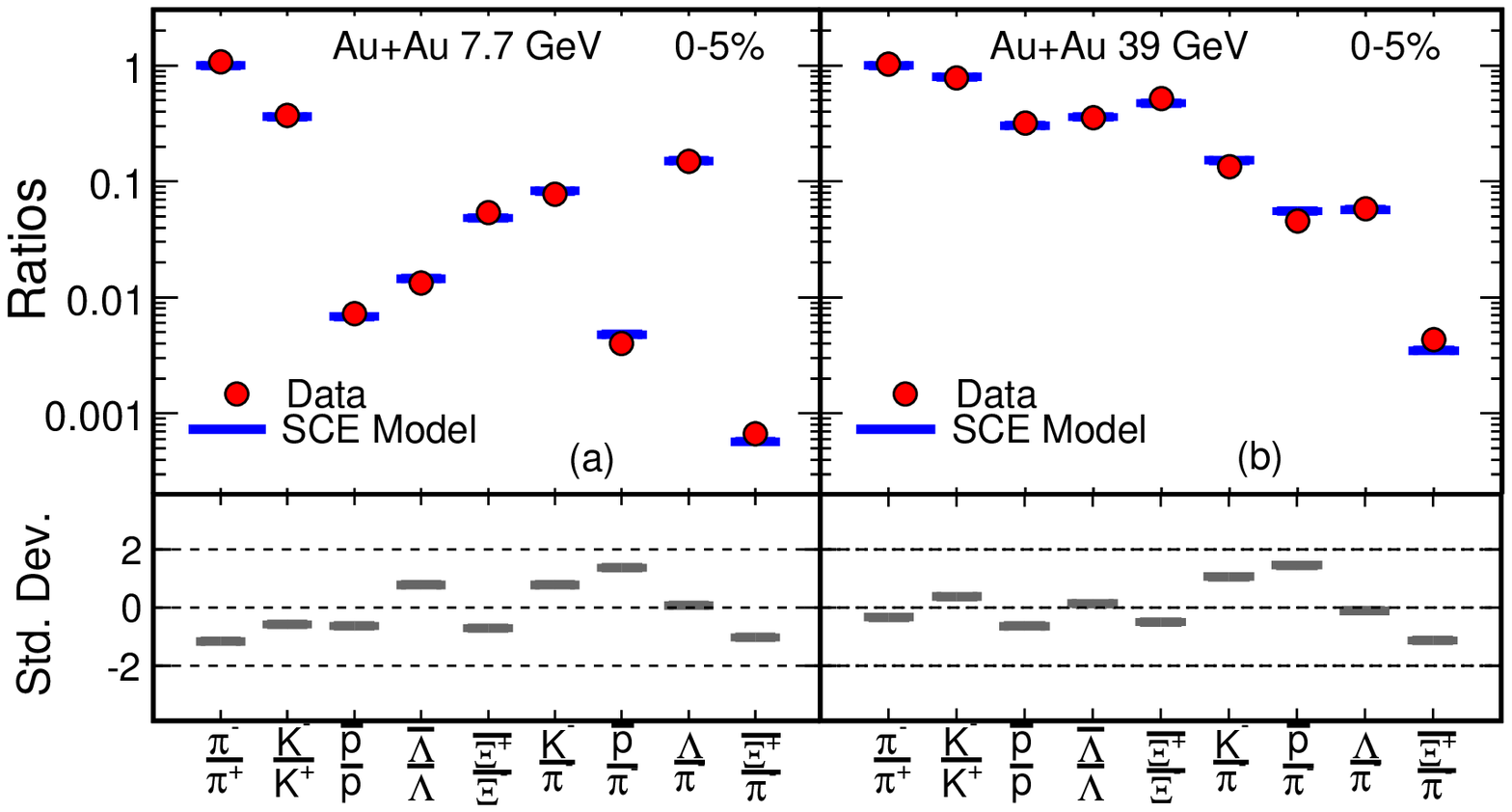}
\vspace{-0.6cm}
\caption{(Color online) The SCE model fits shown along with standard
  deviations for (a) Au+Au 7.7
  and (b) Au+Au 39 GeV in 0--5\% central collisions. Top panels are
  for the particle yields fit and lower panels are for particle
  ratios fit. Uncertainties on experimental data represent statistical and
  systematic uncertainties added in quadrature. Here, the uncertainties
  are smaller than the symbol size.
}
\vspace{-0.5cm}
\label{fig:chem_fits2}
\end{center}
\end{figure}
\begin{figure*}
\begin{center}
\includegraphics*[width=15.cm]{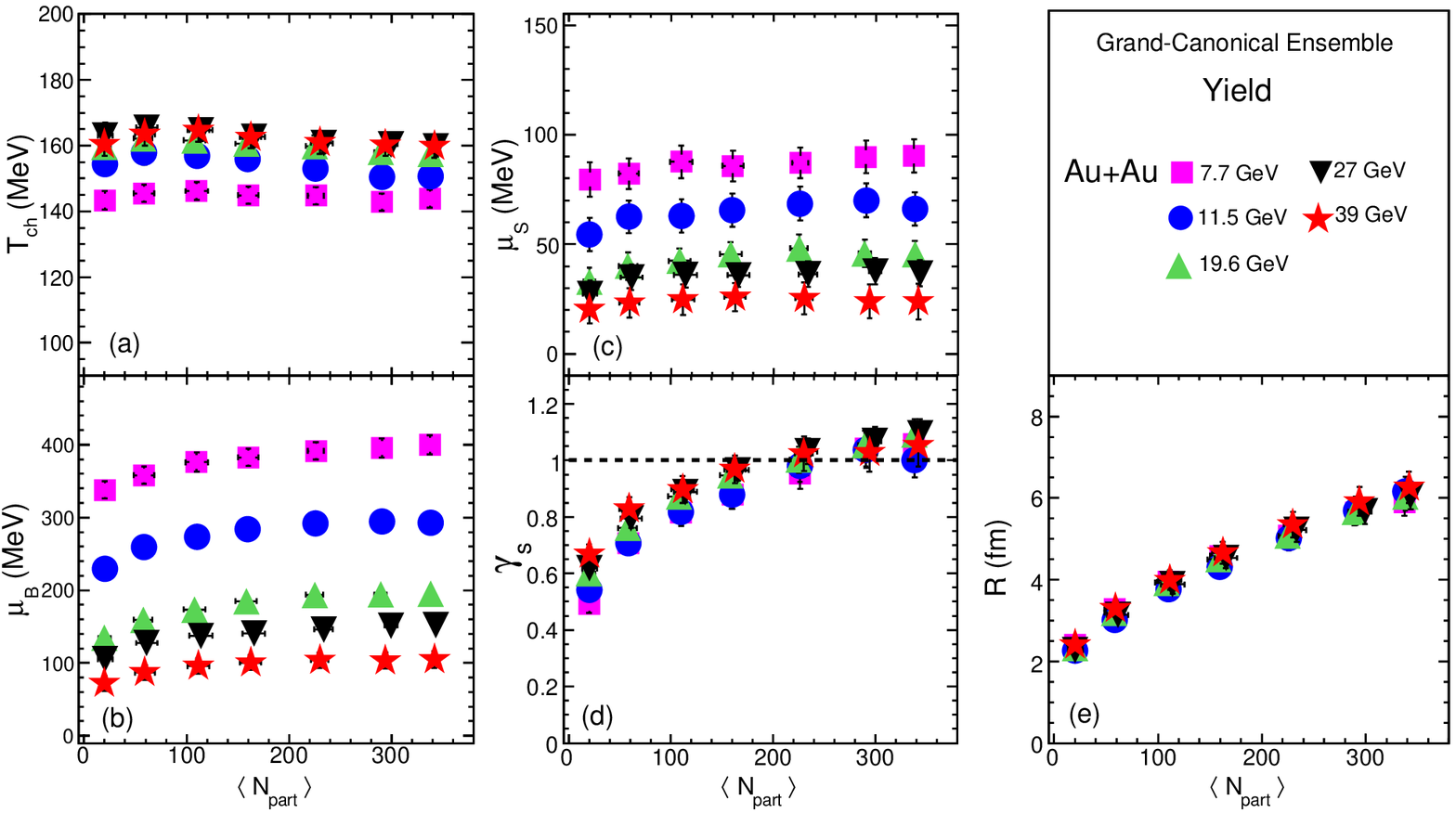}
\vspace{-0.4cm}
\caption{(Color online) Chemical freeze-out parameters (a) $T_{\rm{ch}}$,
  (b) $\mu_B$, (c) $\mu_S$, (d) $\gamma_S$, and (e) $R$ plotted versus $\langle
  N_{\rm{part}} \rangle$
  in GCE for particle yields fit.  Uncertainties represent systematic errors.
}
\vspace{-0.5cm}
\label{fig:gce_yr}
\end{center}
\end{figure*}
\begin{figure*}
\begin{center}
\includegraphics*[width=12.cm]{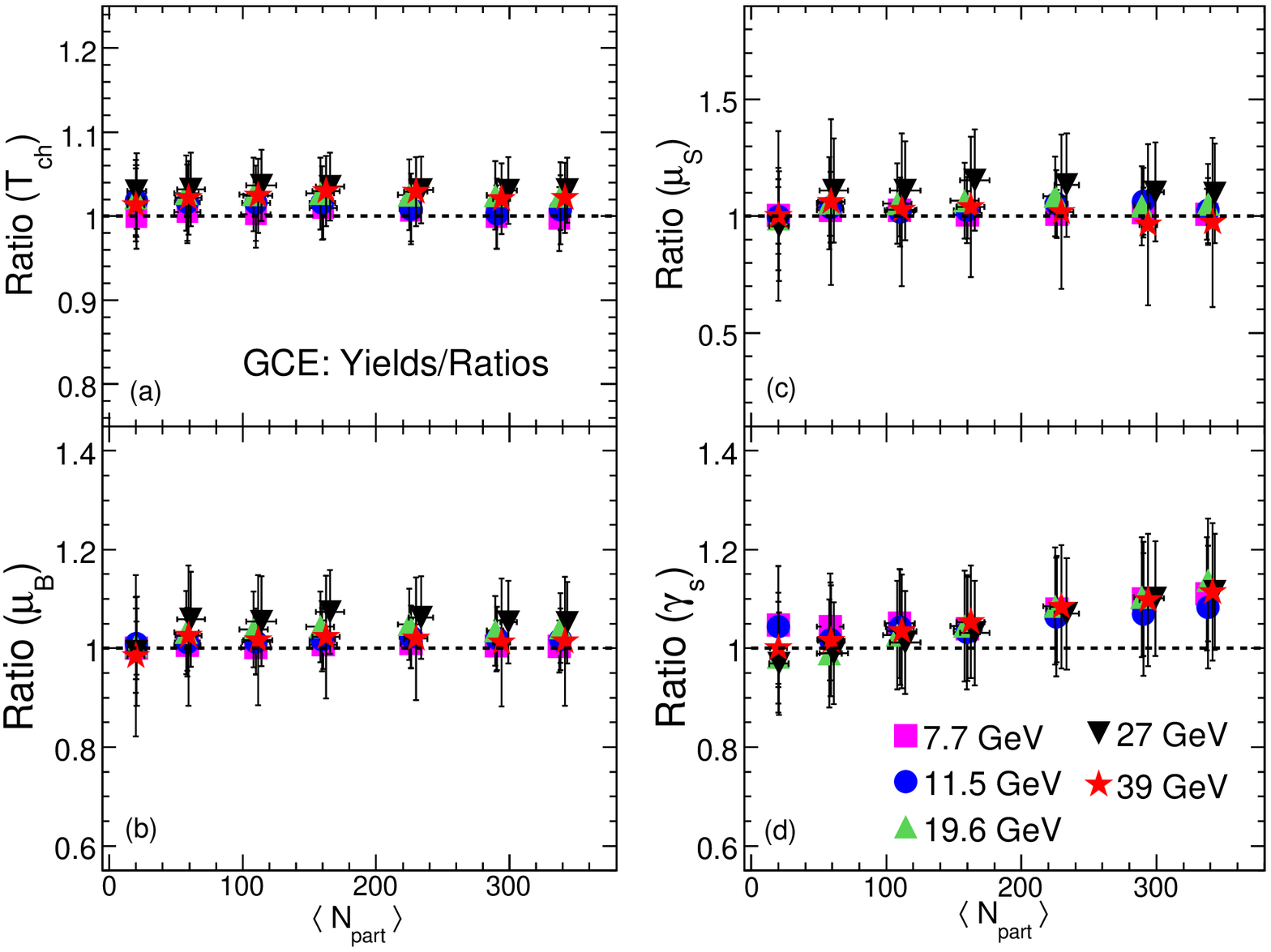}
\vspace{-0.4cm}
\caption{(Color online) Ratio of chemical freeze-out parameters (a) $T_{\rm{ch}}$,
  (b) $\mu_B$, (c) $\mu_S$, and (d) $\gamma_S$ between results from particle yield
  fits to particle ratio fits in GCE plotted versus $\langle
  N_{\rm{part}} \rangle$. Uncertainties represent  systematic errors.
}
\vspace{-0.5cm}
\label{fig:gce_yr_r}
\end{center}
\end{figure*}
The integrated invariant yields and $p_T$ spectra of hadrons 
provide information about the system at freeze-out. Two types of
freeze-out are commonly discussed in heavy-ion collisions ---
chemical freeze-out and kinetic freeze-out. 
The state
when the inelastic interactions among the particles stop is referred
to as chemical freeze-out. 
The yields of the produced particles become fixed at chemical freeze-out.
Statistical thermal models have successfully described the chemical
freeze-out stage with system parameters such as chemical
freeze-out temperature, $T_{\rm{ch}}$, and baryon chemical potential,
$\mu_B$~\cite{Adams:2005dq,Abelev:2008ab,Andronic:2008ev,Wheaton:2004qb,Tawfik:2014dha}. 

After chemical freeze-out, elastic interactions among the particles are
still ongoing which lead to changes in the momenta of
the particles. When the average inter-particle
distance becomes so large that elastic interactions stop, the
system is said to have undergone kinetic freeze-out. At this stage, the
transverse momentum spectra of the produced particles become
fixed. Hydrodynamics inspired models such as the Blast Wave Model~\cite{Adams:2005dq,Schnedermann:1993ws,Abelev:2008ab}
have described the kinetic freeze-out scenario with a common temperature
$T_{\rm{kin}}$ and average transverse radial flow velocity $\langle \beta
\rangle$ which reflects the expansion in the transverse direction. In the
following subsections, we  discuss these freeze-out parameters in detail.

\subsection{Chemical Freeze-out}
\begin{figure*}
\begin{center}
\includegraphics*[width=14.cm]{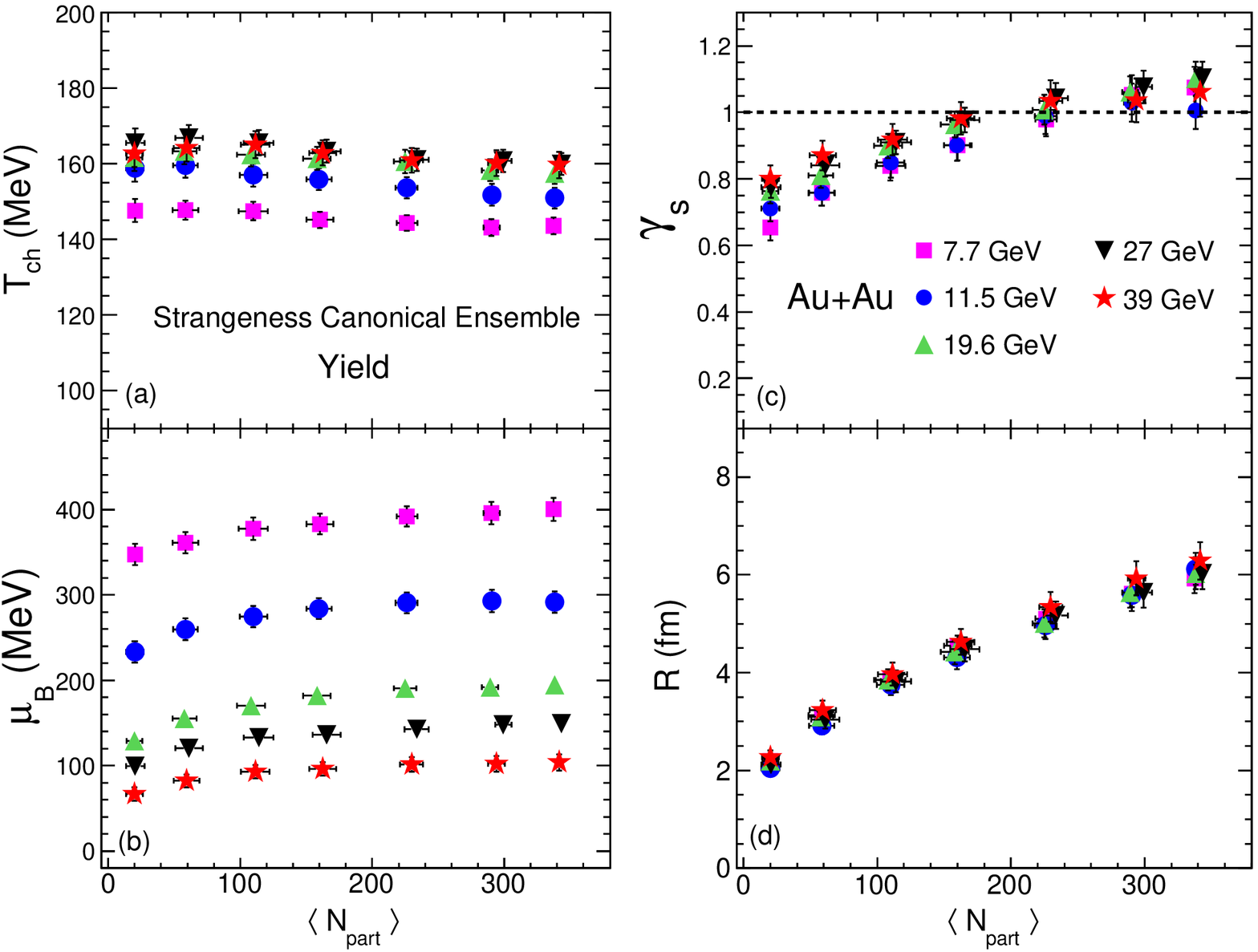}
\vspace{-0.4cm}
\caption{(Color online) Chemical freeze-out parameters (a) $T_{\rm{ch}}$,
 (b) $\mu_B$, (c) $\gamma_S$, and (d) $R$ plotted versus $\langle
  N_{\rm{part}} \rangle$
  in SCE for particle yields fit. Uncertainties represent  systematic errors.
}
\vspace{-0.5cm}
\label{fig:sce_yr}
\end{center}
\end{figure*}
\begin{figure*}
\begin{center}
\includegraphics*[width=17.cm]{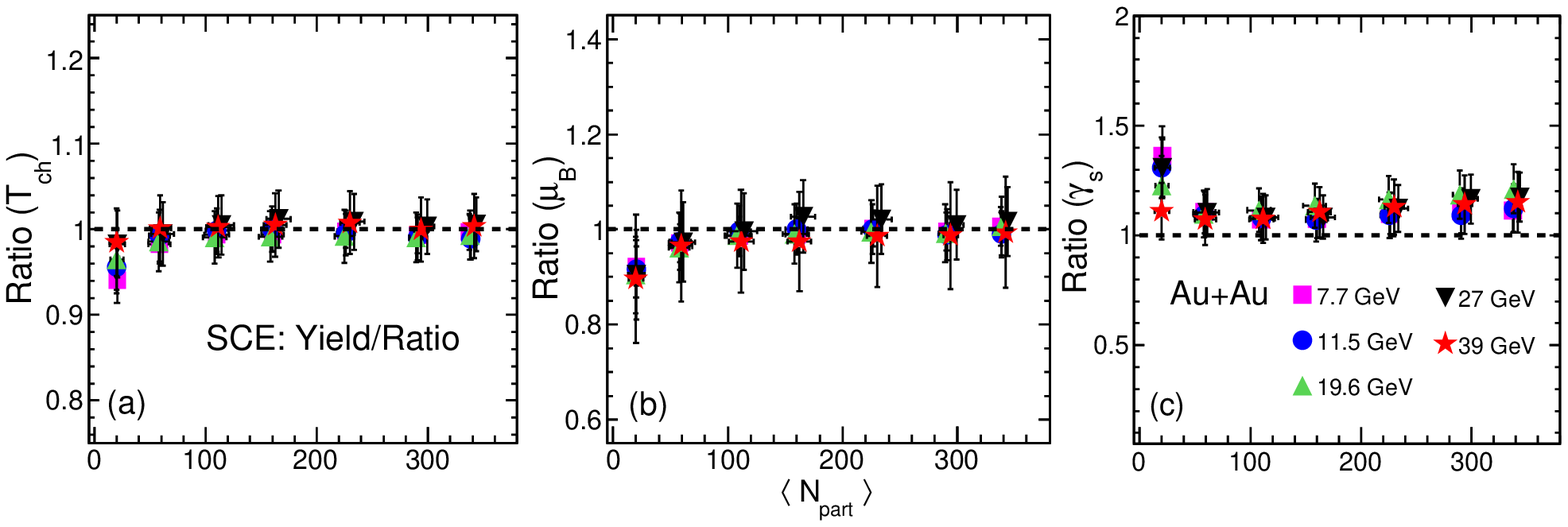}
\vspace{-0.4cm}
\caption{(Color online) Ratio of chemical freeze-out parameters (a) $T_{\rm{ch}}$,
  (b) $\mu_B$, and (c) $\gamma_S$ between yield and ratio fits in SCE plotted versus $\langle
  N_{\rm{part}} \rangle$. Uncertainties represent  systematic errors.
}
\vspace{-0.5cm}
\label{fig:sce_yr_r}
\end{center}
\end{figure*}
\begin{figure*}
\begin{center}
\includegraphics*[width=17.cm]{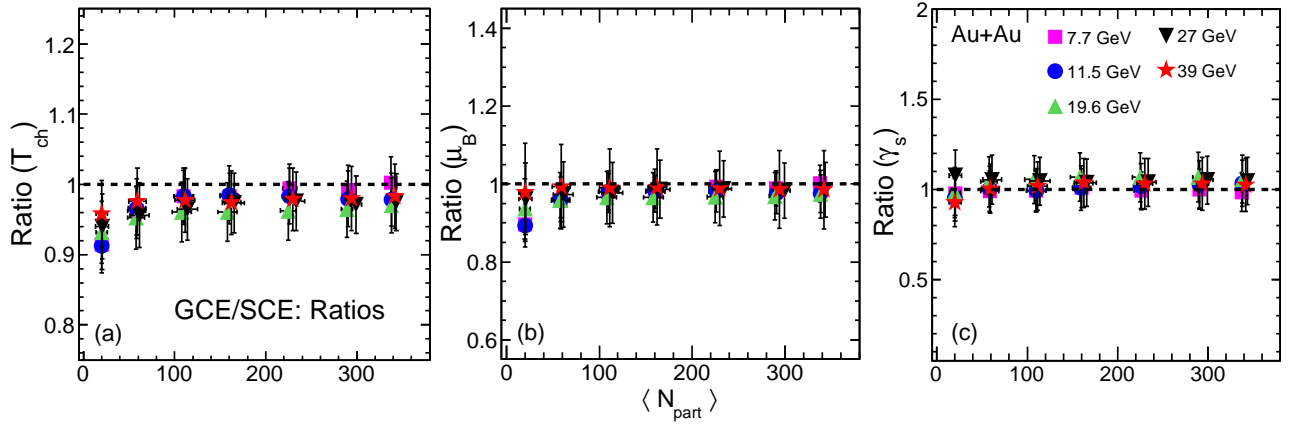}
\vspace{-0.4cm}
\caption{(Color online) Ratio of chemical freeze-out parameters (a) $T_{\rm{ch}}$,
  (b) $\mu_B$, and (c) $\gamma_S$ between GCE and SCE results using particle
  ratios in fits plotted versus $\langle
  N_{\rm{part}} \rangle$. Uncertainties represent  systematic errors.
}
\vspace{-0.5cm}
\label{fig:rat_gs}
\end{center}
\end{figure*}
\begin{figure*}
\begin{center}
\includegraphics*[width=14.cm]{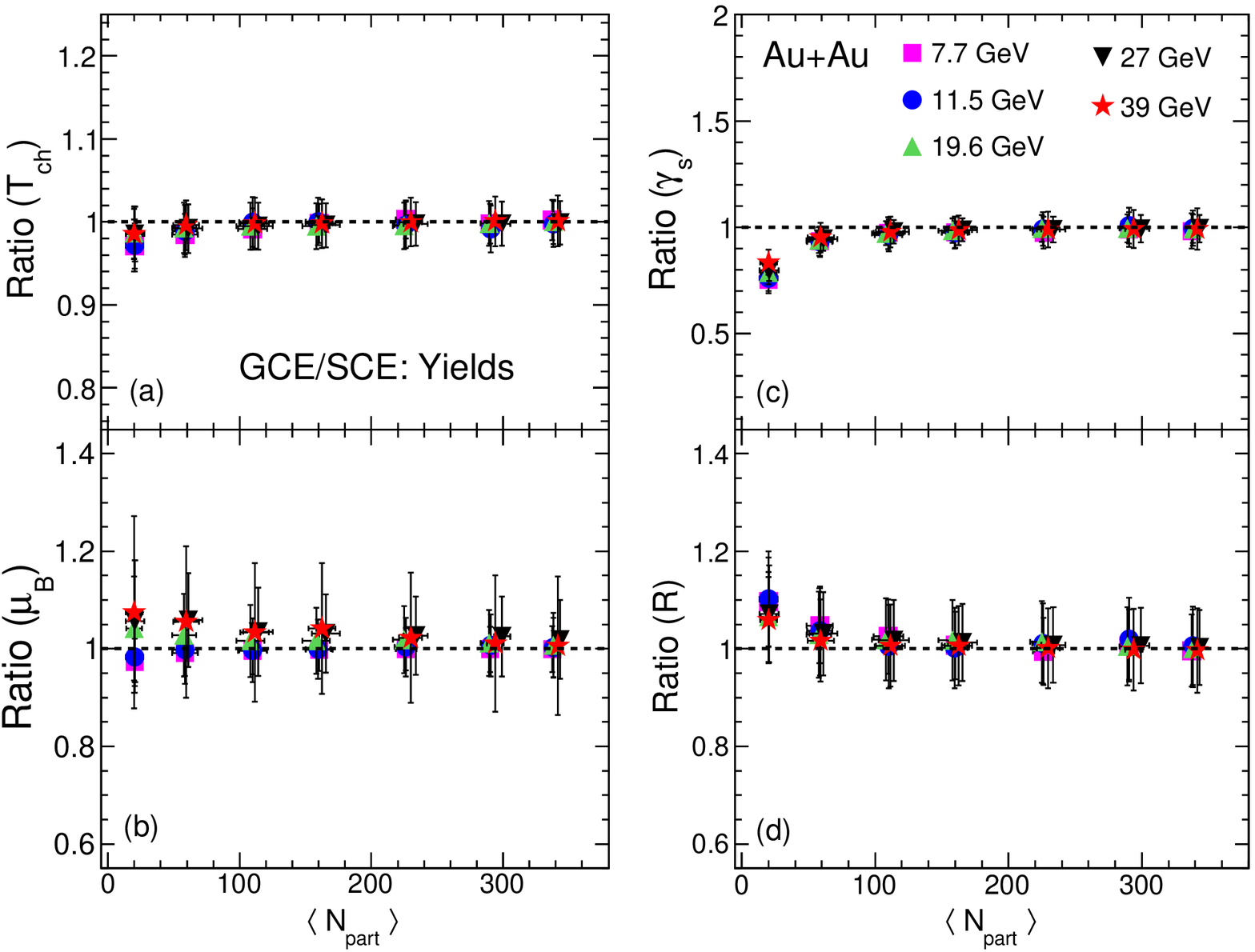}
\vspace{-0.4cm}
\caption{Ratio of chemical freeze-out parameters (a) $T_{\rm{ch}}$,
  (b) $\mu_B$, (c) $\gamma_S$, and (d) $R$ between GCE and SCE results
  using particle yields in fits  plotted versus $\langle
  N_{\rm{part}} \rangle$. Uncertainties represent  systematic errors.
}
\vspace{-0.5cm}
\label{fig:yld_gs}
\end{center}
\end{figure*}

The chemical freeze-out parameters are obtained 
from statistical thermal model 
analyses of the produced particles 
using the THERMUS package~\cite{Wheaton:2004qb}. 
Two approaches are used to obtain the
chemical freeze-out parameters: Grand-Canonical Ensemble (GCE) and
Strangeness Canonical Ensemble (SCE). 
In the GCE, the energy and quantum numbers, or particle numbers, are
conserved on average through the temperature and chemical
potentials. This is reasonable if the number of particles carrying the quantum number is large.
GCE is widely used in high-energy heavy-ion collisions. For the SCE,
the strangeness ($S$) in the system is fixed exactly
by its initial value of $S,$ while the baryon and charge contents are
treated grand-canonically. At lower energies, low production of
strange particles requires a canonical treatment of strangeness~\cite{Cleymans:1997sw}.
Since the BES data cover a wide range of energies from low to high,
both GCE and SCE approaches are studied here. 

\begin{figure}
\begin{center}
\includegraphics*[width=8.5cm]{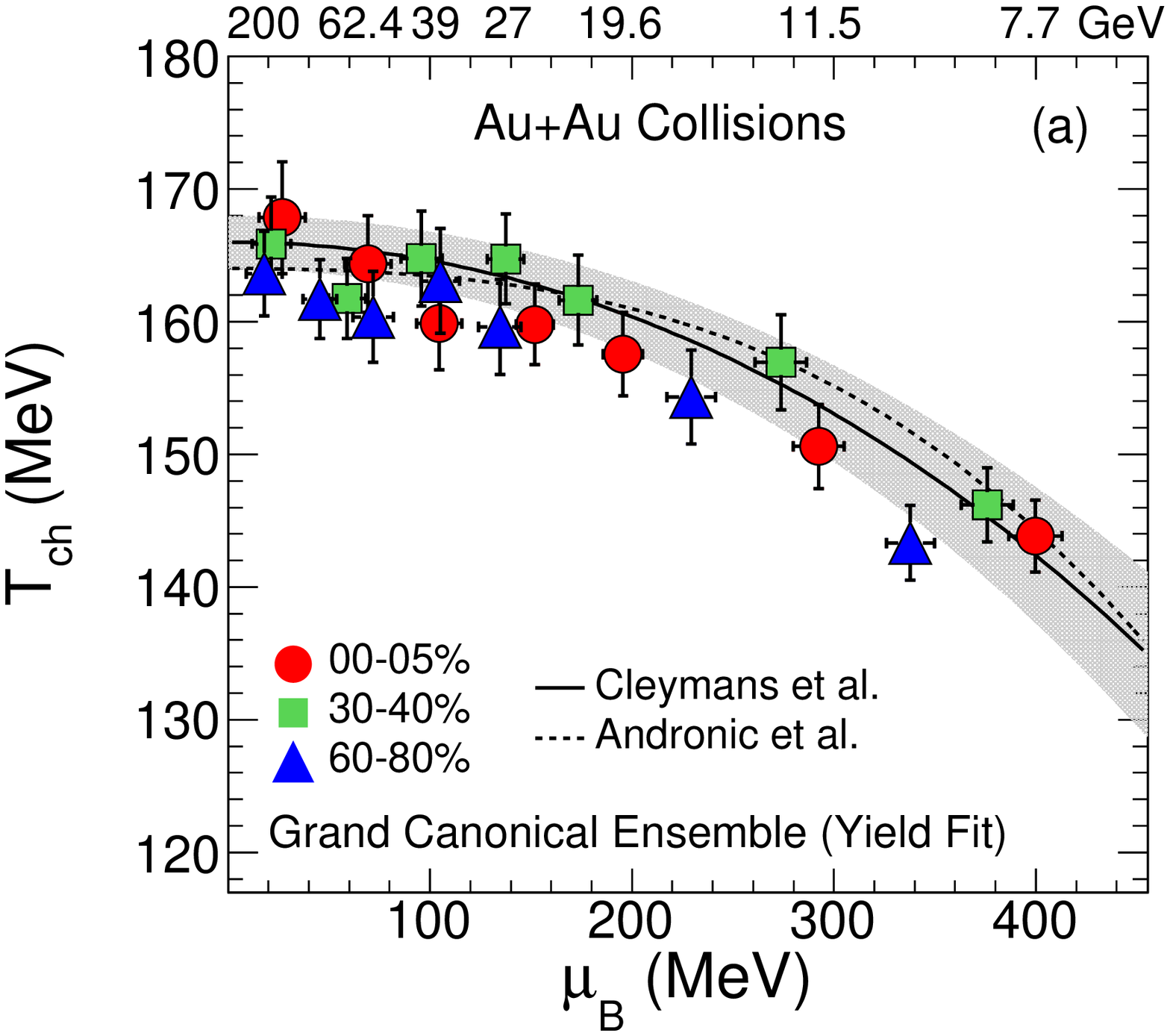}
\includegraphics*[width=8.5cm]{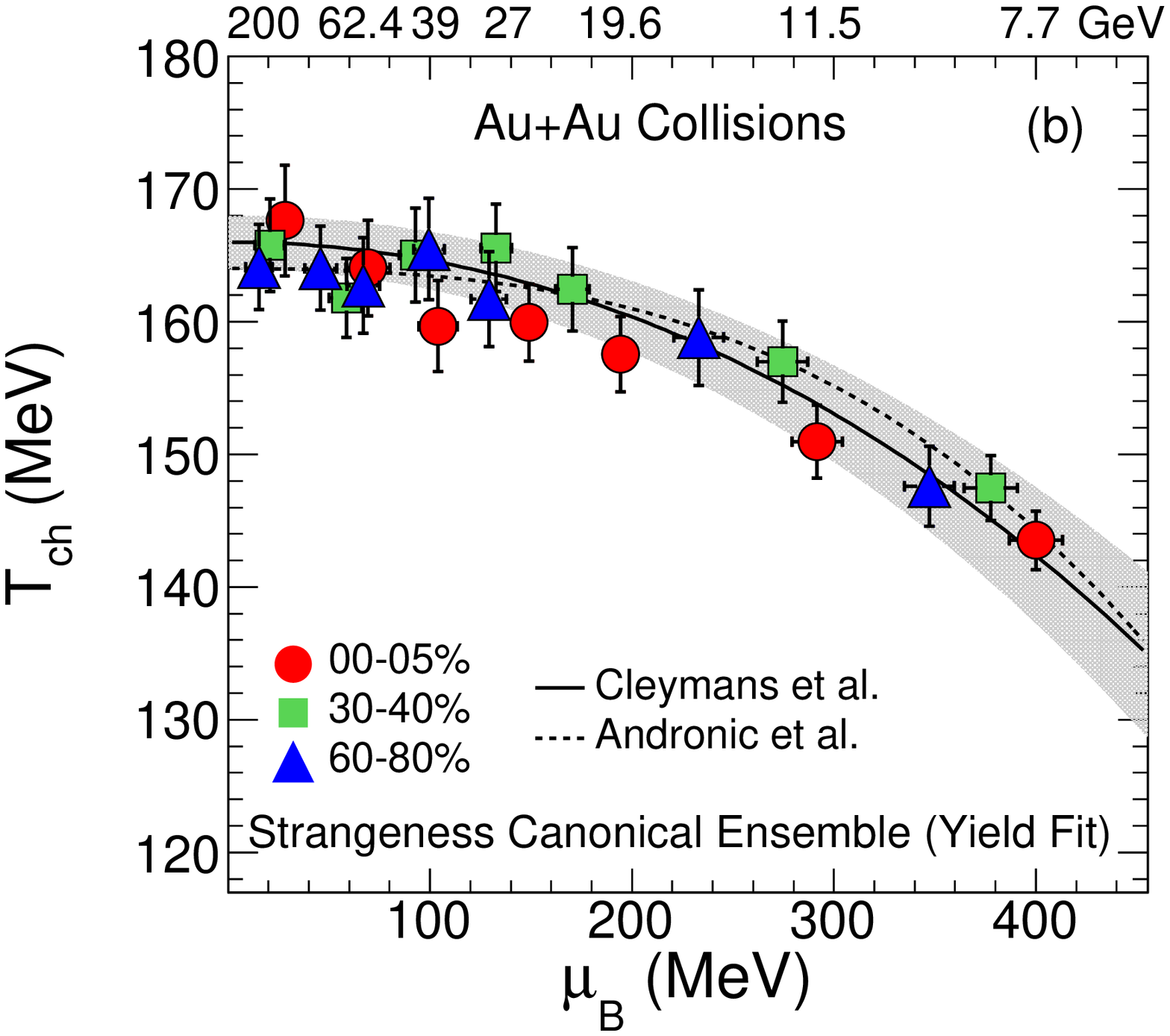}
\vspace{-0.6cm}
\caption{(Color online)  Extracted chemical freeze-out temperature versus baryon chemical
  potential for (a) GCE and (b) SCE cases using particle yields as input for
  fitting. Curves represent two model
  predictions~\cite{Cleymans:2005xv,Andronic:2009jd}. The grey bands
  represent the theoretical prediction ranges of the Cleymans {\it et
    al.} model~\cite{Cleymans:2005xv}.
Uncertainties represent  systematic errors.
}
\vspace{-0.5cm}
\label{fig:tmub_all}
\end{center}
\end{figure}

In addition, different approaches have been proposed to fit the data,
i.e. whether particle yields or the particle
ratios should be used in the fit. The fitting of particle ratios leads to the
cancellation of a volume
factor, thus getting rid of an extra parameter. However, 
a possible disadvantage is the use of a common particle to construct different ratios, leading to correlated uncertainties. We investigate the difference between
these two approaches by fitting both the particle ratios and particle
yields in THERMUS.  

Since the freeze-out parameters represent collision system properties,
it is better to also include the other strange particles in the THERMUS
fitting. The results presented here for particle yields are obtained
using yields of $\pi^{\pm}$, $K^{\pm}$, $p$, $\bar{p}$, 
$\Lambda$, $\bar{\Lambda}$, $\Xi$, and $\overline{\Xi}$. The corresponding
results for particle ratios are obtained by using the ratios
$\pi^-/\pi^+$, $K^-/K^+$, $\bar{p}/p$, $\bar{\Lambda}/\Lambda$,
$\overline{\Xi}/\Xi$, $K^-/\pi^-$, $\bar{p}/\pi^-$, $\Lambda/\pi^-$,
and $\overline{\Xi}/\pi^-$. 
The dN/dy of $\Lambda$, $\bar{\Lambda}$, $\Xi$ and $\bar{\Xi}$ are
obtained from the measured $p_{T}$ spectra within $| y| <$ 0.5, and a follow-up paper on the $p_{T}$ spectra of these particles is in preparation (the technical details are currently available in Ref.~\cite{fortheSTAR:2013gwa}).
As mentioned earlier, the (anti-)
proton yields reported here by STAR are inclusive. The corresponding yields in the
THERMUS model are treated in the same manner as in data  i.e. all inclusive. The fraction
of weak-decay  feed-down contribution (from $\Lambda$, $\Sigma$, and
$\Xi$) to the proton yield from THERMUS is found to be 18\% at 7.7 GeV and
up to 29\% at 39 GeV. The weak-decay feed-down contribution to anti-proton
yield is found to be
up to 50\% at 7.7 GeV and 37\% at 39 GeV.
It may be noted that the strange
particle yields ($\Lambda$, $\bar{\Lambda}$, $\Xi$, and
$\overline{\Xi}$) used here are measured for $|y|<$ 0.5 while the
light hadron yields ($\pi^{\pm}$, $K^{\pm}$, $p$, and $\bar{p}$) are
measured for $|y|<$ 0.1. 
The uncertainty due to this
difference is not considered in the
extraction of chemical freeze-out parameters.

\begin{figure*}
\begin{center}
\includegraphics*[width=15cm]{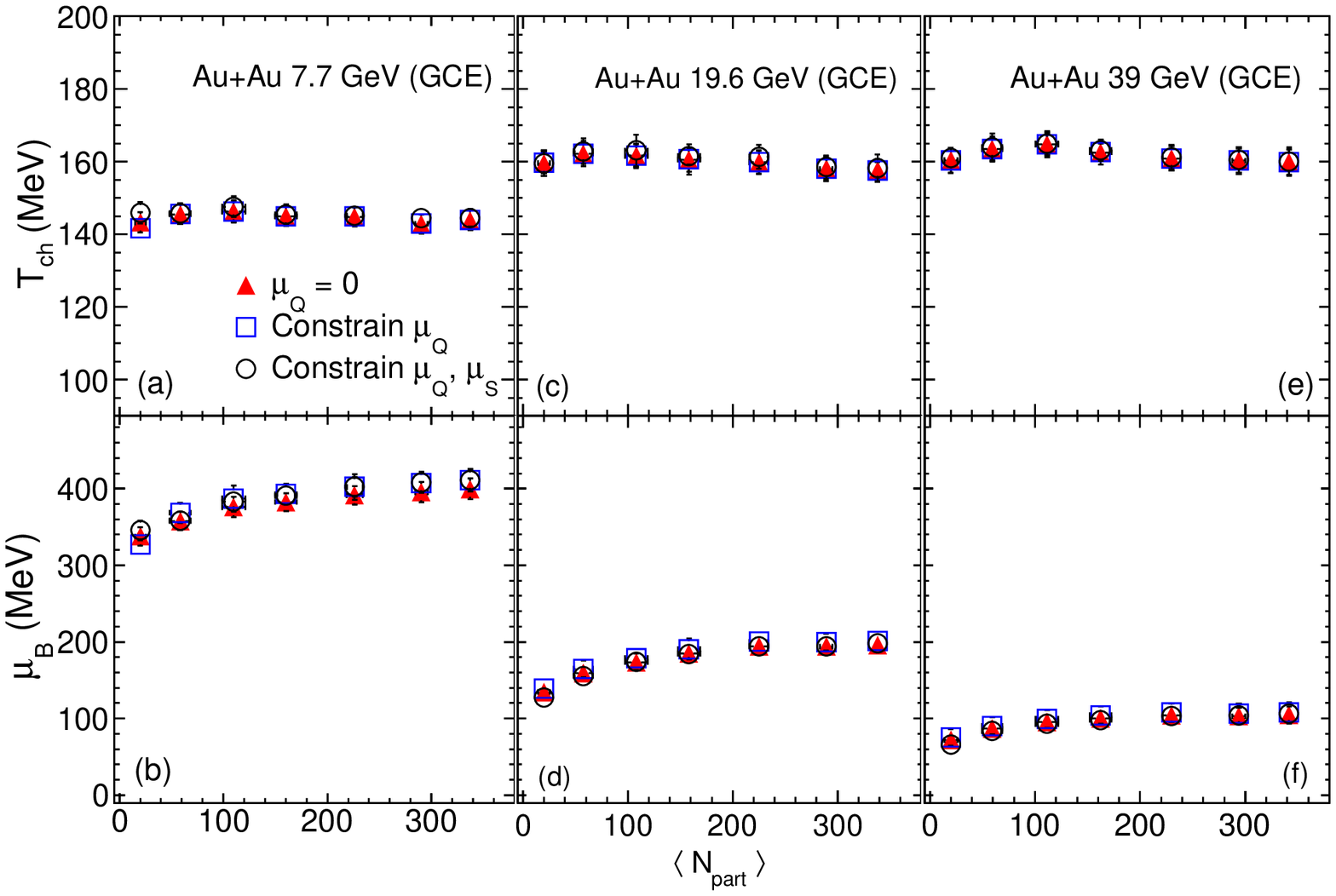}
\vspace{-0.4cm}
\caption{{\it Choice on constraints:} (Color online)
Extracted chemical freeze-out temperature shown in (a), (c), and (e), and baryon
chemical  potential shown in (b), (d), and (f)  for GCE using particle yields as
input for  fitting, respectively for 
Au+Au collisions at $\sqrt{s_{NN}}=$7.7, 19.6, and 
39 GeV. Results are compared for three initial conditions:
  $\mu_Q=0$, $\mu_Q$ constrained to $B/2Q$ value, and $\mu_Q$
  constrained to $B/2Q$ along with $\mu_S$ constrained to 0.
Uncertainties represent  systematic errors.
}
\vspace{-0.5cm}
\label{fig:muq_const}
\end{center}
\end{figure*}
\begin{figure*}
\begin{center}
\includegraphics*[width=6.8cm]{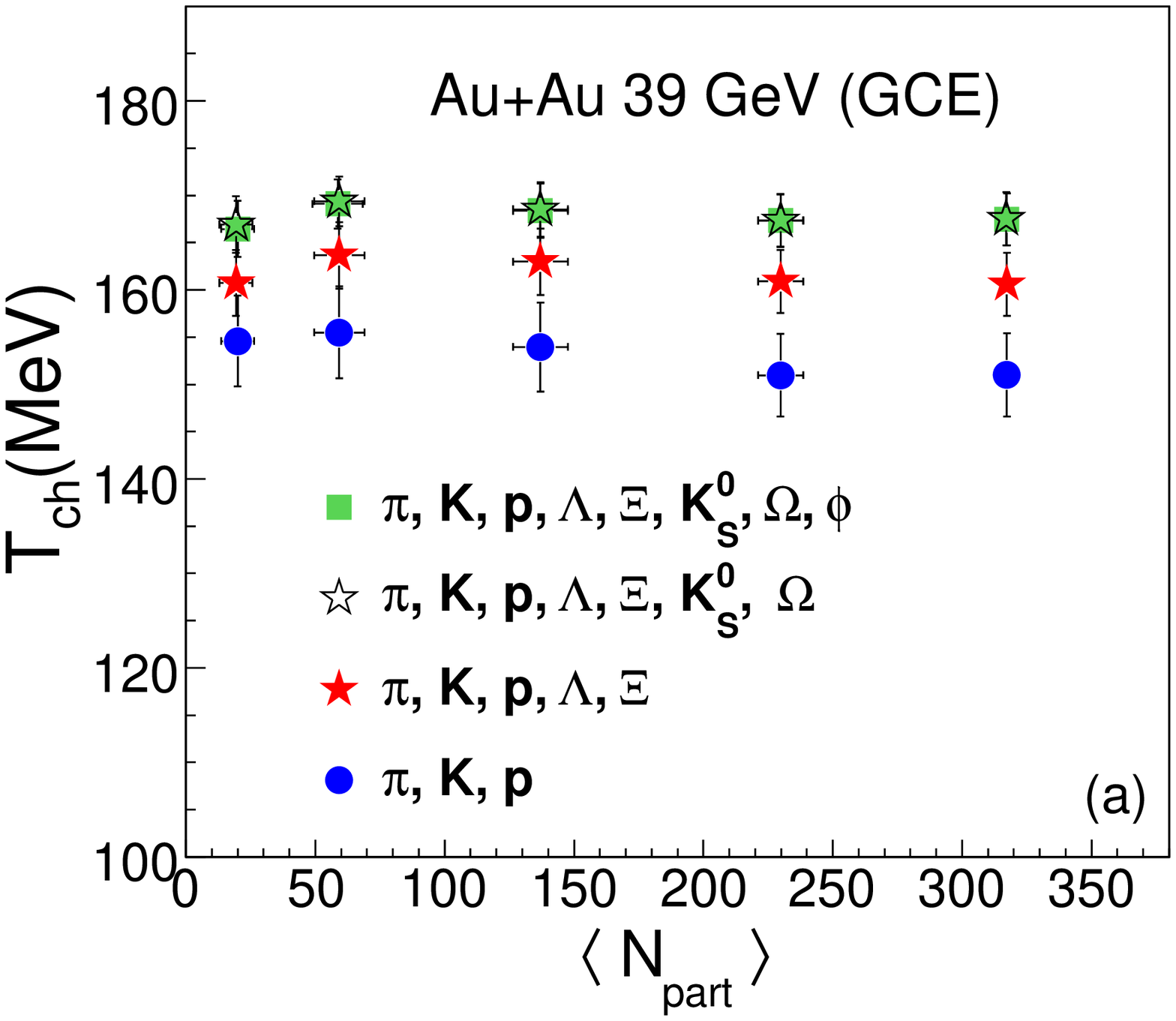}
\includegraphics*[width=6.8cm]{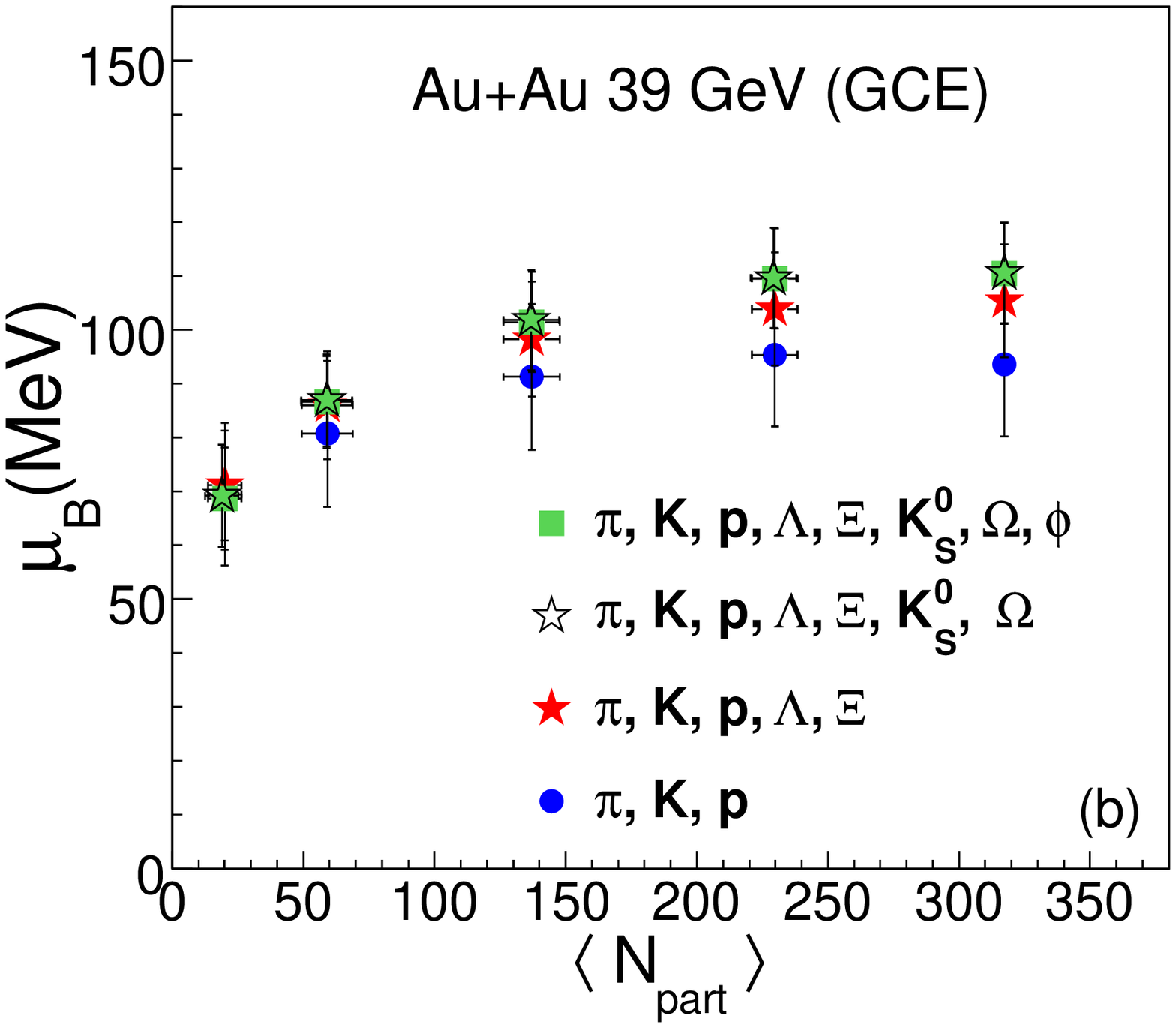}
\includegraphics*[width=6.8cm]{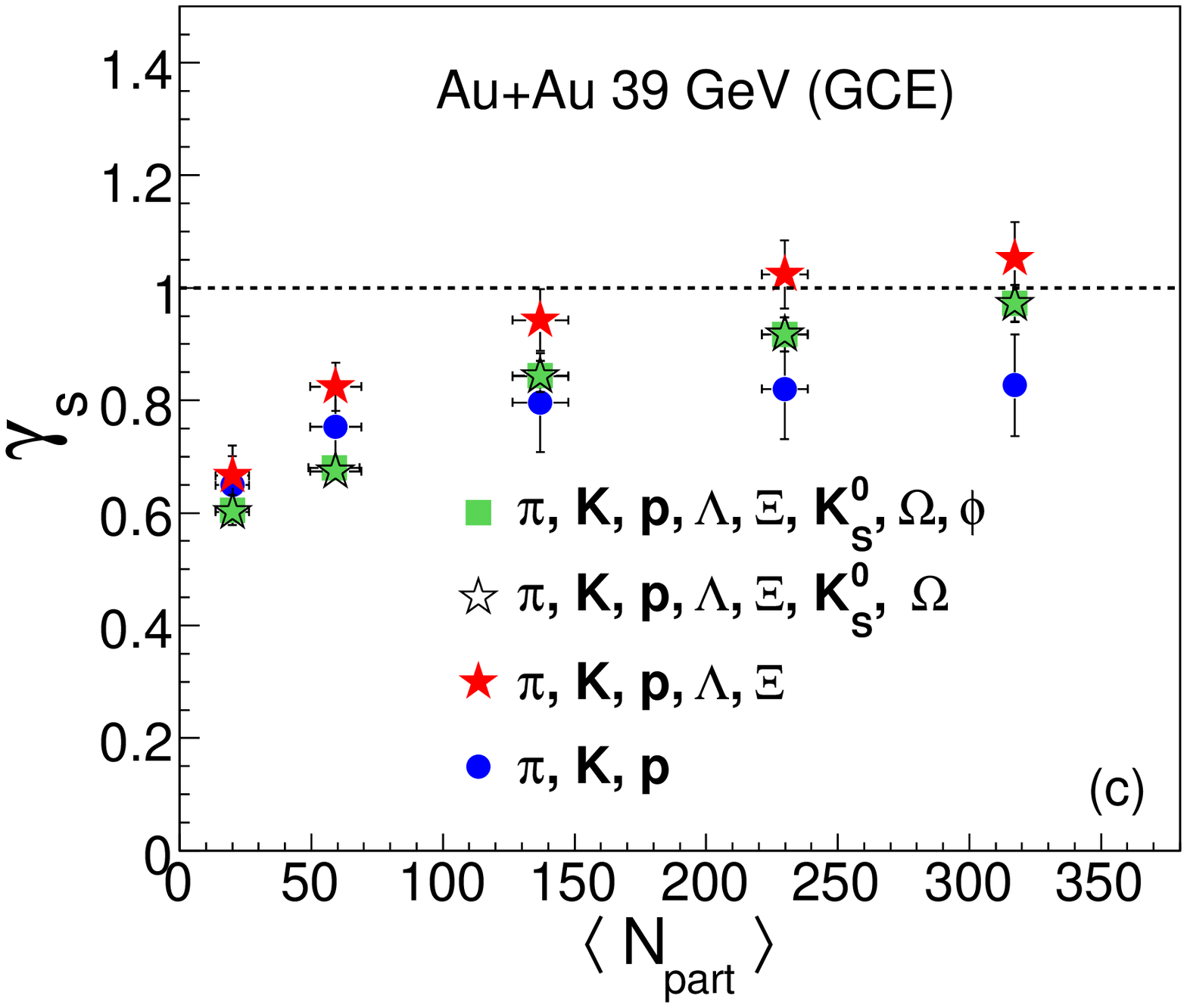}
\includegraphics*[width=6.8cm]{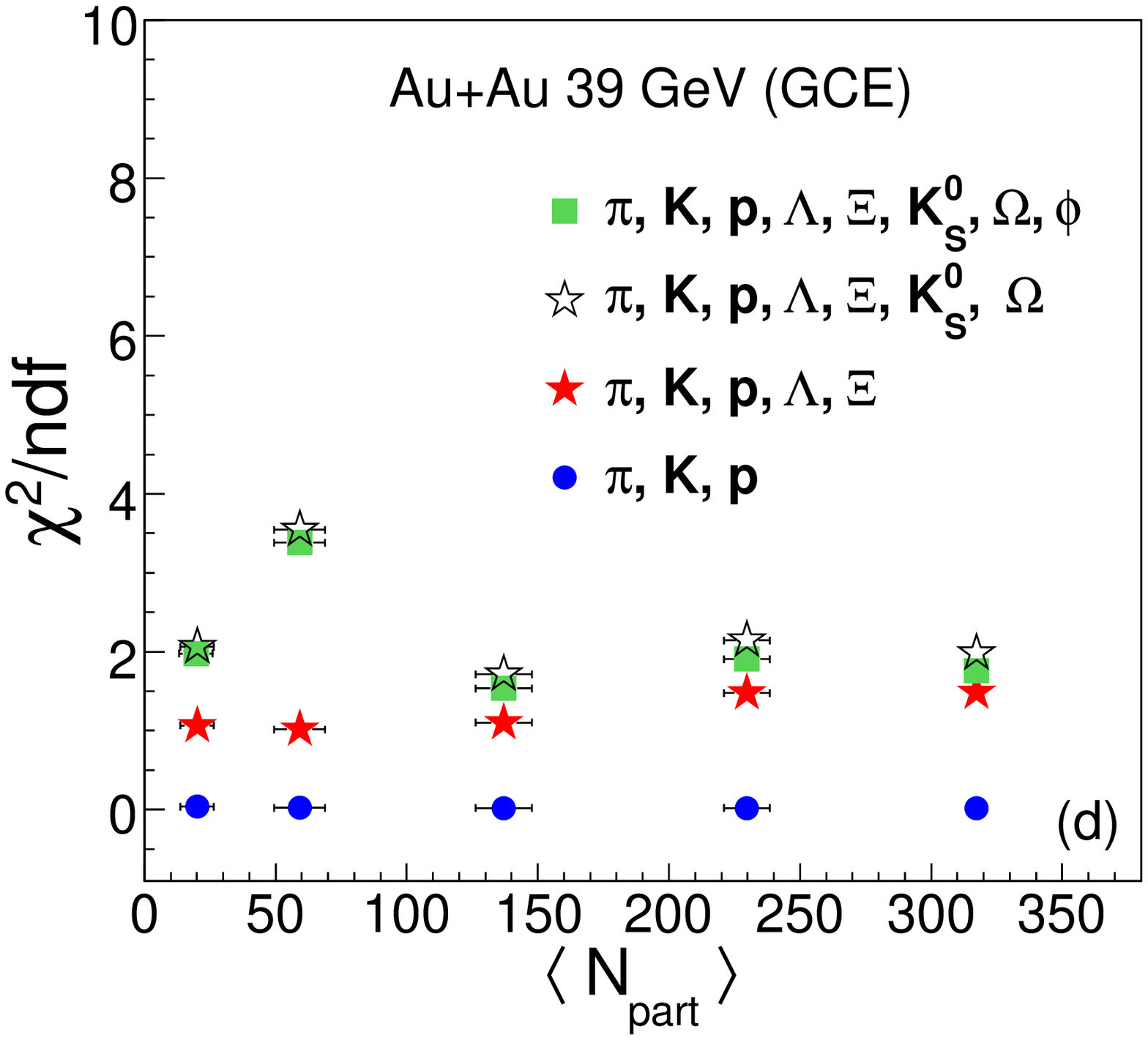}
\vspace{-0.4cm}
\caption{{\it Choice on including more particles:} (Color online) Extracted chemical freeze-out
parameters (a) $T_{\rm{ch}}$,  (b) $\mu_B$, and (c) $\gamma_S$ along with (d) $\chi^2/\rm{ndf}$
for GCE using particle yields as input for
  fitting. Results are compared for Au+Au collisions at
  $\sqrt{s_{NN}}=~$39 GeV for four different sets of 
particle yields used in fitting. Uncertainties represent  systematic errors.
}
\vspace{-0.5cm}
\label{fig:diff_part}
\end{center}
\end{figure*}

Considering the grand canonical case, for a hadron gas of volume $V$ and temperature $T$, the logarithm of
the total partition function is given by~\cite{Wheaton:2004qb},
\begin{eqnarray}
\ln Z^{GC}(T,V,\{\mu_i\})=\sum_{\rm{species}~i} \frac{g_iV}{(2\pi)^3}
\int  d^3p \ln(1\pm \nonumber \\ e^{-\beta(E_i-\mu_i)})^{\pm1}
\end{eqnarray}
where, $g_i$ and $\mu_i$ are degeneracy and chemical potential of
hadron species $i$ respectively, $\beta=1/T$, and $E_i=\sqrt{p^2+m_i^2}$, $m_i$
being the mass of particle. The plus sign corresponds to fermions and
minus sign to bosons. The chemical potential for particle species $i$ in this case is given by
\begin{eqnarray}
\mu_i = B_i \mu_B + Q_i \mu_Q + S_i \mu_S, 
\end{eqnarray}
where $B_i$, $S_i$, and $Q_i$ are the baryon number, strangeness, and
charge number, respectively, of hadron species $i$, and $\mu_B$,
$\mu_Q$, and $\mu_S$ are the respective chemical  potentials. The
particle multiplicities are given by
\begin{eqnarray}
N_i^{GC} = T \frac{\partial \ln Z^{GC}}{\partial \mu_i} =
\frac{g_iV}{2\pi^2}  \sum_{k=1}^{\infty} (\mp1)^{k+1}
\frac{m_i^2T}{k}K_2 \left(\frac{km_i}{T} \right) \nonumber \\ \times e^{\beta k\mu_i}
\end{eqnarray}
where $K_2$ is the Bessel function of second order. 
In the strangeness or mixed canonical ensemble, the partition function
for a Boltzmann hadron gas is given by
\begin{eqnarray}
Z_S = \frac{1}{2\pi} \int_{-\pi}^{\pi} d\phi_S e^{-iS\phi_S} 
\exp[\sum_{\rm{hadrons}~i} \frac{g_iV}{(2\pi)^3}  \int  d^3p
\nonumber \\ e^{-\beta(E_i-\mu_i)}  e^{iS_i\phi_S}] 
\end{eqnarray}
In this case, the chemical potential of hadron species $i$ is given by
\begin{eqnarray}
\mu_i = B_i \mu_B + Q_i \mu_Q 
\end{eqnarray}
and particle multiplicities are given by
\begin{eqnarray}
N_i^S = 
\left. \left(  \frac{Z_{S-S_i}}{Z_S} \right) N_i^{GC} \right|_{\mu_S=0}
\end{eqnarray}

The main fit parameters obtained are the chemical freeze-out
temperature $T_{\rm{ch}}$, baryon chemical potential $\mu_B$, strange
chemical potential $\mu_S$, strangeness suppression factor $\gamma_S$
(to account for observed deviation from chemical equilibrium in
the strangeness sector)~\cite{Becattini:1997ii,Becattini:2003wp,Bearden:2002ib,Cleymans:2001at,Cleymans:2002xu,Cleymans:2003yp,Tawfik:2014dha},
and (canonical) radius parameter ($R_C$) $R$. For fitting in
strangeness canonical ensemble, we have fixed $R_C=R$.
The results presented here are obtained with fixed $\mu_Q=$0.
Tables~\ref{tab:fit_results_gce} and \ref{tab:fit_results_sce} show the fit parameters obtained
in Au+Au collisions at $\sqrt{s_{NN}}=$ 7.7, 11.5, 19.6, 27, 39, 62.4,
and 200 GeV, in various centralities for GCE
and SCE, respectively. 

Figures~\ref{fig:chem_fits} and \ref{fig:chem_fits2}  show the GCE and SCE model fits along
with 
number of standard deviations in the difference between data and model
 for Au+Au 7.7
  and 39 GeV in 0--5\% central collisions, respectively.
Upper panels are for the particle yields and lower panels are for
particle ratios fit. The plots show that fits for particle yields and
ratios are within 2 standard deviations.

Figure~\ref{fig:gce_yr} shows the extracted chemical freeze-out parameters ($T_{\rm{ch}}$,
  $\mu_B$, $\mu_S$, $\gamma_S$, and $R$) plotted versus $\langle
  N_{\rm{part}} \rangle$
  in GCE for particle yields fit. 
The results are
  shown for 7.7, 11.5, 19.6, 27, and 39 GeV. 
We observe that $T_{\rm{ch}}$
  increases from 7.7 to 19.6 GeV and then  remains almost
  constant. For a given energy, the value of  $T_{\rm{ch}}$ is almost
the same for all centralities.
Baryon chemical potential  $\mu_B$ decreases with increasing
  energy and shows centrality dependence for a
  given energy.
The centrality dependence of $\mu_B$ is more significant at lower
energies (7.7--19.6 GeV).
The $\mu_B$  increases from peripheral to central collisions.  This
  behavior is likely due to the stronger baryon stopping at lower energies which
  may also be centrality dependent.
The strangeness chemical potential $\mu_S$ decreases with increasing
energy and also shows a weak increase from peripheral to
central collisions. 
The strangeness suppression factor $\gamma_S$ accounts for the
possible deviations of strange particle abundances from chemical
equilibrium; $\gamma_S$ equal to unity means chemical equilibration
of strange particles. 
The strangeness suppression factor $\gamma_S$ 
for central collisions is almost the same and close to unity for 
  all the energies. However, for peripheral collisions, it is less
  than unity and 
 shows a slight energy dependence, i.e. decreases with decreasing energy. For
  a given energy, it increases from peripheral to central
  collisions.
The radius parameter $R$ is related to the volume of the fireball at
 chemical freeze-out and is obtained for the yield fit case. For the BES energy
 range, the radius parameter shows no energy dependence. 
We note a similar energy dependence of the volume at chemical
freeze-out per unit of rapidity $dV/dy$ for the energy range similar
to BES, as discussed in Ref.~\cite{Andronic:2005yp}. For higher
energies, the $dV/dy$ increases.
The radius parameter shows centrality dependence for a given energy,
increasing from peripheral to central  collisions.

Figure~\ref{fig:gce_yr_r} shows the
ratio of chemical freeze-out parameters ($T_{\rm{ch}}$,
  $\mu_B$, $\mu_S$, $\gamma_S$, and $R$) between results from yield
  fits to ratio fits in GCE plotted versus $\langle
  N_{\rm{part}} \rangle$.
We observe that the extracted freeze-out parameters for GCE
using ratio and yield fits are consistent with each other within uncertainties. 
We found that the results using particle ratios in the fits
have large uncertainties compared to those using particle yields. This
may be
because the particle ratios used for fitting are constructed mostly using common particle
yields, say e.g. pions, which leads to correlated uncertainties, but
we treated all the ratio uncertainties as independent in our fit. 

 Figure~\ref{fig:sce_yr} shows the chemical freeze-out parameters ($T_{\rm{ch}}$,
  $\mu_B$, $\gamma_S$, and $R$) plotted versus $\langle
  N_{\rm{part}} \rangle$
  in SCE for particle yields fit. 
The behavior of the  freeze-out parameters is generally similar to
what we discussed  above for GCE. 
However, $T_{\rm{ch}}$ in SCE seems to be higher in  peripheral collisions, but the centrality dependence is still weak.
Figure~\ref{fig:sce_yr_r}  shows the ratio of chemical freeze-out parameters ($T_{\rm{ch}}$,
  $\mu_B$, and $\gamma_S$) between yield and ratio fits in SCE plotted versus $\langle
  N_{\rm{part}} \rangle$.
We observe that within uncertainties, the results using yield and ratio fits are
similar except for $\gamma_S$ in the most peripheral collision.

Figure~\ref{fig:rat_gs} shows the ratio of chemical freeze-out parameters ($T_{\rm{ch}}$,
  $\mu_B$, and $\gamma_S$) between GCE and SCE results
  obtained using the particle ratio fit plotted versus $\langle
  N_{\rm{part}} \rangle$. Similarly, 
Fig.~\ref{fig:yld_gs} shows the ratio of chemical freeze-out parameters ($T_{\rm{ch}}$,  $\mu_B$, $\gamma_S$, and $R$) between GCE and SCE results
  obtained using particle yields fit plotted versus $\langle  N_{\rm{part}} \rangle$.
 We observe that the results are consistent within uncertainties for
 GCE and SCE using
both the  ratio and yield fits, except  for $\gamma_S$ in the most
peripheral collision in case of yields fit.

Figure~\ref{fig:tmub_all} shows the variation of chemical freeze-out
temperature with baryon chemical potential at various energies and for
three centralities 0--5\%, 30--40\% and 60--80\%. For 62.4 GeV, the
three centralities shown are 0--5\%, 20--40\% and 60--80\%.  The results are
shown for both GCE and SCE cases obtained using particle yields fit.
The curves represent two  model predictions~\cite{Cleymans:2005xv,Andronic:2009jd}.
In general, the behavior is the same for the two cases, i.e. a centrality dependence of baryon chemical potential is observed which is significant at lower energies. 

Next, we test the robustness of our results by comparing to results
obtained with different constraints and using more particles in the fit.

\subsubsection{Choice on Constraints}
The results presented here are obtained assuming $\mu_Q=0$. 
However, we have checked the results by constraining
$\mu_Q$ to the initial baryon-to-charge ratio for
Au+Au collisions, i.e. $B/2Q$=1.25.
We have also checked the results by applying both constraints, i.e. $\mu_Q$
constrained to 1.25 as well as $\mu_S$ constrained to initial strangeness density, i.e. 
0.
Figure~\ref{fig:muq_const} shows the extracted chemical freeze-out
temperature and baryon chemical potential in Au+Au collisions at
  $\sqrt{s_{NN}}=$7.7, 19.6, and 39 GeV for GCE using particle yields
  as input to the fit, for the three conditions mentioned above. It is
  observed that these three different conditions have negligible
  effect ($< 1\%)$ on the final extracted $T_{\rm{ch}}$ and $\mu_B$. The extracted parameters are
similar for these different cases. Similarly, $\mu_S$, the radius parameter,
$\gamma_S$, and $\chi^2$/NDF (plots not shown here), all show similar results for the three
cases discussed above. The same exercise was repeated for the SCE case
and the conclusion remains the same. 

\subsubsection{Choice on Including More Particles}
For the default results discussed above, the particles included in the
THERMUS fit are: $\pi$, $K$, $p$, $\bar{p}$, $\Lambda$, and $\Xi$. 
It is interesting to compare the freeze-out parameters extracted using different
particles sets in the thermal fit.
Figure~\ref{fig:diff_part} shows the comparison
of extracted freeze-out parameters in Au+Au collisions at
$\sqrt{s_{NN}}=$ 39 GeV for GCE using yields as input to the fit. Results
are compared for four different sets of particle yields used as input for
fitting. When only $\pi$, $K$ and $p$ yields are used in fit (as in Ref.~\cite{Abelev:2008ab}), the
temperature obtained is lower compared to other sets that include
strange hadron yields. 
Also, $\gamma_S$ is less than unity, even for central collisions. 
It can be seen that for all other cases, the
results are similar within uncertainties. 
However, the $\chi^2$/NDF increases with increasing
number of particles used for fitting. 

\subsection{Kinetic Freeze-out}
The kinetic freeze-out parameters are obtained by fitting the spectra
with a blast wave model.
The model assumes that the  particles are locally
thermalized at a kinetic freeze-out temperature and are moving with a
common transverse collective flow velocity~\cite{Schnedermann:1993ws,Abelev:2008ab}. Assuming a radially boosted
thermal source, with a kinetic freeze-out temperature $T_{\rm{kin}}$
and a transverse radial flow velocity $\beta$, the $p_T$ distribution of the particles is given by~\cite{Schnedermann:1993ws} 
\begin{eqnarray}
\frac{dN}{p_T \, dp_T}  \propto  \int_0^R r \, dr \, m_T
I_0\left(\frac{p_T \sinh\rho(r)}{T_{\rm{kin}}}\right)  \nonumber  \\
\times K_1\left(\frac{m_T \cosh\rho(r)}{T_{\rm{kin}}}\right),
\end{eqnarray}
where $m_T$ is the transverse mass of a hadron, 
$\rho(r)=\rm{tanh}^{-1}\beta$, and $I_0$ and  $K_1$ are the modified
Bessel functions. We use a radial flow velocity
  profile of the form 
\begin{eqnarray}
\beta=\beta_S(r/R)^n,
\end{eqnarray}
where $\beta_S$ is the
surface velocity, $r/R$ is the relative radial position in the
thermal source, and $n$ is the exponent of flow velocity profile. Average transverse radial flow velocity $\langle \beta \rangle$ can then be obtained from $\langle \beta \rangle =  \frac{2}{2+n} \beta_S$.
Usually $\pi^{\pm}$, $K^{\pm}$, $p$, and $\bar{p}$ particle
spectra are fitted simultaneously with the blast-wave model. Including more
particles such as multi-strange hadrons in the fit would amount to forcing all
the species to freeze-out at the same time which may not be true. 
It has been shown that at top RHIC energy the spectra of multi-strange
particles reflect a higher kinetic freeze-out
temperature~\cite{Xu:2001zj,Adams:2005dq}. This can be interpreted as
diminished hadronic interactions with the expanding bulk matter after chemical
freeze-out. For the results presented here for kinetic freeze-out, we
use $\pi^{\pm}$, $K^{\pm}$, $p$, and  $\bar{p}$ spectra in the
blast-wave model fit. We also note the recent study of separate
fit of positively and negatively charged particles $v_2$ using
a blast wave model~\cite{Sun:2014rda,Adamczyk:2015fum}.

\begin{figure*}
\begin{center}
\includegraphics*[width=15cm]{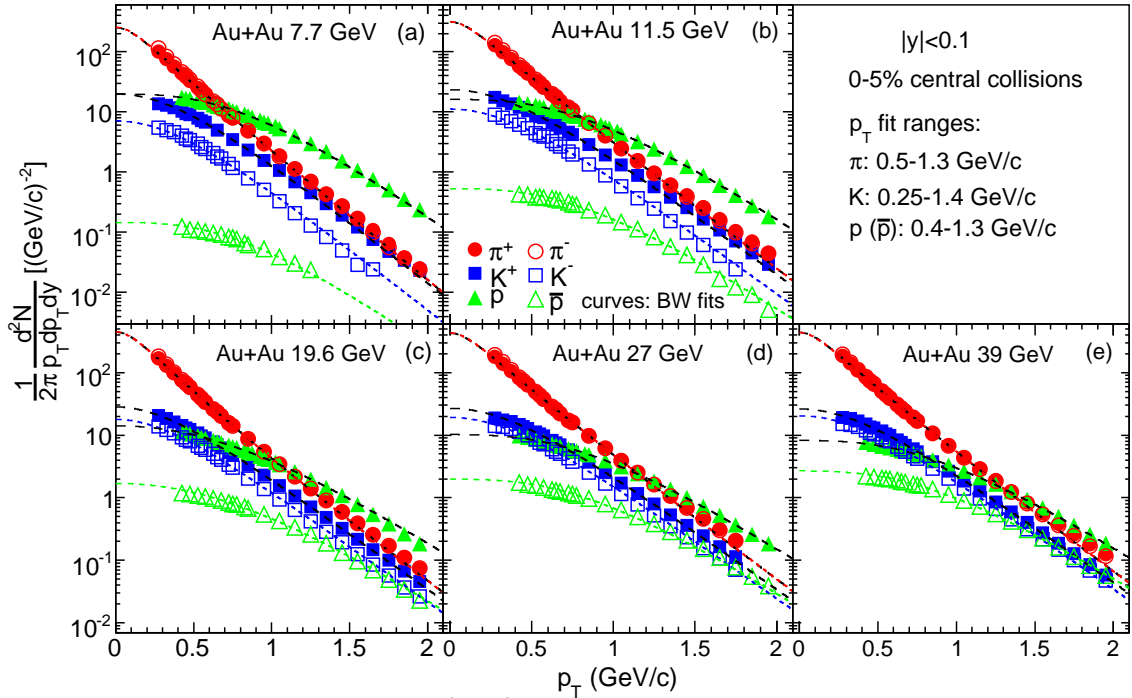}
\vspace{-0.6cm}
\caption{(Color online) Blast wave model fits of $\pi^{\pm}$, $K^{\pm}$,      $p$
  and $\bar{p}$ $p_T$ spectra in 0--5\% central Au+Au collisions at
  $\sqrt{s_{NN}}=$ (a) 7.7 GeV, (b) 11.5 GeV, (c) 19.6 GeV, (d) 27 GeV, and (e) 39 GeV. Uncertainties on
  experimental data represent statistical and systematic uncertainties added in quadrature. Here, the uncertainties are smaller than the symbol size.
}
\vspace{-0.5cm}
\label{fig:pt_spectra}
\end{center}
\end{figure*}
\begin{figure}
\begin{center}
\includegraphics[width=9.cm]{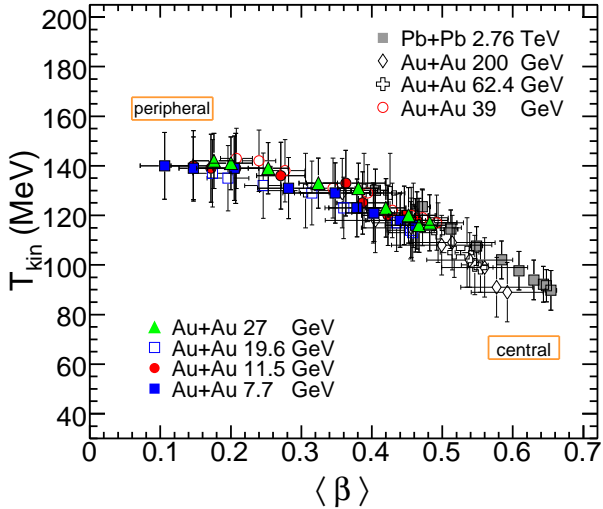}
\vspace{-0.8cm}
\caption{
(Color online)  Variation of $T_{\rm{kin}}$ with $\langle  \beta \rangle$
for different energies and centralities. The centrality increases from
left to right for a given energy. The data points other than BES
energies are taken from Refs.~\cite{Abelev:2008ab,Abelev:2013vea}.
Uncertainties represent  systematic uncertainties.}
\label{fig:tkin_beta}
\vspace{-0.8cm}
\end{center}
\end{figure}
\begin{figure}
\begin{center}
\includegraphics[width=9cm]{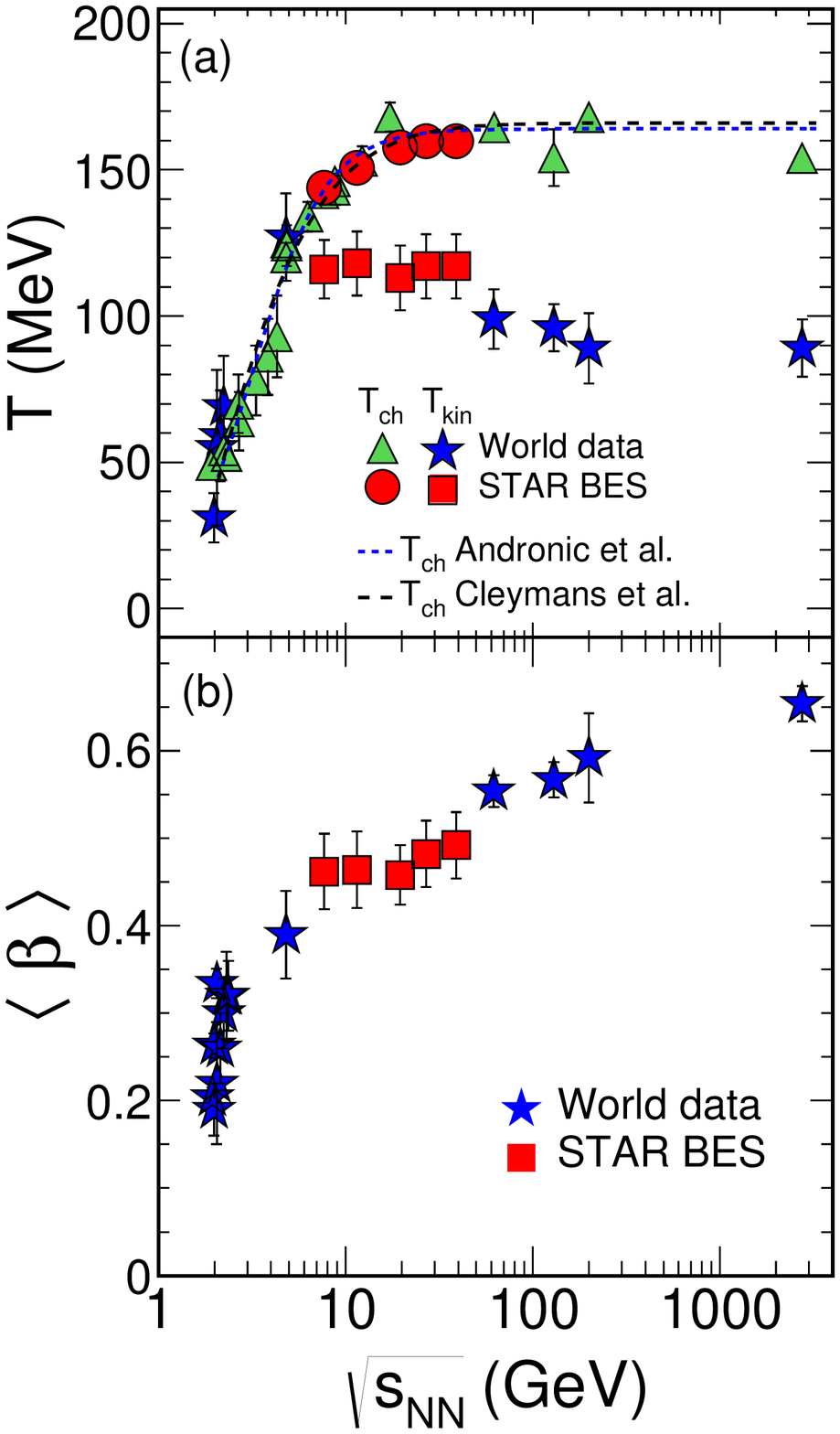}
\vspace{-1.2cm}
\caption{(Color online) (a) Energy dependence of kinetic and
chemical freeze-out temperatures for central heavy-ion collisions. The curves represent various
theoretical predictions~\cite{Cleymans:2005xv,Andronic:2009jd}.
(b) Energy dependence of average transverse radial flow
velocity for central heavy-ion collisions. The data points other than
BES energies are taken from 
Refs.~\cite{Ahle:2000wq,Ahle:1999uy,Klay:2001tf,Barrette:1999ry,Akiba:1996xf,Ahle:1999in,Ahle:1999va,Ahle:1998jc,Afanasiev:2002mx,Alt:2007aa,Alt:2006dk,Anticic:2004yj,Abelev:2008ab,Abelev:2013vea}
and references therein. The BES data points are for 0--5\% central
collisions, AGS energies are mostly for 0--5\%, SPS energies for mostly
0--7\%, and top RHIC and LHC energies for 0--5\% central collisions.
Uncertainties represent  systematic uncertainties.
}
\vspace{-0.4cm}
\label{fig:tb_en}
\end{center}
\end{figure}
Figure~\ref{fig:pt_spectra} shows the 
blast wave model fits of $\pi^{\pm}$, $K^{\pm}$, and
      $p$ and  ($\bar{p}$) $p_T$ spectra in 0--5\% central Au+Au collisions at
      $\sqrt{s_{NN}}=$ 7.7, 11.5, 19.6, 27, and 39 GeV.
The model describes well the $p_T$ spectra of $\pi^\pm, K^\pm, p$, and $\bar{p}$ at
all energies studied. The fit parameters are $T_{\rm{kin}}$, $\langle  \beta \rangle$, and $n$. 
The low $p_T$ part of the pion spectra is affected by resonance
decays, and consequently the pion spectra are fitted only for 
$p_T>$~0.5~GeV/$c$. 
The blast wave model is hydrodynamics-motivated which 
provides a good description of data at low $p_T$, but is not suited for
describing hard processes at high $p_T$~\cite{Wilk:1999dr}. 
Thus the blast wave model results are sensitive to the $p_T$
fit ranges used for fitting~\cite{Abelev:2013vea}. The results
presented here
use similar values of 
low $p_T$ as
were used in previous studies by STAR and ALICE~\cite{Abelev:2008ab,Abelev:2013vea}.  We keep consistent
$p_T$ ranges for simultaneous fitting of the 
$\pi^\pm$, $K^\pm$, $p$, and $\bar{p}$ spectra across all the BES
energies as shown in Fig.~\ref{fig:pt_spectra}. 
The extracted kinetic freeze-out parameters for the BES energies are
listed in Table~\ref{tab:model_pars}.

Figure~\ref{fig:tkin_beta} shows the variation of $T_{\rm{kin}}$ with $\langle  \beta \rangle$
for different energies and centralities. 
The $\langle  \beta \rangle$ decreases from central to peripheral collisions indicating more rapid
expansion in central collisions. On the other hand, 
$T_{\rm{kin}}$ increases from central to peripheral collisions, 
consistent with the expectation of a shorter-lived fireball in
peripheral collisions~\cite{Heinz:2004qz}.
Furthermore, we observe
that these parameters show a two-dimensional
anti-correlation band. Higher values of $T_{\rm{kin}}$ correspond to
lower values of $\langle  \beta \rangle$ and vice-versa. 

Figure~\ref{fig:tb_en} (a) shows the energy dependence of kinetic and
chemical freeze-out temperatures for central heavy-ion collisions. We observe that the values of kinetic and chemical
freeze-out temperatures are similar around $\sqrt{s_{NN}}=$4 -- 5 GeV. If
the collision energy is increased, the chemical freeze-out temperature
increases and becomes constant after  $\sqrt{s_{NN}}=$11.5
GeV. On the other hand,  $T_{\rm{kin}}$ is almost constant around the
7.7--39 GeV and then decreases up to LHC
energies. The separation between $T_{\rm{ch}}$ and  $T_{\rm{kin}}$
increases with increasing energy. This might
suggest the effect of increasing hadronic interactions between
chemical and kinetic freeze-out at higher energies~\cite{Adams:2005dq}.
Figure~\ref{fig:tb_en} (b) shows the average transverse
radial flow velocity plotted as a function of  $\sqrt{s_{NN}}$. 
The $\langle \beta \rangle$ shows a rapid increase 
at very low energies, 
then a steady increase up to LHC energies. The $\langle \beta \rangle$
is almost constant for the lowest three BES energies.



\section{Summary and Conclusions}
We have presented measurements of identified particles 
$\pi,K,p$, and $\bar{p}$ at midrapidity ($|y|<0.1$) in Au+Au collisions at $\sqrt{s_{NN}}=$~7.7, 11.5, 19.6, 27, and 39
GeV from the beam energy scan program at RHIC. 
The transverse momentum spectra of pions, kaons, protons, and anti-protons are presented
for 0--5\%, 5--10\%, 10--20\%, 20--30\%, 30--40\%, 40--50\%, 50--60\%,
60--70\%, and 70--80\% collision centrality classes.
The bulk properties are studied by measuring the identified hadron $dN/dy$, 
$\langle p_{T} \rangle$, particle ratios, and freeze-out
parameters. The results are compared with corresponding published
results from other energies and  experiments. 

The yields of charged pions, kaons, and anti-protons decrease with
decreasing collision energy. However, the yield of protons is higher for
the lowest energy of 7.7 GeV which suggests high baryon stopping at
mid-rapidity at lower energies. 
The yields decrease from central to peripheral collisions
for $\pi^{\pm}$, $K^{\pm}$, and $p$. However, the centrality dependence
of yields for $\bar{p}$ is weak.
The energy dependence of pion yields changes slope as a function of beam energy.
The slope above 19.6 GeV is different when compared to
that at lower energies. 
This may suggest a change in particle production
mechanism below 19.6 GeV. 

The $\pi^{-}$/$\pi^{+}$ ratio is close to unity for most of the
energies. 
The lowest energy of 7.7 GeV has a greater $\pi^{-}/\pi^{+}$ ratio
than at other energies due to isospin and significant contributions
from resonance decays (such as $\Delta$ baryons).
The $K^{-}$/$K^{+}$ ratio increases with increasing
energy, and shows very little centrality dependence. The increase in
$K^{-}$/$K^{+}$ ratio with energy shows the increasing contribution to
kaon production due to pair production.
The $K^{+}$/$\pi^{+}$ ratio
shows a maximum at 7.7 GeV and then decreases with increasing energy. This
is due to the associated production dominance at lower energies as the
baryon stopping is large. 
This maximum corresponds to the maximum baryon density
predicted to be achieved in heavy-ion collisions.
The centrality dependence is similar at all energies, 
increasing from 
peripheral to central collisions. 
The $\bar{p}$/$p$ ratio increases with increasing energy. The ratio
increases from central to peripheral collisions. 
The results reflect the large baryon stopping at mid-rapidity at lower
energies in central collisions.
The $p$/$\pi^{+}$ ratio decreases with increasing energy and is larger
at $\sqrt{s_{NN}} = $ 7.7 GeV. This is again a consequence of the higher degree of baryon stopping for the
collisions at lower energies compared to $\sqrt{s_{NN}} = $ 62.4 and 200 GeV. 

The $\langle m_{T}\rangle - m$
values increase with $\sqrt{s_{NN}}$ at lower AGS energies, stay
independent of $\sqrt{s_{NN}}$ at the SPS and BES energies,
then tend to rise further with 
increasing $\sqrt{s_{NN}}$ at the higher beam energies at RHIC. 
The constant value 
of $\langle m_{T}\rangle - m$ vs. $\sqrt{s_{NN}}$ around BES energies could be
interpreted as reflecting 
the formation of a mixed phase of a QGP and hadrons during the evolution of the heavy-ion
system.

The chemical freeze-out 
parameters are extracted
from a thermal model fit to the data at midrapidity. 
The GCE and SCE approaches are
studied by fitting the particle yields as well as the particle
ratios. 
The results for particle yield fits compared to particle ratio fits
are consistent within uncertainties for both GCE and SCE. The GCE
and SCE results are also consistent with each other for either ratio
or yield fits. The SCE results obtained by fitting particle yields
seem to give slightly higher temperature towards peripheral collisions compared
to that in 0-5\% central collisions. 
The chemical freeze-out parameter $T_{\rm{ch}}$
  increases from 7.7 to 19.6 GeV; after that it remains almost
  constant. For a given energy, the value of  $T_{\rm{ch}}$ is similar for all centralities. In all the cases studied, a
centrality dependence of baryon chemical potential is observed which
is significant at lower energies. 

The kinetic freeze-out parameters are extracted from a blast-wave model fit to pion, kaon, 
proton, and anti-proton $p_{T}$ spectra.  
$T_{\rm{kin}}$ increases
from central to peripheral collisions suggesting a longer lived
fireball in central collisions, while $\langle  \beta \rangle$
decreases from central to peripheral collisions suggesting stronger expansion in central collisions. Furthermore, we observe
that these parameters show a two-dimensional
anti-correlation band. Higher values of $T_{\rm{kin}}$ correspond to
lower values of $\langle  \beta \rangle$ and vice-versa. 
The separation between $T_{\rm{ch}}$ and  $T_{\rm{kin}}$
increases with increasing energy. This might
suggest the effect of increasing hadronic interactions between
chemical and kinetic freeze-out at higher energies.
The $\langle \beta \rangle$ shows a rapid increase 
at very low energies, then a slow increase across the BES energies, after
which it again increases steadily up to LHC energies. 


In conclusion we have studied the bulk properties of matter in the
Beam Energy Scan program at RHIC. The BES program covers the energy
range from 7.7 GeV to 39 GeV which along with top RHIC energy
corresponds to the baryon chemical potential region of 20--400 MeV.
The mid-rapidity yields of identified hadrons have been presented. They show the expected
signatures of a high-baryon stopping region at lower energies. At high
energies, the pair production mechanism dominates the particle production. At intermediate energies there is clearly a
transition between these two regions, which is explored by the BES  
Program. 

The data have been used to analyse both chemical and kinetic freeze-out parameters. The chemical freeze-out
was studied using both GCE and SCE approaches, and the fits were
performed using both particle yields and particle ratios. The results show no
significant difference between these approaches, but indicate
in heavy-ion collisions a clear centrality dependence of the baryon chemical potential at lower
energies. The centrality dependence of the freeze-out parameters provides
an opportunity for the BES program at RHIC to enlarge the ($T,\mu_B$) region of the
phase diagram to search for the QCD critical point. The difference
between chemical and kinetic freeze-out increases with increasing
energy suggesting increasing hadronic interactions after chemical
freeze-out at higher energies. 
We thank the RHIC Operations Group and RCF at BNL, the NERSC Center at
LBNL, and the Open Science Grid consortium for providing resources and
support. This work was supported in part by the Office of Nuclear
Physics within the U.S. DOE Office of Science, the U.S. National
Science Foundation, the Ministry of Education and Science of the
Russian Federation, National Natural Science Foundation of China,
Chinese Academy of Science, the Ministry of Science and Technology of
China and the Chinese Ministry of Education, the National Research
Foundation of Korea, GA and MSMT of the Czech Republic, Department of
Atomic Energy and Department of Science and Technology of the
Government of India; the National Science Centre of Poland, National
Research Foundation, the Ministry of Science, Education and Sports of
the Republic of Croatia, RosAtom of Russia and German
Bundesministerium fur Bildung, Wissenschaft, Forschung and Technologie
(BMBF) and the Helmholtz Association.

\bibliography{bes_pid}


\begin{table*}
\caption{Extracted $dN/dy$ values for $|y|<0.1$ in Au+Au collisions at
  $\sNN = $ 7.7, 11.5, 19.6, 27, and 39 GeV. Quoted errors in parenthesis are the combined statistical and systematic uncertainties.}
\begin{center}
\label{tab:dndy}
\begin{tabular}{|c|c|c|c|c|c|c|c|}
\hline
$\sqrt{s_{NN}}$ (GeV) & \% cross-section & $\pi^{+}$ & $\pi^{-}$ & $K^{+}$&$K^{-}$ & $p$ & $\bar{p}$\\

\hline
&00--05     & 93.4 (8.4) &100 (9.0)  &20.8 (1.7) &7.7 (0.6) &54.9 (6.1) &0.39 (0.05)\\
&05--10   & 76.8 (6.9) &81.9 (7.4) &17.3 (1.4) &6.4 (0.5) &45.4 (5.0) &0.32 (0.04)\\
&10--20 & 58.7 (5.3) &62.9 (5.7) &12.4 (1.0) &4.7 (0.4) &33.4 (3.7) &0.26 (0.03)\\
&20--30 &40.5 (3.7)  &43.3 (3.9) &8.6 (0.7)   &3.2 (0.3) &23.2 (2.6) &0.19 (0.02)\\
7.7 
&30--40 &26.9 (2.4) &29.1 (2.6) &5.3 (0.4)      &2.1 (0.2)      &15.8 (1.7) &0.14 (0.02)\\
&40--50 &17.6 (1.6) &18.8 (1.7) &3.2 (0.3)      &1.3 (0.1)      &9.3 (1.0) &0.09 (0.01)\\
&50--60 &10.9 (0.9) &11.8 (1.1) &1.8 (0.1)      &0.71 (0.06)  &5.4 (0.6) &0.06 (0.01)\\
&60--70 &6.1 (0.6)   &6.6 (0.6)   &0.82 (0.07)  & 0.32 (0.03) &2.8 (0.3) &0.033 (0.004)\\
&70--80 &3.1 (0.3)   &3.4 (0.3)   & 0.33 (0.03) & 0.13 (0.01) &1.4 (0.2) &0.018 (0.002)\\[+1mm]
\hline
&&&&&&&\\
&00--05   & 123.9 (12.4) &129.8 (13.0) &25.0 (2.5) &12.3 (1.2) &44.0 (5.3) &1.5 (0.2)\\
&05--10  & 97.1 (9.7) &102.3 (10.3) &20.6 (2.1) &10.2 (1.0) &35.2 (4.2) &1.2 (0.2)\\
&10--20 & 73.4 (7.4) &77.0 (7.7) &14.8 (1.5) &7.5 (0.7) &26.1 (3.1) &0.9 (0.1)\\
&20--30 &49.5 (4.9) &52.0 (5.2) &9.6 (1.0) &4.9 (0.5) &17.8 (2.1) &0.7 (0.1)\\
11.5 
&30--40 &33.9 (3.4) &35.7 (3.6) &6.1 (0.6) &3.2 (0.3) &11.8 (1.4) &0.5 (0.1)\\
&40--50 &21.3 (2.1) &22.5 (2.3) &3.7 (0.4) &1.9 (0.2) &7.3 (0.9) &0.33 (0.04)\\
&50--60 &12.9 (1.3) &13.6 (1.4) &1.9 (0.2) &1.0 (0.1) &4.2 (0.5) &0.21 (0.03)\\
&60--70 &7.6 (0.8) &7.9 (0.8) &0.98 (0.09) & 0.53 (0.05) &2.1 (0.3) &0.13 (0.02)\\
&70--80 &3.9 (0.4) &4.2 (0.4) &0.46 (0.05) & 0.25 (0.03) &1.0 (0.1) &0.07 (0.01)\\[+1mm]
\hline
&&&&&&&\\
&00--05   & 161.4 (17.8) &165.8 (18.3) &29.6 (2.9) &18.8 (1.9) &34.2 (4.5) &4.2 (0.5)\\
&05--10  & 130.3 (14.4) &133.7 (14.7) &24.3 (2.4) &15.5 (1.6) &29.3 (3.8) &3.4 (0.4)\\
&10--20 & 99.3 (10.9) &102.1 (11.3) &18.0 (1.8) &11.6 (1.2) &21.9 (2.9) &2.7 (0.4)\\
&20--30 & 67.1 (7.4) &68.8 (7.6) &12.3 (1.2) &7.9 (0.8) &14.6 (1.9) &1.9 (0.3)\\
19.6 
&30--40 & 44.8 (4.9) &46.0 (5.1) &7.8 (0.8) &5.2 (0.5) &9.2 (1.2) &1.4(0.2)\\
&40--50 & 28.1 (3.1) &28.9 (3.2) &4.7 (0.5) &3.2 (0.3) &5.8 (0.8) &0.95 (0.1)\\
&50--60 & 17.1 (1.9) &17.6 (1.9) &2.7 (0.3) &1.8 (0.2) &3.3 (0.4) &0.6 (0.1)\\
&60--70 & 9.5 (1.0) &9.7 (1.1) &1.3 (0.1) &0.9 (0.1) &1.8 (0.2) &0.35 (0.05)\\
&70--80 & 5.0 (0.6) &5.2 (0.6) &0.65 (0.06) &0.45 (0.04) &0.8 (0.1) &0.18 (0.02)\\[+1mm]

\hline
&&&&&&&\\
&00--05   & 172.9 (19.1) &177.1 (19.5) &31.1 (2.8) &22.6(2.0) &31.7 (3.8) &6.0 (0.7)\\
&05--10  & 144.3 (15.9) &147.5 (16.3) &25.8 (2.3) &18.7 (1.7) &26.5 (3.2) &5.1 (0.6)\\
&10--20 & 109.4 (12.1) &111.6 (12.3) &19.4 (1.8) &14.5 (1.3) &19.4 (2.3) &4.0 (0.5)\\
&20--30 & 74.3 (8.2) &75.9 (8.4) &12.9 (1.2) &9.8 (0.9) &12.9 (1.5) &2.9 (0.3)\\
27 
&30--40 & 48.8 (5.4) &49.9 (5.5) &8.3 (0.8) &6.2 (0.6) &8.9 (1.1) &2.0(0.2)\\
&40--50 & 30.7 (3.4) &31.5 (3.5) &5.2 (0.5) &3.9 (0.3) &5.6 (0.7) &1.4 (0.2)\\
&50--60 & 18.6 (2.0) &18.9 (2.1) &2.9 (0.3) &2.2 (0.2) &3.2 (0.4) &0.8 (0.1)\\
&60--70 & 10.4 (1.1) &10.6 (1.2) &1.5 (0.1) &1.1 (0.1) &1.7 (0.2) &0.49 (0.05)\\
&70--80 & 5.1 (0.6) &5.3 (0.6) &0.68 (0.06) &0.51 (0.05) &0.8 (0.1) &0.23 (0.03)\\[+1mm]

\hline
&&&&&&&\\
&00--05   & 182.3 (20.1) &185.8 (20.5) &32.0 (2.9) &25.0 (2.3) &26.5 (2.9) &8.5 (1.0)\\
&05--10  & 151.4 (16.7) &155.0 (17.1) &27.0 (2.4) &21.0 (1.9) &22.7 (2.5) &7.4 (0.9)\\
&10--20 & 115.9 (12.8) &118.4 (13.1) &20.3 (1.8) &15.9 (1.4) &17.3 (1.9) &5.4 (0.7)\\
&20--30 & 78.9 (8.7) &80.7 (8.9) &13.6 (1.2) &10.7 (1.0) &11.9 (1.3) &3.9 (0.5)\\
39 
&30--40 & 51.8 (5.7) &52.9 (5.8) &8.8 (0.8) &7.0 (0.6) &7.9 (0.9) &2.8 (0.3)\\
&40--50 & 32.9 (3.6) &33.7 (3.7) &5.4 (0.5) &4.4 (0.4) &4.9 (0.5) &1.8 (0.2)\\
&50--60 & 20.1 (2.2) &20.6 (2.2) &3.2 (0.3) &2.6 (0.2) &2.9 (0.3) &1.2 (0.1)\\
&60--70 & 11.0 (1.2) &11.3 (1.2) &1.6 (0.1) &1.3 (0.1) &1.5 (0.2) &0.64 (0.08)\\
&70--80 & 5.9 (0.7) &6.1 (0.7)   &0.8 (0.07) &0.7 (0.1) &0.7 (0.1) &0.33 (0.04)\\[+1mm]

\hline
\end{tabular}%
\end{center}
\end{table*}

\begin{table*}
\caption{Extracted average transverse momentum $\langle p_T \rangle$
  values for for $|y|<0.1$ in Au+Au collisions at $\sNN = $ 7.7, 11.5, 19.6, 27, and 39
  GeV. Quoted errors in parenthesis are the combined statistical and systematic uncertainties.}
\begin{center}
\label{tab:meanpt}
\begin{tabular}{|c |c |c |c |c |c |c |c|}
\hline
$\sqrt{s_{NN}}$ (GeV) & \% cross-section & $\pi^{+}$ & $\pi^{-}$ & $K^{+}$&$K^{-}$ & $p$ & $\bar{p}$\\ 
\hline
&00--05 & 0.385  (0.019) & 0.376  (0.019) &0.576  (0.024) &0.539  (0.022) & 0.797  (0.064) &0.779  (0.055)\\
&05--10 & 0.381  (0.019) &0.373  (0.019) &0.563  (0.023) &0.532  (0.022) &0.764  (0.061) &0.770  (0.054)\\
&10--20 & 0.380  (0.019) &0.373  (0.019) &0.552  (0.023) &0.521  (0.021) &0.754  (0.060) &0.722  (0.051)\\
&20--30 & 0.374  (0.019) &0.368  (0.019) &0.533  (0.022) &0.506  (0.021) &0.745  (0.060) &0.702  (0.049)\\
7.7 GeV
&30--40 &0.368  (0.019) &0.363 (0.018) &0.528 (0.022) &0.499 (0.021) &0.699 (0.056) &0.657  (0.046)\\
&40--50 &0.357 (0.018) &0.354  (0.018) &0.505 (0.021) &0.470 (0.019) &0.659 (0.053) &0.608 (0.043)\\
&50--60 &0.346 (0.018) &0.344  (0.017) &0.485 (0.020) &0.460 (0.019) &0.617 (0.050) &0.567  (0.040)\\
&60--70 &0.339  (0.017) &0.335  (0.017) &0.472  (0.019) & 0.438  (0.018) &0.585  (0.047) &0.541  (0.038)\\
&70--80 &0.325  (0.016) &0.326  (0.017) & 0.457  (0.019) & 0.427  (0.018) &0.520 (0.042) &0.486  (0.034)\\[+1mm]

\hline
&&&&&&&\\
&00--05 & 0.389  (0.020) &0.382  (0.019) &0.585 (0.030) &0.556  (0.028) &0.798  (0.056) &0.798  (0.064)\\
&05--10 & 0.387  (0.020) &0.380  (0.019) &0.572 (0.029) &0.551  (0.028) &0.794 (0.056) &0.781 (0.063)\\
&10--20 & 0.385  (0.020) &0.380  (0.019) &0.564 (0.029) &0.540 (0.028) &0.766 (0.054) &0.757  (0.061)\\
&20--30 & 0.384  (0.019) &0.379  (0.019) &0.557 (0.028) &0.532 (0.027) &0.755  (0.053) &0.726  (0.059)\\
11.5 GeV
&30--40 & 0.379  (0.019) &0.375  (0.019) &0.550 (0.028) &0.527 (0.027) &0.717 (0.050) &0.688  (0.055)\\
&40--50 &0.372 (0.019) &0.368  (0.019) &0.526 (0.027) &0.503 (0.026) &0.670 (0.047) &0.644  (0.052)\\
&50--60 &0.362  (0.018) &0.360  (0.018) &0.512 (0.027) &0.489 (0.025) &0.636 (0.045) &0.595  (0.048)\\
&60--70 &0.351  (0.018) &0.351  (0.018) &0.495 (0.026) &0.474  (0.024) &0.600 (0.042) &0.559  (0.045)\\
&70--80 &0.343  (0.017) &0.343 (0.017) &0.480 (0.025) &0.447  (0.023) &0.568 (0.040) &0.526  (0.042)\\[+1mm]
\hline
&&&&&&&\\
&00--05 & 0.397  (0.024) & 0.392 (0.024) &0.590 (0.036) & 0.571  (0.035) &0.812  (0.049) &0.834 (0.076)\\
&05--10 & 0.395  (0.024) & 0.391 (0.024) &0.578 (0.035) &  0.562 (0.034) &0.811 (0.049) &0.810 (0.073)\\
&10--20 & 0.395  (0.024) & 0.391 (0.024) &0.575 (0.035) & 0.559 (0.034) &0.787 (0.047) &0.789 (0.071)\\
&20--30 & 0.390  (0.024) & 0.388 (0.023) &0.565 (0.034) &  0.543 (0.033) &0.772  (0.047) &0.758  (0.069)\\
19.6 GeV
&30--40 & 0.385  (0.023) &0.383 (0.023) &0.557 (0.034) & 0.537 (0.033) &0.733 (0.044) &0.732 (0.066)\\
&40--50 & 0.380  (0.023) &0.379 (0.023) &0.533 (0.032) &0.519  (0.032) &0.700 (0.042) &0.692  (0.063)\\
&50--60 & 0.370  (0.022) &0.373 (0.023) &0.520 (0.032) &0.501 (0.030) &0.659 (0.040) &0.647 (0.059)\\
&60--70 & 0.360  (0.022) &0.366 (0.022) &0.502 (0.031) &0.483 (0.029) &0.637 (0.038) &0.610 (0.055)\\
&70--80 & 0.352  (0.021) &0.354 (0.021) &0.490 (0.030) &0.469 (0.029) &0.599 (0.036) &0.577 (0.052)\\[+1mm]
\hline
&&&&&&&\\
&00--05 & 0.409  (0.025) &0.407  (0.025) &0.603 (0.037) &0.581  (0.035) &0.841  (0.051) &0.838  (0.076)\\
&05--10 & 0.406  (0.025) &0.403  (0.024) &0.596 (0.036) &0.575 (0.035) &0.836  (0.050) &0.833 (0.075)\\
&10--20 & 0.404  (0.024) &0.399 (0.024) &0.594 (0.036) &0.567 (0.035) &0.787 (0.047) &0.810 (0.073)\\
&20--30 & 0.401 (0.024) &0.396 (0.024) &0.586 (0.036) &0.556  (0.034) &0.755 (0.046) &0.777  (0.070)\\
27 GeV
&30--40 & 0.400 (0.024) &0.393  (0.024) &0.575  (0.035) &0.553 (0.034) &0.742 (0.045) &0.723 (0.065)\\
&40--50 & 0.393 (0.024) &0.385 (0.023) &0.553 (0.034) &0.535 (0.033) &0.726 (0.044) &0.696 (0.063)\\
&50--60 & 0.380 (0.023) &0.378 (0.023) &0.547 (0.033) &0.524 (0.032) &0.666 (0.040) &0.678  (0.061)\\
&60--70 & 0.372 (0.023) &0.368  (0.022) &0.523  (0.032) &0.506 (0.031) &0.631 (0.038) &0.627  (0.057)\\
&70--80 & 0.363 (0.022) &0.362 (0.022) &0.505 (0.031) &0.488 (0.030) &0.589  (0.036) &0.588  (0.053)\\[+1mm]

\hline
&&&&&&&\\
&00--05 & 0.417  (0.025) &0.413  (0.025) &0.613  (0.037) &0.608  (0.037) &0.860  (0.052) &0.867  (0.096)\\
&05--10 & 0.414  (0.025) &0.410  (0.025) &0.610  (0.037) &0.599  (0.036) &0.838  (0.051) &0.842  (0.093)\\
&10--20 & 0.411  (0.025) &0.408  (0.025) &0.607  (0.037) &0.597  (0.036) &0.828  (0.050) &0.832  (0.092)\\
&20--30 & 0.408  (0.025) &0.405  (0.025) &0.599  (0.036) &0.588  (0.036) &0.812  (0.049) &0.799  (0.088)\\

39 GeV
&30--40 & 0.405  (0.025) &0.403  (0.024) &0.590  (0.036) &0.580  (0.035) &0.766   (0.046) &0.776  (0.086)\\
&40--50 & 0.400  (0.024) &0.394  (0.024) &0.569  (0.035) &0.562  (0.034) &0.750   (0.045) &0.739  (0.082)\\
&50--60 & 0.389  (0.024) &0.387  (0.023) &0.559  (0.034) &0.548  (0.033) &0.704   (0.042) &0.691  (0.076)\\
&60--70 & 0.379  (0.023) &0.378  (0.023) &0.548  (0.033) &0.534  (0.032) &0.665   (0.040) &0.654  (0.072)\\
&70--80 & 0.370  (0.022) &0.370  (0.022) &0.537  (0.033) &0.518  (0.031) &0.633   (0.038) &0.617  (0.068)\\[+1mm]

\hline
\end{tabular}%
\end{center}
\end{table*}


\begin{table*}
\setlength{\tabcolsep}{4pt}
\caption{Extracted chemical freeze-out parameters for Grand Canonical
  Ensemble using both yield (GCEY) and ratio (GCER) fits at different
  centralities in Au+Au collisions at $\sqrt{s_{NN}}=$7.7, 11.5, 19.6,
  27, 39, 62.4, and 200 GeV. Errors in parenthesis  are systematic uncertainties.}
\label{tab:fit_results_gce}
\vspace{-.3cm}
\begin{center}
\resizebox{2.1\columnwidth}{!}{%
\begin{tabular}{c|c|c|c|c|c|c|c|c|c|c|c|c}
\hline
\hline
$\sqrt{s_{NN}}$ & \% cross &\multicolumn{2}{c|}{$T_{\rm {ch}}$~(MeV)} &\multicolumn{2}{c|}{$\mu_B$~(MeV)}&\multicolumn{2}{c|}{$\mu_S$~(MeV)}&\multicolumn{2}{c|}{$\gamma_S$}& \multicolumn{1}{c|}{R~(fm)} &\multicolumn{2}{c}{$\chi^2/{\rm NDF}$}\\[0.1cm]
\cline{3-13}
(GeV) & section & GCER &GCEY & GCER & GCEY & GCER &GCEY & GCER & GCEY & GCEY & GCER & GCEY  \\ [0.05cm]
\hline
7.7 
 & 00--05    & 144.3 (4.8)   &   143.8 (2.7)  &  398.2 (16.4)  &399.8 (13.3)   & 89.5 (6.0) & 90.2 (7.6) & 0.95 (0.08) & 1.05 (0.06) &5.89 (0.33)   &1.4  & 1.3 \\ [0.15cm]
 & 05--10    & 143.0 (4.7)   &   142.9 (2.6)  &  393.5 (15.6)  &395.6 (13.0)   & 88.5 (5.7) & 89.8 (7.5) & 0.95 (0.08) & 1.04 (0.06) &5.65 (0.31)   &1.2  & 1.0 \\ [0.15cm]
 & 10--20    & 143.8 (4.6)   &   144.7 (2.6)  &  388.0 (14.9)  & 391.6 (12.1)   & 86.4 (5.4) & 87.1 (7.0) & 0.88 (0.07) & 0.95 (0.05) & 5.08 (0.27) &0.9  & 1.1 \\ [0.15cm]
 & 20--30    & 143.5 (4.5)   &   144.9 (2.6)  &  379.5 (14.4)  & 382.4 (11.8)   & 85.2 (5.2) & 85.7 (7.0) & 0.85 (0.07) & 0.88 (0.05) & 4.58 (0.24) &0.6  & 1.0 \\ [0.15cm]
 & 30--40    & 145.9 (4.9)   &   146.2 (2.8) &   375.4 (15.3)  & 376.1 (12.9)   & 85.5 (5.7) & 87.6 (7.4) & 0.78 (0.07) & 0.82 (0.05) & 3.95 (0.22) &0.9  & 0.6 \\ [0.15cm]
 & 40--60    & 144.7 (4.7)   &   145.5 (2.7) &   355.6 (13.9)  & 357.8 (12.0)   & 80.3 (5.2) & 82.2 (7.0) & 0.68 (0.06) & 0.71 (0.04) & 3.28 (0.17) &0.7  & 0.9 \\ [0.15cm]
 & 60--80    & 143.4 (4.7)  &   143.3 (2.8)  &   337.5 (13.7)   & 337.8 (12.0)   & 79.3 (5.5) & 79.5 (8.0) & 0.47 (0.04) & 0.49 (0.03) & 2.40 (0.13)&1.0  & 0.7 \\ [0.15cm]
\hline

11.5 
& 00--05  & 149.4 (5.2)  &  150.6 (3.2)   & 287.3 (12.5)   & 292.5 (12.6)   & 64.5 (4.7) &  66.0 (7.6)   & 0.92 (0.09)  &  1.00 (0.06)  &  6.16 (0.36) &1.0    &  1.1   \\ [0.15cm]
& 05--10  & 150.1 (5.4)  &  150.5 (3.2)   & 288.9 (12.9)   & 294.6 (13.1)   & 65.8 (4.9) &  70.0 (7.8)   & 0.96 (0.09)  &  1.04 (0.06)  & 5.69 (0.34) &1.4    &  1.3   \\ [0.15cm]
& 10--20  & 151.8 (5.4)  &  153.1 (3.2)   & 284.9 (12.7)   & 291.6 (12.4)   & 65.1 (4.9)   & 68.6 (7.7)  & 0.92 (0.09)  & 0.98 (0.06)   & 5.02 (0.30) &1.2     & 1.3    \\ [0.15cm]
& 20--30  & 153.5 (5.7)  &  155.9 (3.4)   & 278.7 (12.8)   & 283.6 (12.3)   & 63.9 (5.0)   & 65.6 (7.5)  & 0.85 (0.08)  & 0.88 (0.05)   & 4.31 (0.27) &0.7     & 1.2    \\ [0.15cm]
& 30--40  & 154.6 (5.8)  &  156.9 (3.6)   &  270.1 (12.8)  & 273.8 (12.7)   & 61.9 (5.0)  &  62.9 (7.6)  & 0.78 (0.08)  &  0.82 (0.05)  & 3.76 (0.24) &0.7     &  1.2   \\ [0.15cm]
& 40--60  & 155.3 (5.9)  &  157.9 (3.7)   &  256.0 (12.4)  & 259.2 (12.6)   & 60.2 (5.0)  &  62.5 (7.6)  & 0.69 (0.07)  &  0.71 (0.04)  & 3.02 (0.19) &0.7     &  1.3   \\ [0.15cm]
& 60--80  & 151.6 (5.4)  &  154.3 (3.5)   & 227.3 (10.8)   & 229.4 (12.2)  & 54.6 (4.4)  & 54.6 (7.6)   & 0.52 (0.05)  & 0.54 (0.03)    & 2.26 (0.14) &0.5     &  0.8   \\ [0.15cm]

\hline
19.6 
& 00--05    & 153.9 (5.2) &  157.5 (3.1)    & 187.9 (8.6)& 195.6 (9.7)   & 43.2 (3.8)  & 45.3 (6.3)   & 0.96 (0.09)   &  1.09 (0.05)    & 6.04 (0.35) & 1.3     &1.9  \\ [0.15cm]
& 05--10    & 154.2 (5.3) &  158.0 (3.2)    & 187.2 (8.6)& 193.9 (9.7)   & 43.9 (3.8)  & 45.8 (6.3)   & 0.95 (0.09)   &  1.05 (0.05)    & 5.67 (0.33) & 0.9     &1.4  \\ [0.15cm]
& 10--20    & 155.9 (5.6)  & 159.8 (3.3)    &184.9 (8.8) & 193.9 (9.7)   & 44.4 (3.9)  & 48.1 (6.2)    & 0.92 (0.09)   &   1.00 (0.05)   & 5.08 (0.30) &1.0      &1.4 \\ [0.15cm]
& 20--30    & 156.4 (5.7)  & 160.6 (3.3)    &177.2 (8.5) & 184.9 (9.0)   & 42.6 (3.7)  & 45.5 (5.6)    & 0.91 (0.09)   &   0.95 (0.04)   & 4.49 (0.27) &0.7      &1.2 \\ [0.15cm]
& 30--40    & 157.5 (5.9)  & 161.6 (3.4)    &166.9 (8.5)  & 173.3 (9.3)   & 40.3 (3.7)  & 42.4 (5.7)    &  0.85 (0.08)   &  0.87 (0.04)  & 3.93 (0.24) &0.7      & 1.2 \\ [0.15cm]
& 40--60    & 157.9 (6.0)  & 162.2 (3.5)    &154.4 (8.2)  & 159.4 (9.8)   & 38.0 (3.8)  & 40.1 (6.3)    &  0.77 (0.08)   &  0.76 (0.04)  & 3.19 (0.19) &0.4      & 1.2 \\ [0.15cm]
& 60--80    & 156.2 (5.9)  & 159.6 (3.6)    &133.7 (7.7)  & 134.6 (10.4)   & 33.3 (3.6)  & 32.9 (6.4)    & 0.61 (0.06)    &  0.60 (0.03)  & 2.33 (0.14) & 0.3     & 0.9 \\ [0.15cm]
\hline

27.0 
& 00--05  & 155.0 (5.1)  & 159.8 (3.0)     & 144.4 (7.2)   & 151.9 (9.3)   & 33.5 (3.6)  &36.7 (6.0)  &  0.98 (0.09)   &  1.09 (0.05)   & 6.05 (0.33) &1.3     & 1.7 \\ [0.15cm]
& 05--10  & 155.6 (5.2)  & 160.4 (3.1)     & 143.9 (7.2)   & 151.6 (9.3)   & 34.1 (3.6)  &37.6 (6.0)  &  0.97 (0.09)   &  1.07 (0.05)   & 5.67 (0.31) &1.3     & 1.7 \\ [0.15cm]
& 10--20  & 155.8 (5.2)  & 160.7 (3.0)    & 137.7 (7.0)   & 146.3 (8.8)   & 32.0 (3.6)  & 36.3 (5.8)   & 0.96 (0.09)    & 1.03 (0.05)    & 5.22 (0.29) &1.2    & 1.6 \\ [0.15cm]
& 20--30  & 157.1 (5.4)  & 162.7 (3.1)    & 131.0 (6.9)   & 140.6 (8.3)   & 31.0 (3.5)  & 35.8 (5.4)   & 0.94 (0.09)    & 0.97 (0.04)    & 4.53 (0.25) &1.2    & 1.7 \\ [0.15cm]
& 30--40  & 158.9 (5.7)  & 164.7 (3.4)    & 130.3 (7.2)   & 137.4 (9.1)   & 32.4 (3.6)  & 35.9 (5.6)   & 0.88 (0.09)    &  0.89 (0.04)   & 3.89 (0.23) &1.0     & 1.4 \\ [0.15cm]
& 40--60  & 160.4 (5.9)  & 165.5 (3.5)    & 120.4 (7.1)   & 127.5 (8.9)   & 31.4 (3.6)  & 34.9 (5.7)   & 0.80 (0.08)    &  0.79 (0.03)   & 3.13 (0.18) &0.6     & 1.2 \\ [0.15cm]
& 60--80  & 158.3 (5.8)  & 163.1 (3.9)    & 105.8 (6.8)   & 105.2 (9.5)   & 28.6 (3.4)  & 27.4 (5.9)   & 0.64 (0.06)    &  0.62 (0.03)   & 2.27 (0.15) &0.4    & 1.5 \\ [0.15cm]
\hline

39.0 
& 00--05  & 156.4 (5.4)  &  159.9 (3.5)   & 103.2 (7.4)  & 104.7 (11.2) &  24.5 (3.8) & 23.8 (8.1)    &  0.94 (0.10)     &   1.05 (0.07)    & 6.27 (0.39)  &0.9    & 1.6 \\ [0.15cm]
& 05--10  & 157.0 (5.5)  &  160.3 (3.4)   & 101.9 (7.2)  & 103.1 (10.9) &  24.8 (3.7) & 23.9 (7.8)    &  0.94 (0.10)     &   1.03 (0.07)    & 5.92 (0.35)  &0.7    & 1.2 \\ [0.15cm]
& 10--20  & 156.3 (5.3)  & 160.9 (3.4)    & 101.9 (6.9)  & 103.8 (10.5)  &  24.9 (3.7) & 25.3 (7.3)    &  0.94 (0.09)     &  1.02 (0.06)    & 5.35 (0.31) &0.8    &  1.5  \\ [0.15cm]
& 20--30  & 157.9 (5.5)  & 162.6 (3.4)    & 98.2 (6.7)  & 100.5 (10.1)  &  24.9 (3.6) & 25.8 (6.5)    &  0.92 (0.09)     &  0.97 (0.05)    & 4.65 (0.27) &0.8    &  1.4  \\ [0.15cm]
& 30--40  & 160.8 (6.0) &  164.8 (3.6)    & 94.2 (6.9)   & 95.8 (10.3)   & 24.0 (3.7)  &  24.7 (6.9)   &  0.87 (0.09)     &  0.90 (0.05)    & 3.99 (0.24) &0.5    &  0.9   \\ [0.15cm]
& 40--60  & 160.0 (5.9) &  163.5 (3.5)    & 84.6 (6.6)   & 86.8 (9.9)   & 21.9 (3.6)  &  23.2 (6.7)   &  0.82 (0.08)     &  0.83 (0.04)    & 3.29 (0.19) &0.4    &  1.1   \\ [0.15cm]
& 60--80  & 158.3 (5.9) &  160.4 (3.4)    & 73.0 (6.5)   & 71.9 (10.0)  & 20.3 (3.5)  &  20.3 (6.5)   &  0.67 (0.07)      &  0.67 (0.03)    & 2.41 (0.14) &0.3    &  1.1   \\ [0.15cm]
\hline

62.4  
& 00--05   &  160.3 (4.9) & 164.3 (3.6)     & 69.8 (5.6)  & 69.2 (11.4) & 16.7 (3.3)  & 15.8 (6.8)  & 0.86 (0.06)   & 0.91 (0.05)    &  6.62 (0.36) &2.1      & 3.7    \\ [0.15cm]
& 05--10   &  158.4 (4.4) & 160.0 (3.2)     & 66.1 (5.3)  & 63.8 (9.9)  & 15.7 (3.4)  & 16.1 (6.8)  & 0.87 (0.06)   & 0.91 (0.05)    &  6.62 (0.34) &1.7      & 2.9    \\ [0.15cm]
& 10--20   &  159.0 (4.3) &  161.4 (3.1)    & 65.4 (5.2)  & 63.7 (9.3)   & 15.4 (3.3)   & 13.6 (6.3)& 0.84 (0.06)   & 0.92 (0.05)    & 5.84 (0.29)  &1.8     &  3.4  \\ [0.15cm]
& 20--40   &  159.8 (4.2) &  161.7 (2.9)    & 60.7 (5.2)  & 58.9 (9.1)   & 15.3 (3.2)   & 13.7 (6.3) & 0.84 (0.06)  & 0.91 (0.05)    & 4.86 (0.24) &2.1      &  3.1   \\ [0.15cm]
& 40--60   &  158.1 (4.3) &  160.1 (2.8)    & 54.1 (5.2)  & 53.7 (8.0)   & 12.1 (3.2)   & 10.1 (6.3) & 0.76 (0.06)  & 0.84 (0.04)    & 3.72 (0.19) &1.8      &  3.8   \\ [0.15cm]
& 60--80   &  157.4 (4.2) &  161.7 (2.9)    & 44.6 (5.9)  & 45.4 (8.3)   & 10.3 (3.2)  &  11.5 (6.2) & 0.69 (0.05)  & 0.74 (0.04)    & 2.49 (0.13)  &1.6     &  4.1 \\ [0.15cm]
\hline

200 
& 00--05  &  164.3 (5.3) &  167.8 (4.2)   &  28.4 (5.8)  &  27.0 (11.4)   & 5.6 (3.9)  & 5.6 (8.3)   & 0.93 (0.08)    &  0.95 (0.06)   &7.13 (0.46) &  1.2     &2.7 \\ [0.15cm]
& 05--10  &  163.5 (4.9) &  168.5 (4.0)   &  28.4 (5.5)  &  25.7 (10.9)   & 5.0 (3.6)  & 4.2 (7.5)   & 0.95 (0.08)    &  0.97 (0.05)   &6.50 (0.41) &  1.4     &2.9 \\ [0.15cm]
& 10--20  & 162.4 (4.4)  &  167.8 (3.8)   &  27.7 (5.1)  &  23.2 (10.2)  & 5.9 (3.2)  &  3.0 (6.8)  &  0.94 (0.07)    & 0.99 (0.05)    &5.91 (0.35) &  2.0     &3.9 \\ [0.15cm]
& 20--30  & 163.9 (4.3)  &  167.5 (3.5)   &  27.4 (4.9)  &  23.3 (9.5)  & 6.4 (2.9)  &  4.1 (5.8)  &  0.90 (0.06)     & 0.95 (0.04)    &5.28 (0.29) &  1.8     &3.4 \\ [0.15cm]
& 30--40  & 161.6 (3.9)  &  165.9 (3.5)   &  23.9 (4.8)  &  21.5 (9.7)   &  6.0 (3.1) &  5.6 (6.3)  &  0.90 (0.06)    &  0.93 (0.04)   &4.73 (0.26) & 1.9      &3.2 \\ [0.15cm]
& 40--60  & 162.3 (3.9)  &  165.8 (3.3)   &  22.9 (4.9)  &  21.3 (9.2)   &  5.8 (3.2) &  4.8 (6.6)  &  0.84 (0.06)    &  0.88 (0.04)   &3.85 (0.21) & 1.2      &2.0 \\ [0.15cm]
& 60--80  & 161.3 (3.8) &   163.6 (3.2)   &  18.2 (4.5)  &  18.0 (8.9)  &  5.4 (3.3) & 6.3 (6.1)   &   0.76 (0.05)    & 0.76 (0.03)    &2.81 (0.14) &  0.7     &1.1 \\ [0.15cm]
\hline
\hline
\end{tabular}%
 }
\end{center}

\end{table*}

\begin{table*}
\setlength{\tabcolsep}{4pt}
\caption{Extracted chemical freeze-out parameters for Strangeness Canonical
  Ensemble using both ratio (SCER) and yield (SCEY) fit at different
  centralities in Au+Au collisions at $\sqrt{s_{NN}}=$7.7, 11.5, 19.6,
  27, 39, 62.4, and 200 GeV. Errors in parenthesis are systematic uncertainties.}
\label{tab:fit_results_sce} 
\vspace{-.3cm}
\begin{center}
\resizebox{2.1\columnwidth}{!}{%
\begin{tabular}{c|c|c|c|c|c|c|c|c|c|c}
\hline
\hline
$\sqrt{s_{NN}}$ & \% cross &\multicolumn{2}{c|}{$T_{\rm {ch}}$~(MeV)} &\multicolumn{2}{c|}{$\mu_B$~(MeV)}&\multicolumn{2}{c|}{$\gamma_S$}& \multicolumn{1}{c|}{R~(fm)} &\multicolumn{2}{c}{$\chi^2/{\rm NDF}$}\\[0.1cm]
\cline{3-11}
(GeV) & section & SCER & SCEY & SCER &SCEY & SCER & SCEY & SCEY & SCER & SCEY  \\ [0.01cm]
\hline
7.7 
& 00--05         & 143.9 (2.0)   &   143.5 (2.2)  &  397.5 (8.9)  &400.1 (13.2)   & 0.97 (0.07)    & 1.08 (0.06)    &5.92 (0.29)&  1.4      &1.0 \\[0.08cm]
& 05--10         & 144.2 (2.1)   &   143.2 (2.2)  &  397.7 (8.8)  &395.9 (13.1)   & 0.95 (0.07)    & 1.05 (0.06)    &5.62 (0.28)&  1.2      &0.9 \\ [0.08cm]
& 10--20         & 144.6 (2.0)   &   144.3 (2.1)  &  390.8 (8.4)  &391.8 (12.0)  &  0.89 (0.06)   &  0.98 (0.05)    &5.11 (0.23)&   0.9    &0.9 \\[0.08cm]
& 20--30         & 146.0 (1.9)   &   145.2 (2.1)  &  387.3 (8.0)  &383.0 (11.9)  &  0.84 (0.05)   &  0.91 (0.05)    &4.55 (0.21)&   0.7    &0.8 \\ [0.08cm]
& 30--40         & 148.4 (2.3)   &    147.5 (2.4) &  383.0 (8.8)  &377.6 (13.1)   &  0.78 (0.06)   &  0.84 (0.04)    &3.85 (0.20)&  0.9      &0.6 \\[0.08cm]
& 40--60         & 150.4 (2.4)   &    147.7 (2.5) &  371.7 (8.7)  &361.0 (12.3)   &  0.68 (0.05)   &  0.76 (0.04)    &3.13 (0.16)&  0.9      &0.8 \\ [0.08cm]
& 60--80         & 156.9 (3.2)  &   147.6 (3.0)  &   376.5 (10.2)  &347.4 (12.3)  &  0.48 (0.04)   &  0.65 (0.04)    &2.19 (0.13)&   2.6     &1.5 \\ [0.1cm]
\hline

11.5 
& 00--05  &          152.7 (2.4)  &  151.0 (2.7)   &  294.6 (7.1)&291.7 (12.5)  & 0.90 (0.07)    & 1.0 (0.06)    &6.12 (0.33)&  1.1      &0.9 \\[0.08cm]
& 05--10  &           153.2 (2.9) &  151.8 (2.9)   &  295.8 (9.2)&292.7 (13.1)  & 0.94 (0.08)    & 1.03 (0.06)  &5.58 (0.32)&  1.5      &1.2 \\ [0.08cm]
& 10--20  &         154.3 (2.9)   &  153.6 (2.8)   &  290.4 (9.0)&290.6 (12.3)   &  0.91 (0.07) &  0.99 (0.05) &4.97 (0.28)&     1.2   &1.1 \\[0.08cm]
& 20--30  &         155.9 (3.0)   &  155.9 (2.9)   &  283.6 (9.0)&283.8 (12.1)   &  0.84 (0.06) &  0.90 (0.05) &4.31 (0.23)&     0.7   &0.9 \\ [0.08cm]
& 30--40  &        157.2 (2.6)    &  157.0 (3.1)   &  275.5 (7.5)&274.6 (12.6)   &  0.79 (0.06) &  0.85 (0.05) &3.75 (0.21)&     0.7   &0.9 \\[0.08cm]
& 40--60  &        160.8 (3.5)    &  159.6 (3.4)   &  266.5 (9.3)&259.7 (12.8)   &  0.69 (0.06) &  0.76 (0.04) &2.91 (0.18)&     0.9   &0.9 \\ [0.08cm]
& 60--80  &         166.1 (3.9)   &  158.8 (3.6)   &  254.3 (9.8)&233.2 (12.4)   & 0.54 (0.05) &   0.71 (0.04) &2.05 (0.13)&     2.2   &1.3 \\ [0.1cm]
\hline

19.6 
& 00--05    & 158.6 (3.5) & 157.6 (2.8)    & 192.9 (8.2)   & 194.2 (8.4)   &  0.91 (0.07)   &  1.1 (0.05)   &6.03 (0.33)&  1.3      &1.6 \\ [0.08cm]
& 05--10    & 159.8 (3.7) & 158.3 (2.9)    & 193.3 (8.3)   & 191.7 (8.4)   &  0.89 (0.07)   &  1.06 (0.05)   &5.64 (0.31)&  1.0      &1.1 \\ [0.08cm]
& 10--20    & 161.9 (4.0) & 160.6 (3.0)    & 191.3 (9.2)   &190.3 (8.7)     & 0.87 (0.07)    & 1.00 (0.04)    &5.01 (0.29)& 1.1       &1.3 \\ [0.08cm]
& 20--30    & 162.8 (3.8) & 161.4 (3.0)    & 183.5 (8.3)   & 181.8 (8.2)    & 0.85 (0.07)    & 0.96 (0.04)    &4.42 (0.24)& 0.9       &1.0 \\ [0.08cm]
& 30--40    & 163.9 (4.0) & 162.4 (3.1)    & 172.8 (8.3)   &170.6 (8.4)     &  0.81 (0.06)  &   0.90 (0.04)  &3.86 (0.22)&   0.8     &1.0 \\ [0.08cm]
& 40--60    & 165.7 (4.6) & 163.3 (3.4)    & 161.2 (8.4)   & 155.1 (8.6)   &  0.74 (0.06)  &   0.81 (0.04)  &3.1 (0.19)&   0.8     &0.9 \\ [0.08cm]
& 60--80    & 167.6 (4.9) & 161.7 (3.6)    & 142.8 (8.4)   & 129.1 (8.7)   &  0.62 (0.06)  &   0.76 (0.04)  &2.19 (0.14)&   1.3     &0.7 \\ [0.1cm]
\hline

27.0 

& 00--05  & 159.0 (4.1) & 160.0 (2.9)    & 146.6 (7.4)   & 149.1 (7.5)&  0.94 (0.08)   & 1.1 (0.05)    &6.03 (0.32)&  1.5      &1.4 \\ [0.08cm]
& 05--10  & 160.2 (4.2) & 160.8 (3.0)    & 146.5 (7.5)   & 147.9 (7.7)   &  0.92 (0.08)   & 1.08 (0.05)    &5.64 (0.30)&  1.6      &1.4 \\ [0.08cm]
& 10--20  & 159.6 (4.2) & 161.1 (3.0)    & 139.5 (7.3)   & 142.5 (7.1)&  0.93 (0.08)   & 1.04 (0.05)    &5.17 (0.28)&  1.4      &1.4 \\ [0.08cm]
& 20--30  & 161.4 (4.3) & 163.3 (3.0)    & 132.8 (7.3)   & 136.4 (6.8)    &  0.90 (0.08)   & 0.98 (0.04)    &4.48 (0.25)&  1.4      &1.5 \\ [0.08cm]
& 30--40  & 164.7 (4.6) & 165.6 (3.3)    & 133.3 (7.7)   & 132.8 (7.6)&  0.84 (0.07)   & 0.91 (0.04)    &3.82 (0.22)&  1.4      &1.3 \\ [0.08cm]
& 40--60  & 167.7 (5.2) & 166.8 (3.4)    & 123.7 (7.8)   & 120.4 (7.0)    &  0.76 (0.07)   & 0.84 (0.03)    &3.04 (0.18)&  1.2      &1.1 \\ [0.08cm]
& 60--80  & 168.4 (5.5) & 165.5 (3.8)    & 110.0 (7.8)   & 99.5 (7.6)    &  0.59 (0.05)   & 0.78 (0.03)    &2.12 (0.14)&  1.8      &0.9 \\ [0.1cm]
\hline
39.0 

& 00--05  & 159.1 (4.9) & 159.6 (3.4)    & 104.6 (7.6)   & 104.0 (9.6)&  0.92 (0.09)   & 1.06 (0.07)    &6.29 (0.38)&  1.1       &1.3 \\ [0.08cm]
& 05--10  & 160.2 (5.0) & 160.2 (3.4)    & 103.3 (7.5)   & 102.0 (9.0)&  0.91 (0.09)   & 1.04 (0.07)    &5.93 (0.35)&  0.9       &0.9 \\ [0.08cm]
& 10--20  & 159.7 (4.8) & 160.9 (3.3)    & 102.9 (7.2)   & 101.5 (8.4)& 0.92 (0.08)   & 1.03 (0.06)      &5.34 (0.31)&  1.1      &1.2 \\ [0.08cm]
& 20--30  & 162.1 (5.0) & 162.9 (3.3)    & 99.0 (7.1)   & 96.4 (7.7)& 0.89 (0.08)   & 0.98 (0.05)   &4.63 (0.26)&  1.2      &1.2 \\[0.08cm]
& 30--40  & 164.5 (4.7) & 165.0 (3.5)    & 95.1 (6.6)    &92.7 (7.9)& 0.85 (0.07) &   0.92 (0.05)  &3.96 (0.23)&    0.7    &0.8 \\ [0.08cm]
& 40--60  & 164.0 (5.3) & 164.1 (3.5)    & 85.3 (6.6)    & 82.3 (7.6)&0.81 (0.08) &    0.87 (0.04)  &3.24 (0.19)&    0.8    &0.9 \\ [0.08cm]
& 60--80  & 165.2 (5.5) & 162.7 (3.6)    & 74.6 (6.9)    &66.9 (7.9)&0.72 (0.08) &  0.80 (0.04)&2.27 (0.14)&       1.3 &0.8 \\ [0.1cm]
\hline

62.4  
& 00--05  &161.6 (4.4)  &164.1 (3.6)     & 70.3 (5.7)   &  69.2 (10.9)& 0.86 (0.06)    & 0.92 (0.05)    &6.65 (0.36)&  2.1       &3.7 \\ [0.08cm]
& 05--10  &159.5 (4.0)  &160.0 (3.2)     & 66.4 (5.4)   &  62.6 (9.5) & 0.87 (0.06)    & 0.92 (0.05)    &6.63 (0.34)&  1.8       &2.9 \\ [0.08cm]
& 10--20  &160.1 (3.9)  &161.3 (3.0)     & 65.6 (5.3)   &  63.9 (8.8)&0.85 (0.05)  &   0.93 (0.05)  &5.85 (0.29)&    1.8    &3.4 \\ [0.08cm]
& 20--40  &161.3 (3.9)  &161.7 (2.9)     & 61.0 (5.3)   &  58.5 (8.5)&0.84 (0.05)   &  0.92 (0.05)   &4.85 (0.24)&   2.2     &3.1  \\ [0.08cm]
& 40--60  &159.1 (4.0)  &160.5 (2.9)     & 54.4 (5.2)   &  54.5 (7.4)&  0.78 (0.06)   &  0.87 (0.05)   &3.69 (0.19)& 1.8     &3.8  \\ [0.08cm]
& 60--80  &159.3 (4.0)  &164.0 (3.1)     & 45.6 (6.1)   &  45.8 (7.9)& 0.73 (0.05)  &   0.84 (0.04)   &2.37 (0.13)&  1.9     &4.1  \\ [0.1cm]
\hline

200 
& 00--05    &         163.8 (5.2)  & 167.6 (4.2)    & 28.9 (5.5)&28.1 (8.4)    &  0.94 (0.08)   &  0.95 (0.05)   &7.15 (0.46)&   0.9     &2.2 \\[0.08cm]
& 05--10    &         162.9 (4.8)  & 168.2 (4.0)    & 29.2 (5.1)   &28.5 (7.9)    &  0.97 (0.08)   & 0.98 (0.05)   &6.53 (0.41)&   1.1     &2.5 \\ [0.08cm]
& 10--20  &           162.2 (4.3)  & 167.4 (3.7)    & 27.8 (4.8)&26.9 (7.5)    &  0.95 (0.07)   &  0.99 (0.05)   &5.94 (0.35)&   1.6     &3.2 \\[0.08cm]
& 20--30  &           163.9 (4.2)  & 167.2 (3.5)    & 27.2 (4.6)&25.9 (6.6)    &  0.91 (0.06)   &  0.96 (0.04)   &5.29 (0.29)&   1.5     &2.8 \\[0.08cm]
& 30--40  &           161.8 (3.9)  & 165.8 (3.5)    & 23.6 (4.7)&20.7 (7.4)    &  0.91 (0.06)   &  0.94 (0.04)   &4.73 (0.26)&   1.6     &2.6 \\[0.08cm]
& 40--60  &           162.6 (3.8)  & 164.5 (3.5)    & 22.7 (4.9)   &25.9 (8.1)    &  0.86 (0.06)   &  0.91 (0.04)   &3.91 (0.21)&   1.0     &1.4 \\ [0.08cm]
& 60--80  &           162.2 (3.7)  & 164.1 (3.2)    & 17.4 (4.2)   &15.3 (6.6)    &  0.80 (0.05)   &  0.83 (0.04)   &2.76 (0.14)&   0.8     &0.9 \\ [0.1cm]
\hline
\hline
\end{tabular}%
 }
\end{center}

\end{table*}


\begin{table*}{}
\caption{Extracted kinetic freeze-out parameters in Au+Au 
  collisions at $\sNN = $ 7.7, 11.5, 19.6, 27, and 39 GeV. Quoted
  errors  in parenthesis are the combined statistical and systematic
  uncertainties. 
}
\label{tab:model_pars}
\begin{center}
\begin{tabular}{c|c|c|c|c|c} 
\hline
\hline
$\sNN$ (GeV)  &~~~\% cross-section ~~~&~~~~~$\Tkin$ (MeV) ~~~~~&~~~~~~~~$\langle\beta\rangle c$~~~~~~~~~&~~~~~~~~$n$~~~~~~~~~~&~~~~~$\chi^{2}$/{\sc ndf}~~~~~\\ \hline 
7.7& 00-05   & 116 (11) & 0.462 (0.043) & 0.5     (0.3)  & 0.52\\ 
     & 05-10   & 118 (11) & 0.440 (0.048) & 0.5     (0.3)  & 0.46  \\ 
     & 10-20   & 121 (12) & 0.403 (0.040) & 0.8     (0.3)  & 0.39  \\ 
     & 20-30   & 123 (12) & 0.379 (0.040) & 0.9     (0.3)  & 0.53  \\ 
     & 30-40   & 129 (12) & 0.348 (0.049) & 0.8     (0.4)  & 0.61  \\ 
     & 40-50   & 131 (12) & 0.282 (0.044) & 1.6     (0.6)  & 0.74   \\ 
     & 50-60   & 139 (13) & 0.205 (0.053) & 2.0     (1.4)  & 1.25   \\ 
     & 60-70   & 139 (13) & 0.147 (0.020) & 5.0     (4.8)  & 0.76  \\ 
     & 70-80   & 140 (13) & 0.106 (0.035) & 5.0     (3.4)  & 0.89  \\ 
\hline 
11.5  & 00-05    & 118 (12)          &0.464 (0.044)         & 0.5 (0.3)          &0.26   \\ 
         & 05-10    & 120 (12)          &0.446 (0.046)         & 0.6 (0.3)          &0.24  \\  
         & 10-20    & 120 (12)          &0.423 (0.038)         & 0.9 (0.3)          &0.23  \\ 
         & 20-30    & 125 (13)          &0.387 (0.037)         & 1.0 (0.3)          &0.21  \\    
         & 30-40    & 133 (13)          &0.363 (0.056)         & 0.8 (0.4)          &0.22  \\    
         & 40-50    & 136 (13)          &0.271 (0.034)         & 2.3 (0.5)          &0.27  \\    
         & 50-60    & 139 (14)          &0.207 (0.033)         & 4.1 (1.0)          &0.33  \\    
         & 60-70    & 139 (14)          &0.172 (0.032)         & 5.0 (0.5)          &0.32  \\    
         & 70-80    & 140 (14)          &0.147 (0.032)         & 5.0 (0.3)          &0.78  \\    
\hline 
19.6  & 00-05   &  113 (11)    &0.458 (0.034)         &  0.9 (0.2)         &   0.19     \\ 
         & 05-10   &  114 (12)    &0.455 (0.033)         &  0.9 (0.2)         &0.38       \\  
         & 10-20   &  117 (12)    &0.435 (0.032)         &  1.1 (0.1)         &0.30       \\ 
         & 20-30   &  121 (12)    &0.402 (0.030)         &  1.3 (0.2)         &0.32       \\    
         & 30-40   &  123 (12)    &0.360 (0.026)         &  1.7 (0.2)           &0.40       \\    
         & 40-50   &  129 (13)    &0.315 (0.024)         &  1.9 (0.2)           &0.39       \\    
         & 50-60   &  132 (13)    &0.246 (0.026)         &  3.6 (0.4)           &0.31       \\    
         & 60-70   &  135 (13)    &0.196 (0.029)         &  5.0 (0.2)              &0.51        \\    
         & 70-80   &  137 (14)    &0.174 (0.028)         &  5.0 (0.2)               &1.11         \\    
\hline 
27.0  & 00-05   & 117 (11)   &      0.482 (0.038)&    0.6 (0.2)      &     0.33       \\ 
         & 05-10   & 116 (11)   &      0.467 (0.026)&    0.8 (0.2)         &  0.44          \\  
         & 10-20   & 120 (11)   &      0.452 (0.028)&    0.8 (0.2)         &  0.46         \\ 
         & 20-30   & 123 (12)   &      0.420 (0.028)&    1.1 (0.2)         &  0.34            \\    
         & 30-40   & 131 (10)   &      0.381 (0.029)&    1.2 (0.2)         &  0.28          \\    
         & 40-50   & 133 (10)   &      0.324 (0.027)&    2.0 (0.3)         &  0.22          \\    
         & 50-60   & 139 (10)   &       0.253 (0.028)&   3.3 (0.6)         &  0.13           \\    
         & 60-70   & 141 (11)   &      0.200 (0.031)&    5.0 (0.4)             &  0.17          \\    
         & 70-80   & 142 (11)   &      0.176 (0.029)&    5.0 (0.3)             &  1.01           \\    
\hline 
39.0  & 00-05   & 117 (11)       &0.492 (0.038)       & 0.7 (0.2)          &  0.18\\ 
         & 05-10   & 119 (11)        &0.472 (0.036)      & 0.8 (0.2)             &  0.18  \\  
         & 10-20   & 120 (11)        &0.456 (0.034)      & 1.0 (0.2)             &  0.15  \\ 
         & 20-30   & 122 (11)        &0.429 (0.036)      & 1.2 (0.2)             &  0.14  \\    
         & 30-40   & 129 (11)        &0.394 (0.033)      & 1.4 (0.2)             &  0.11  \\    
         & 40-50   & 131 (12)        &0.345 (0.031)      & 2.0 (0.3)                &  0.11  \\    
         & 50-60   & 138 (13)        &0.277 (0.028)      & 3.1 (0.5)             &  0.10   \\    
         & 60-70   & 142 (12)        &0.240 (0.023)      & 4.0 (0.6)                   &  0.20  \\    
         & 70-80   & 143 (12)        &0.208 (0.022)      & 5.0 (0.3)                 &  0.39  \\
\hline
\hline
\end{tabular} 
\end{center}
\end{table*} 


\end{document}